\documentclass[12pt]{article}
\usepackage{apacite}
\usepackage{threeparttable}
\usepackage{array}
\usepackage{bm}
\usepackage{afterpage}
\usepackage{pdflscape}
\usepackage{graphicx}
\usepackage{longtable}
\usepackage{afterpage}
\usepackage{csquotes}
\usepackage{tikz}
\usepackage{setspace,etoolbox}% http://ctan.org/pkg/{setspace,etoolbox}
\usepackage{amsmath}
\usetikzlibrary{decorations.pathreplacing}
\usepackage{booktabs}
\usepackage{xr}
\usepackage[left=1in,right=1in,bottom=1in,top=0.75in]{geometry}
\begin{document}
\begin{titlepage}
\title{Optimal Model Selection in RDD and Related Settings Using Placebo Zones\thanks{For useful discussions and comments on earlier drafts, we especially thank Marc Chan, Mengheng Li, Timothy J Moore and Akanksha Negi, as well as Josh Angrist, Victoria Baranov, Colin Cameron, Gordon Dahl, Yingying Dong, Denzil Fiebig, Mario Fiorini, Christopher Taber, Thomas Tao Yang, and participants at seminars at Monash University, University of Technology Sydney, University of Sydney, University of New South Wales, the Melbourne Institute, the Society of Labor Economics Meeting, Labour Econometrics Workshop, the Australian Conference of Economists and the Econometric Society Australasian Meeting for helpful feedback and discussions. Any errors are our own.}} 
\smallskip
\author{Nathan Kettlewell\thanks{Email: Nathan.Kettlewell@uts.edu.au} \quad \quad Peter Siminski\thanks{Email: Peter.Siminski@uts.edu.au} \\ University of Technology Sydney} 
\date{\today}
\clearpage\maketitle
\thispagestyle{empty}
\smallskip
\begin{abstract}
We propose a new model-selection algorithm for Regression Discontinuity Design, Regression Kink Design, and related IV estimators. Candidate models are assessed within a `placebo zone' of the running variable, where the true effects are known to be zero. The approach yields an optimal combination of bandwidth, polynomial, and any other choice parameters. It can also inform choices between classes of models (e.g. RDD versus cohort-IV) and any other choices, such as covariates, kernel, or other weights. We outline sufficient conditions under which the approach is asymptotically optimal. The approach also performs favorably under more general conditions in a series of Monte Carlo simulations. We demonstrate the approach in an evaluation of changes to Minimum Supervised Driving Hours in the Australian state of New South Wales. We also re-evaluate evidence on the effects of Head Start and Minimum Legal Drinking Age. Our Stata commands implement the procedure and compare its performance to other approaches.                 
\end{abstract}

%\bigskip
%\textit{This is a draft version of the paper. Please do not cite or circulate without author permission.}
\thispagestyle{empty}
\end{titlepage}

%\raggedright
\newgeometry{left=1.25in,right=1.25in,bottom=1.5in,top=1.5in}
\section{Introduction}\label{sec-intro}

Policy rules frequently create discontinuities in exposure to policies and programs. Regression discontinuity design (RDD) has become a key tool for empirical researchers in these settings \shortcite<see e.g.>[for overviews]{imbens_lemieux_2008,lee_lemieux_2010,cattaneo_idrobo_titiunik_2020}. In the canonical sharp RDD case, the treatment $T$ changes discontinuously from $T=0$ to $T=1$ at some threshold along the running variable $X$. Setting $x=0$ as that threshold, the goal is to estimate the change in the outcome $Y$ at $x=0$:

\begin{equation}\label{eq-tau}
\tau(x) = \underset{x \rightarrow 0^+}{\mathrm{lim}} \ E[Y|X=x] - \underset{x \rightarrow 0^-}{\mathrm{lim}} \ E[Y|X=x]
\end{equation}   

$\tau(x)$ is commonly estimated by local polynomial regression. Researchers select some neighbourhood of observations around $x=0$ (the bandwidth) where $E[Y_{T=0}|X=x]$ and $E[Y_{T=1}|X=x]$ are expected to meet the continuity assumption \shortcite{hahn_etal_2001} and estimate the jump in $Y$ while flexibly controlling for $X$ above and below $x=0$.

RDD is appealing because it facilitates estimation of causal effects under relatively weak assumptions. Moreover, the assumptions for RDD have simple, testable implications \shortcite<see e.g.>{mccrary_2008,cattaneo_jansson_ma_2019}. The ability to visualize RDDs in simple plots of the running variable and outcome \shortcite{calonico_cattaneo_titiunik_2015} also gives an appealing air of transparency to this approach. A number of related estimators extend the basic RDD. The regression kink design (RKD) identifies causal effects by exploiting discontinuous changes in the slope of the running variable under similar assumptions to RDD \shortcite{card_etal_2015}. Fuzzy RDD and RKD deal with situations where only the probability of treatment changes at the threshold, or treatment is a continuous variable. \citeA{dong_2018} suggests a regression probability jump and kink design (RPJKD) for settings where there is both a discontinuous `jump' and/or `kink'. In many related settings it is also possible to fit a global polynomial through the running variable and instrument the treatment using binned means (cohort-IV), as in \citeA{angrist_lavy_1999}.\footnote{The cohort-IV approach can also be used in related situations where there is no clear discontinuity, and yet treatment is a non-smooth function of the running variable. See for example \citeA{imben_van_der_klaaw_1995}, \citeA{bound_turner_2002} and \shortciteA{cousley_etal_2017}.}                
 
While theoretically appealing, when using discontinuity designs researchers face a daunting challenge in selecting a preferred estimator. The choice of bandwidth involves a difficult trade-off between bias and variance. Researchers must also choose what order of polynomial to use, what kernel to use, whether to include covariates, and in some applications what discontinuity model to estimate (e.g. in situations where there is a `jump' and a `kink'). Sometimes it is also useful to adopt different polynomial orders on the left and the right of the threshold, or different bandwidths. Consequently, researchers will typically have thousands of potential estimators to select from, and there is no widely accepted standard for making this choice. In a given application, estimates may vary widely depending on the choices the researcher makes.

Various solutions to model selection have been suggested, but these typically focus on one decision and fix other important decisions. Optimal bandwidth selection has received a lot of attention. In economics, least squares cross validation and plug-in approaches have dominated \cite{imbens_lemieux_2008}. Cross validation methods typically select a bandwidth to minimize the mean squared error of the local polynomial fit. \citeA{ludwig_miller_2005,ludwig_miller_2007} discuss an alternative approach that minimizes error at the boundary, although ultimately reject this method for their application. Early plug-in approaches also focused on the polynomial function's fit, for example the rule-of-thumb approach discussed in \citeA{fan_gijbels_1996} \cite<see also>{lee_lemieux_2010}. In an influential paper, \citeA{imbens_kalyanaraman_2012} (IK) argue that instead of focusing on the global fit of the polynomial function, the bandwidth should minimize the asymptotic mean squared error of the treatment effect (boundary) estimator \cite<see also>{ludwig_miller_2005}. They derive a plug-in algorithm that selects the optimal bandwidth to achieve this. \shortciteA{calonico_cattaneo_titiunik_2014} (CCT) derive a bias correction to IK's method to improve confidence interval estimation. The IK/CCT approach is now popular in applied work.

\shortciteA{card_etal_2017}, however, caution against using the IK/CCT approach as a default and demonstrate through simulations that it may not perform best for a given application.\footnote{\shortciteA{card_etal_2017} suggest that the regularization term used in the IK/CCT plug-in approach may be overly punitive to large bandwidths in practice. The regularization term is used to account for the fact that the curvature parameters for the polynomial fits -- which are parameters themselves in the plug-in formula -- are unknown and must be estimated from the data.} Further, none of these approaches deals with the simultaneous modelling choices researchers need to make. For example, the optimal bandwidth will almost always depend on the polynomial order.\footnote{\citeA{hall_racine_2015} propose a leave-one-out cross validation approach that jointly selects the bandwidth and polynomial order. Their cross validation approach is subject to the issues discussed in IK.} There has been less theoretical development on the question of polynomial choice. \citeA{gelman_imbens_2019} argue that researchers should generally use local linear or quadratic regressions because higher order terms can induce undesirable effects on the estimates.\footnote{\citeA{gelman_imbens_2019} point out that higher order terms can have the practical effect of giving disproportional weighting to certain observations, are generally not selected on the basis of optimizing the objective of boundary estimation, and can lead to misleading inference.} \shortciteA{pei_etal_2020} are less critical of higher order terms and suggest that, conditional on a given bandwidth and other modelling choices, researchers should calculate the implied asymptotic mean squared error for the boundary estimator (similar in spirit to IK/CCT for bandwidth selection). They also show, through a review of recent literature, that most researchers simply default to using local linear estimation.

To understand how researchers are dealing with the challenges of model selection in discontinuity designs, we conducted a review of papers published in leading journals for applied economics research in 2019.\footnote{See \citeA{kettlewell_siminski_2020} Appendix D for further details.} Of the 26 papers we identified, 12 gave no formal rationale for their preferred bandwidth. Of those that did motivate their choice, 13 used IK/CCT; however, many of these merely used the method to `guide' their choice (e.g. by noting that the IK/CCT bandwidth was similar to whatever bandwidth they ultimately used). Almost all studies conducted some kind of sensitivity testing by varying the bandwidth. Only five studies provided any justification for their chosen order of polynomial. Most studies (18) used local linear regression and typically added higher order terms as a robustness check.\footnote{Other common robustness checks included adding covariates, different kernels, `donuts' around the threshold and falsification tests using placebo cut-off points.}

\begin{table}[ht!]
	\caption{Discontinuity studies published in leading journals in 2019}\label{tbl-recent-rd}
	\begin{tabular}{p{7cm} *{3}{p{2.2cm}}}
		\toprule
		 & Sharp RDD & Fuzzy RDD & Cohort-IV   \\
		\midrule
		Papers using this model & 15 & 10 & 2 \\
		\quad \underline{Method for bandwidth choice} & & & \\
		No stated method & 6 & 4 & 2 \\
		IK/CCT & 8 & 6 & 0 \\
		\quad \underline{Method for polynomial choice} & & & \\
		No stated method & 10 & 9 & 2 \\
		Local linear polynomial as baseline & 11 & 8 & - \\
		\quad \underline{Robustness tests} & & & \\
	    Varied bandwidth & 15 & 9  & 1 \\
		Varied polynomial & 13 & 5 & 1 \\
		\bottomrule
	\end{tabular}
	\begin{tablenotes}[para,flushleft]
		\footnotesize
		\item Notes: One paper used both sharp and fuzzy RDD as main specifications, so columns add to more than the sample size ($n=26$). Papers that use spatial or multivariate RDD are included in Sharp RDD or Fuzzy RDD (depending on whether the treatment had complete or partial take-up). No papers used RKD; however, one cohort-IV study did use kink variation as an instrument. See \citeA{kettlewell_siminski_2020} Appendix D for a more detailed overview. 
	\end{tablenotes}
\end{table}

Overall, we surmise that there is no consensus among applied researchers about how to select a preferred model in discontinuity settings. In many cases, researchers seem to be selecting a baseline model either arbitrarily or based on possible defaults like local linear regression. The focus away from emphasizing a preferred model and towards sensitivity analysis may be problematic in certain applications. Further, if there is no clear preferred model, there is also no clarity around the confidence interval for the treatment effect. More generally, the emphasis on robustness tests means that we may be `setting the bar too high' for what constitutes credible evidence in RDD and related contexts.

In this paper we propose a new method for model selection with broad application. Our method allows researchers to select an optimal combination of bandwidth, polynomial, and any other choice parameters he/she wants to consider. It can also be used to choose between competing models (e.g. RDD versus cohort-IV) in certain settings and can accommodate nonlinear models. It relies on using observations of the running variable away from the discontinuity (the placebo zone) as a training ground to assess the performance of candidate models where a `pseudo-treatment' effect is known to be zero. The estimator that minimizes the preferred performance criterion (e.g. lowest root mean squared error) across all pseudo-treatments is then selected as the `best' specification for estimating the actual treatment effect. Our approach is applicable in settings where the point of discontinuity can be reasonably thought of as being randomly chosen from the domain of the running variable.\footnote{While we focus on applications of linear models in the paper, our method naturally extends to nonlinear models like logit and ordered logit \cite<see>[for an alternate IK-type approach to bandwidth selection for categorical dependent variables]{xu_2017}. In nonlinear models the marginal effect for the treatment indicator when $x = 0$ could be used as the target parameter for RMSE minimization, for example.}

We are not the first to recognize the value in placebo zone data. \citeA{imbens_lemieux_2008} suggest testing for jumps at specific psuedo-thresholds as a general test for specification error, a common practice in applied work. \citeA{wing_cook_2013} use the placebo zone to create a kind of differences-in-differences structure, which they argue can improve precision and allow one to learn something about the treatment effect away from the threshold. \citeA{gelman_imbens_2019} use results from the distributions of placebo estimates to inform general advice about higher order terms in RDD studies. Closest to our work is \citeA{ganong_jager_2018}, who suggest a randomization inference approach to hypothesis testing based on the distribution of pseudo-treatment effect estimates (we propose extensions to this procedure). We extend all of this work by using the placebo zone for \textit{ex ante} model selection, which to the best of our knowledge is a new idea.

Our approach also has parallels with studies that use estimates from randomized controlled trials (RCTs) to assess RDD estimators. A prominent example is \shortciteA{hyytinen_etal_2018} who use one such RCT-RDD pair, and conclude that CCT bias-corrected estimators perform well in that application.\footnote{See \shortciteA{chaplin_etal_2018} for a review and meta-analysis of similar studies.} In this literature, as in our approach, the assessment rests on knowing the true parameter that the RDD estimator targets. In that literature, the target is the estimate generated by an RCT. In our case, the target estimate is zero, since there is no actual treatment. Our approach builds on the RCT-RDD approach in three important ways. First, rather than making a single comparison of RDD to RCT estimates, our approach assesses each candidate estimator’s performance repeatedly -- at hundreds or thousands of placebo thresholds throughout the placebo zone. Second, these comparisons serve to inform the choice of estimator to apply within the same context, to estimate the effect of a real treatment using the same data, in a range of the running variable that borders the placebo zone. There is no reason to believe that the best-performing estimator will perform best in other unrelated contexts where the DGP may be completely different, or with other sources of data. Thirdly, the target parameter in the RCT-RDD literature is subject to sampling bias, whereas the placebo-zone target of zero is known with certainty.

Finally, our approach positions point-estimation as a principal goal of model selection in RDD. As emphasised by \citeA{cattaneo_vazquez-bare_2016}, the choice of method for (point) estimation should be seen as distinct from the task of inference, or constructing a valid confidence interval. Our approach is primarily focussed on point estimation, rather than inference. In contrast, a key motivation for using the CCT approach, which has become increasingly popular in applied research, has been that it provides theoretically robust coverage even when there is misspecifcation. Recent work has also been explicit in making confidence interval estimation the primary objective \cite<e.g.>{armstrong_kolesar_2018,armstrong_kolesar_2020}. While inference is important, point estimation is also important. 

Whilst we are primarily interested in point estimation, we also suggest using the placebo zone to assess coverage of associated confidence intervals, and propose a randomization inference procedure which again draws on estimates in the placebo zone. We also assess coverage for our approach in a range of simulations, and obtain favourable results. If desired, one can always use our approach to obtain point estimates and then use other approaches for inference, such as those promoted in the papers above.   

We begin by outlining sufficient conditions under which our selection algorithm is `asymptotically optimal', which is when it converges on selecting the estimator with the lowest mean squared error amongst all candidate estimators as the number of placebo zone repetitions goes to infinity. 

We then conduct Monte Carlo simulations using a number of DGPs. Some of these DGPs are stylized, and others are based on well-known RDD applications. Many of these DGPs depart greatly from the conditions we discuss in our proof of optimality. Our approach has lower RMSE than CCT's estimator in almost every simulation, and always lower RMSE than both CCT and IK estimators with DGPs derived from real data.     

We then demonstrate our approach with a novel evaluation of a policy designed to reduce motor vehicle accidents (MVAs) for young drivers. The policy requires that learner drivers meet a minimum supervised driving hours (MSDH) mandate before being able to drive independently; a common requirement in jurisdictions using graduated driver licencing systems.\footnote{Countries implementing graduated driver licensing include Australia, Canada, New Zealand and the U.S.} We are among the first to causally evaluate the effect of MSDH on MVAs.

In New South Wales, Australia, policy rules created two discontinuities whereby young drivers needed to complete either 0, 50 or 120 MSDH depending on their birth cohort and date of obtaining license. There are apparent discontinuities in both the level and slope of the first-stage relationship between treatment and date of birth. We could estimate a global polynomial model like \citeA{angrist_lavy_1999} (cohort-IV), RKD, RDD, or RPJKD, and it is \textit{a priori} unclear which approach we should adopt. More importantly, within each model type, we need to make important functional form and bandwidth choices. In this setting, there are also good reasons to consider models with both asymmetric bandwidths and asymmetric polynomial orders. Institutional details prevent long bandwidths on the left (but not on the right) of the threshold. Institutional details also result in complete compliance on the right, but strong non-linearity on the left, of the threshold. In total we consider almost 10,000 different estimators considering model type, functional form and bandwidth. 

Somewhat surprisingly, our `best' estimator is a month-of-birth cohort-IV with linear trend. A mixed order RPJKD also performs well, and indeed performs best when asymmetric bandwidths are allowed. Strikingly, the root mean squared error is about five times greater across the placebo zone if we use the bandwidths suggested by CCT rather than the best performing RDD or RKD bandwidths. In a different application, \shortciteA{card_etal_2017} come to a similar conclusion, drawing on Monte Carlo simulations. When comparing across model types, our `best' estimator is a month-of-birth cohort-IV with linear trend. A mixed order RPJKD also performs well, and indeed performs best when asymmetric bandwidths are allowed.

We find that going from 0 to 50 MSDH lowers the probability of an MVA in the first year of independent driving by 1.4 percentage points (21\%). This estimate is highly statistically significant and robust to the randomization inference procedure that we propose in this paper.

We also re-evaluate evidence on the effect of Head Start on child mortality \cite{ludwig_miller_2007} and the minimum legal drinking age on drinking behavior \shortcite{lindo_siminski_yerokhin_2016}. For both applications, the best performing model is linear RDD with a relatively long bandwidth (much longer than in the original papers) and CCT estimators perform considerably worse than our best models in the placebo zone. 

Our Stata command -pzms- implements the placebo zone model selection algorithm, and our proposed approach to randomization inference.\footnote{The program can be installed by typing ``ssc install pzms'' into the Stata command window.} We recommend that researchers consider using our approach whenever the researcher has access to a sufficiently wide placebo zone. Our companion Stata command -pzms\_sim- can be used to easily compare the performance of our approach to CCT in simulations using DGPs tailored to any specific application.\footnote{pzms\_sim can be downloaded at: \\ https://drive.google.com/file/d/1rtNn2McXKbrY\_lQybpDhHujWN8bk7X\_S/view.}  

The remainder of the paper is structured as follows. In Section \ref{sec-theory} we outline sufficient conditions under which our method is asymptotically optimal. Section \ref{sec-sims} presents results from Monte Carlo simulations. In Section \ref{sec-model-selection} we illustrate in detail how to apply the placebo-zone approach in our evaluation of MSDH laws. Section \ref{sec-inference} discusses using the placebo zone for randomization inference. Section \ref{sec-results} presents results for our MSDH evaluation which adopt the chosen estimators.  Section \ref{sec-further-apps} presents a re-evaluation of Head Start and minimum legal drinking age studies. Section \ref{sec-discussion} concludes and discusses practical considerations and recommendations for using the placebo zone approach. 

\section{Asymptotically Optimal Model Selection Using a Placebo Zone}\label{sec-theory}
Following standard sharp-RDD notation, consider a random sample ($Y_i(0)$, $Y_i(1)$, $X_i$), $i = 1, 2,\dots , n$, where $Y_i(0)$,$ Y_i(1)$ are potential outcomes, with and without treatment. Treatment ($T$) is determined by the forcing variable exceeding a threshold at $X = 0$ so that $T_i = \textbf{1}(X_i \geq 0)$. The observed sample is therefore ($Y_i$, $X_i$), where $Y_i = (1- T_i )Y_i(0) + T_i Y_i(1)$.

The parameter of interest is the average treatment effect at the threshold $\tau = E[Y_i(1) - Y_i(0)|X_i = 0]$.

There are many candidate estimators of $\tau$. Amongst the set of such candidates, assume the existence of a single estimator which has the lowest MSE($\hat{\tau}$) of all candidate estimators. For each candidate estimator, MSE($\hat{\tau}$) is not observed. Intuitively, our approach will be useful if the set of placebo estimates are informative of MSE($\hat{\tau}$) for each candidate specification.

In Appendix A, we prove that our model selection approach is `asymptotically optimal' under a set of sufficient conditions. By `asymptotically optimal', we mean that it converges on selecting the estimator with the lowest MSE amongst all candidate estimators of $\tau$, as the number of placebo zone repetitions, $m$, gets large, keeping constant the distance between consecutive placebo thresholds. Perhaps the most important of these conditions is that the conditional expectations of $Y_i(0)$ and $Y_i(1)$ are continuous and have zero fourth derivatives with respect to $X$. In other words, that the global DGP is cubic, or a lower order polynomial. The proof also requires homoskedasticy and uniformly distributed $X$.

While these sufficient conditions are restrictive, they are not intended to be realistic. Rather, it serves as a baseline case. More importantly, as we show with simulations in Section \ref{sec-sims}, the approach continues to perform favourably under major violations of these assumptions, and with a finite number of placebo estimates.

\section{Monte Carlo simulations}\label{sec-sims}
In this section we present the results of Monte Carlo simulations, in which we illustrate the performance of our approach under various conditions, and compare this to popular bandwidth selection algorithms. While our approach can be used to choose between candidate estimators that vary on a range of dimensions, its main use is likely to be as an aid for choosing bandwidth and polynomial order, for RDD estimators. We therefore focus on these designs and choices in this section. First we consider a range of stylised DGPs, and then turn to realistic DGPs, based on well-known applications.   

\subsection{Stylized DGPs}
We commence with some simple DGPs. In each case the sample size is 900 observations. For observation $i$, the running variable $x$ is equal to $i - 100.5$. The running variable is therefore uniformly distributed across the range (-100, 800).\footnote{This structure is common for RDD applications with sample sizes of this order. In particular, this structure arises when a larger original data set has been collapsed into a smaller data set in which each observation represents the mean of $y$ within a particular range of $x$.} The outcome variable is given by 

\begin{equation}
y = 0.3(x>0) + f(x) + \epsilon
\end{equation}

where 0.3 is the discontinuity at $x = 0$, and $\sigma^2 = 0.1^2$ (representing `large' error variance), or $0.03^2$ (`small' error variance).\footnote{The size of the discontinuity has no bearing on the results. 0.3 was chosen for presentational purposes.} $f(x)$ is either: 

Linear: $f(x) = x/400$

Quadratic: $f(x) = (x/400)^2$

Cubic: $f(x) = (x/400)^3$

Sine: $f(x) = \sin(2\pi x/400)/2$

Cosine: $f(x) = \cos(2\pi x/400)/2$

Linear, quadratic and cubic DGPs were chosen because these are theoretically ideal conditions for our approach. The sine and cosine functions were chosen to represent DGPs that do not satisfy assumption (3) in Section \ref{sec-theory} (zero fourth order derivative) and are dissimilar near the treatment threshold to the placebo zone. More specifically, the 3rd derivative of the sine curve takes its maximum value at the threshold. In contrast, the 3rd derivative of the cosine curve takes its minimum value at the threshold. 

The resulting ten DGPs are depicted in Figure C1. Each panel shows the deterministic component of $f(x)$ as well as a scatter plot with one entire simulated data set (the data generated in the first iteration of the simulation). The vertical lines at $x=400$ show where the placebo zone is truncated when we test the performance of each approach with a smaller placebo zone. Given that our approach relies on asymptotics for the number of placebo replications, we expect it to perform better when we use a longer placebo zone.

The results of the simulations are shown in Panels A-E of Table \ref{tbl-sim-stylized-dgps}. For each DGP, we show the RMSE of the estimated discontinuity across 1,000 iterations for our approach (labelled KS), as well as CCT and IK with polynomial order 1.\footnote{We focus on polynomial order 1 since this is the typical specification used with these methods. Moreover, we found that across all our specifications, CCT with order 1 always outperformed CCT with order 2 (results available on request). We also ran versions of CCT and IK without the regularisation term. The RMSEs from these versions were similar but usually slightly higher than those shown for the DGPs in Panels A-E. The exception is for the cosine function, where the IK RMSE was much higher without the regularisation term.} We also show the average `optimal' bandwidth across iterations, as selected by each approach. For our approach, we also show the share of iterations in which the `optimal' estimator is linear (as opposed to quadratic). For our approach, we set the maximum bandwidth to 300 for the `long' placebo zone trials, and 200 for the `short' placebo zone trials. In each candidate model, the bandwidths are set to be symmetrical, until reaching 100 units (which is equal to the full support on the left side of the discontinuity). At higher (right side) bandwidths, the left bandwidth is fixed at 100 units. We discuss the choice of maximum bandwidth further in the conclusion.
\begin{singlespace}
	
	\begin{longtable}[t]{p{4cm}*{7}{p{1.1cm}}}
		\caption{Results of Monte Carlo simulations}\label{tbl-sim-stylized-dgps} \\
		\toprule
		& \multicolumn{3}{c}{KS}       & \multicolumn{2}{c}{CCT} & \multicolumn{2}{c}{IK}        \\
		\cmidrule(lr){2-8}
		& RMSE   & mean BW & linear (\%) & RMSE       & mean BW    & RMSE       & mean BW \\
		& \multicolumn{7}{c}{A: Linear DGP} \\
		\cmidrule(lr){2-8}
		Baseline DGP                          & \textbf{0.0241} & \textbf{250.19} & \textbf{1.000} & 0.0413          & 66.42          & 0.0295          & 149.85          \\
		Small error variance                  & \textbf{0.0072} & \textbf{250.19} & \textbf{1.000} & 0.0092          & 83.01          & 0.0079          & 201.92          \\
		Small placebo zone                    & \textbf{0.0292} & \textbf{151.53} & \textbf{0.974} & 0.0601          & 34.70          & 0.0319          & 120.94          \\
		Small placebo zone and error variance & 0.0088          & 151.53          & 0.974          & 0.0180          & 34.69          & \textbf{0.0087} & \textbf{149.19} \\
		& \multicolumn{7}{c}{B: Quadratic DGP} \\
		\cmidrule(lr){2-8}
		Baseline DGP                          & \textbf{0.0285} & \textbf{114.30} & \textbf{0.999} & 0.0413          & 66.37          & 0.0331          & 137.67          \\
		Small error variance                  & \textbf{0.0088} & \textbf{98.63}  & \textbf{0.995} & 0.0124          & 65.87          & 0.0166          & 149.54          \\
		Small placebo zone                    & \textbf{0.0318} & \textbf{118.92} & \textbf{0.969} & 0.0601          & 34.70          & 0.0332          & 115.63          \\
		Small placebo zone and error variance & \textbf{0.0096} & \textbf{98.54}  & \textbf{0.953} & 0.0180          & 34.70          & 0.0116          & 117.95          \\
		& \multicolumn{7}{c}{C: Cubic DGP} \\
		\cmidrule(lr){2-8}
		Baseline DGP                          & \textbf{0.0301} & \textbf{93.59}  & \textbf{1.000} & 0.0414          & 66.38          & 0.0325          & 114.81          \\
		Small error variance                  & \textbf{0.0104} & \textbf{79.49}  & \textbf{1.000} & 0.0125          & 65.94          & 0.0109          & 89.25           \\
		Small placebo zone                    & \textbf{0.0321} & \textbf{100.39} & \textbf{0.953} & 0.0601          & 34.69          & 0.0340          & 106.87          \\
		Small placebo zone and error variance & \textbf{0.0110} & \textbf{82.66}  & \textbf{0.919} & 0.0180          & 34.66          & 0.0111          & 83.62           \\
		& \multicolumn{7}{c}{D: Sine DGP} \\
		\cmidrule(lr){2-8}
		Baseline DGP                       & 0.0481 & 59.0    & 0.857     & 0.0605     & 34.1       & \textbf{0.0443}     & \textbf{60.8}    \\
		Small error variance                   & \textbf{0.0172} & \textbf{41.6}     & \textbf{0.886}     & 0.0187     & 30.8       & 0.0174     & 56.5   \\
		Small placebo zone                          & 0.0633 & 79.9    & 0.729     & \textbf{0.0605}     & \textbf{34.1}       & \textbf{0.0443}     & \textbf{60.8}   \\
		Small placebo zone and error variance & 0.0207 & 58.6    & 0.673     & \textbf{0.0187}     & \textbf{30.8}       & \textbf{0.0174}     & \textbf{56.5}   \\
		& \multicolumn{7}{c}{E: Cosine DGP} \\
		\cmidrule(lr){2-8}
		Baseline DGP                                & 0.0446 & 83.9    & 0.640     & 0.0601     & 34.7       & \textbf{0.0377}     & \textbf{75.1}    \\
		Small error variance                    & 0.0147 & 42.1    & 0.876     & 0.0180     & 34.7       & \textbf{0.0114}     & \textbf{71.0}    \\
		Small placebo zone                          & 0.0437 & 82.5    & 0.579     & 0.0601     & 34.7       & \textbf{0.0377}     & \textbf{75.1}    \\
		Small placebo zone and error variance & 0.0153 & 41.8    & 0.778     & 0.0180     & 34.7       & \textbf{0.0114}     & \textbf{71.0}    \\
		& \multicolumn{7}{c}{F: Head Start DGP} \\
		\cmidrule(lr){2-8}
		Mortality                                   & \textbf{0.6794} & \textbf{13.86}   & \textbf{0.999}     & 1.4265     & 7.93       & 0.9853     & 14.87   \\
		& & & & & & & \\
		& \multicolumn{7}{c}{G: Political incumbency DGP} \\
		\cmidrule(lr){2-8}
		Wins                                        & \textbf{0.0111} & \textbf{21.59}   & \textbf{0.988}     & 0.0125     & 22.76      & 0.0119     & 30.35   \\
		& & & & & & & \\
		& \multicolumn{7}{c}{H: MLDA DGP} \\
		\cmidrule(lr){2-8}
		Ever Drinks                                 & \textbf{0.0271} & \textbf{3.81}    & \textbf{0.981}     & 0.0591     & 0.96       & 0.0347     & 2.50    \\
		Drinks Regularly                            & \textbf{0.0340} & \textbf{3.96}    & \textbf{0.983}     & 0.0750     & 0.96       & 0.0425     & 2.89    \\
		Proportion of Days Drinks                   & \textbf{0.0154} & \textbf{3.96}    & \textbf{0.983}     & 0.0339     & 0.96       & 0.0189     & 3.15 \\
		\bottomrule   
	\end{longtable}
	\begin{tablenotes}[para,flushleft]
		\footnotesize
		\item Notes: For panels A-D, the sample sizes are set to 900 observations with the running variable uniformly distributed between (-100,800). For the `small placebo zone' we truncate at 400. The outcome variable is given by the equation $y = 0.3(x>0) + f(x) + \epsilon$ with $\sigma^2 = 0.1^2$ (representing `large' error variance), or $0.03^2$ (`small' error variance). For panel A: $f(x) = x/400$. For panel B: $f(x) = (x/400)^2$. For panel C: $f(x) = (x/400)^3$. For panel D: $f(x) = sin(2\pi x/400)/2$. For panel E: $f(x) = cos(2\pi x/400)/2$. For our approach (KS), we set the maximum bandwidth to 300 for the `long' placebo zone trials, and 200 for the `short' placebo zone trials. In each candidate model, the bandwidths are set to be symmetrical, until reaching 100 units. At higher (right side) bandwidths, the left bandwidth is fixed at 100 units. For panels F-H, the simulated datasets are constructed by fitting a 5th order global polynomial through the support of the original data, allowing a discontinuity and a kink at the threshold, and then fitting a beta distribution to the same data to summarise the distribution of the running variable. In each iteration, the sample size is the same as the original sample, and the variance of the error term is the same as variance of the residual from the regression in the first step. The column `linear' shows the fraction of iterations where KS selects linear (rather than quadratic) as the best estimator. For each DGP we conduct 1,000 Monte Carlo simulations. The results for the estimator with the lowest RMSE for each DGP are shown in bold font.
	\end{tablenotes}
\end{singlespace}

A main feature of Table \ref{tbl-sim-stylized-dgps} is that KS attains a lower RMSE than CCT in almost every version of the simulation. The CCT bandwidth (and its performance) are relatively insensitive to most of the parameters of the simulations: $f(x)$, the error variance, and the size of the placebo zone.

The relative performance of KS is best in the linear, quadratic and cubic DGPs, especially with large SEs and large bandwidth. Here, the RMSE is always considerably lower for KS than CCT, and in all but one case lower than IK. In all of these simulations, our algorithm selects linear functions in a large majority (over 91\%) of the iterations.

For the sine DGP, the results are more mixed. The IK performs best in 3 of the 4 versions. KS also performs quite well, and is actually wins when the placebo zone is short and error variance is low. For the Cosine function, IK performs best and CCT worst. KS performs reasonably well, especially when the error variance is large.

We conclude from this that KS outperforms the alternate approaches when the DGP is linear, quadratic or cubic, even when the placebo zone is not overly long. Even with very unstable DGPs, and relatively small placebo zones, our approach still performs reasonably well.   

\subsection{DGPs based on prominent applications}\label{sub-sec-sims-real-degps}
We now turn to DGPs which are designed to mimic realistic scenarios, drawing on three well known applications -- Head Start \cite{ludwig_miller_2007}, political incumbency \cite{lee_2008}, and Minimum Legal Drinking Age (MLDA) \cite{lindo_siminski_yerokhin_2016}.\footnote{The MLDA context is one of the best known applications of RDD, beginning with \citeA{carpenter_dobkin_2009}. \citeA{carpenter_dobkin_2009} used restricted variables from the NHIS, which are not easily available. Instead, we draw on data from \shortciteA{lindo_siminski_yerokhin_2016}'s corresponding analysis for the Australian state of New South Wales.} In each case, we take the following approach. Using the original data from each application, we fit $f(x)$: a 5th order global polynomial through the support of the data, allowing a discontinuity and a kink at the threshold, which allows the treatment effect to be heterogeneous in a way that satisfies assumption (4) in Section \ref{sec-theory} (i.e., the average treatment effect has a zero second derivative with respect to $X$).\footnote{Other papers which have undertaken similar exercises have taken the approach of fitting 5th order polynomials on either side of the threshold. We do not believe this is appropriate for the present exercise. Fitting 5th order polynomials on either side results in discontinuities in every derivative of $y$ w.r.t. $x$. It can also result in a wildly unstable fit on either side of the threshold -- but a much smoother fit elsewhere \cite<see for example>[Figure 1, Model 2]{calonico_cattaneo_titiunik_2014}. This is only realistic under a violation of the usual assumption of ``smoothness'' of all other determinants of $y$, or when the ATE is a highly non-linear function of $X$. This is not appropriate for testing our approach, which relies on the assumption that data patterns away from the threshold can sometimes be informative about likely patterns near the threshold. Nevertheless, our approach still outperforms the other candidate approaches in a majority of the cases discussed here, even when a 5th order polynomial is fitted on either side (results available on request).} \footnote{ For the political incumbency application, we follow the precedent of IK by first dropping observations where the running variable $< -.99$ or $>.99$.} We then fit a beta distribution to the same data to summarize the distribution of the running variable.

For each iteration of the simulation, the sample size is set equal to the original sample. We randomly draw values of the running variable from the beta distribution. Finally, we set $y = f(x) + \epsilon$, where $\epsilon$ is normally distributed with zero mean and variance equal to the variance of the residuals from the regression in the first step.

The resulting DGPs are depicted in Figure C2. Each panel shows $f(x)$ and a scatter plot with the full data set generated in the first iteration of the simulations.

For the Head Start application, in each iteration we allocate each placebo zone observation randomly into one of two groups. This is to address the fact that the density is approximately twice as large in the placebo zone as the treatment zone. We expand on this in Section \ref{sec-further-apps}.

The results of these simulations are shown in Panels F-H of Table \ref{tbl-sim-stylized-dgps}. The key result is that KS outperforms the other approaches for each application, while CCT is consistently last.\footnote{We also ran versions of CCT and IK without the regularisation term (available on request). The RMSEs from these versions were slightly lower than those shown for the DGPs in Panels E-G, but never enough to change the RMSE rankings of the three approaches.} We conclude from this that the KS approach outperforms the other approaches on `realistic' DGPs based on well known applications.

\subsection{Coverage} \label{sub-sec-coverage}
The primary objective of our approach is optimal point estimation. However, it is worth recognizing that an important contribution of CCT is to pioneer `robust confidence intervals' for RDD designs, since conventional standard errors do not  guarantee correct coverage rates. We therefore report coverage results from our simulations in Table C1, where confidence intervals for the KS method are constructed using conventional asymptotic standard errors clustered at units of the running variable. We also report results for our method using a novel randomization inference approach based on the placebo estimates \cite<building on>{ganong_jager_2018}, which we outline in Section \ref{sec-inference}. 

The results can be summarized as follows. When our approach is combined with the conventional inference procedure, coverage is close to 95\% for each of the 'realistic' DGPs, with smaller average confidence intervals than other methods, especially CCT. Similarly, for each of the stylized DGPs except sine, our approach with conventional inference also achieves coverage close to 95\%, often closer to 95\% than CCT. Our confidence intervals are also markedly shorter than CCT (sometimes less than half the length). The only DGP where conventional inference does not perform well with our approach is the sine DGP. The randomization inference procedure also performs well, with similar coverage to CCT for most DGPs, and better coverage in the majority of the DGPs considered. Overall, our approach generally performs better or at least similarly on coverage using either conventional inference or randomization inference than CCT, while also having much shorter confidence intervals.    
 
\section{An application: minimum supervised driving hours}\label{sec-model-selection}
We now demonstrate our method in detail, in the context of a novel application -- estimating the effectiveness of learners' permit policy changes in New South Wales.\footnote{Our study received ethics approval from the UTS Human Research Ethics Committee (Application number ETH17-1547).} Online Appendix B describes institutional details, including the two policy changes and key features of the data. The policy reforms provide exogenous variation in the probability of being subject to MSDH, as a function of date of birth (DOB).What makes this application so interesting is there are many classes of estimators (e.g., RDD, RKD, cohort IV) that can be used to estimate treatment effects in this setting, as we will show. We use our method to compare the performance of estimators within each class, as well as between classes.

We begin by describing the many credible candidate models which could be applied to estimate the effect of the policy changes. We then describe the `placebo zone'. This is a set of 2,556 consecutive DOBs (from 1 July 1984 to 30 June 1991). Within this zone, there is no reason to suspect any systematic relationship between DOB and the outcome variable (MVAs within 1 year of obtaining a provisional license). It therefore provides an opportunity for testing the performance of candidate models in estimating the true treatment effect within this zone (which is zero). Next, we summarize the performance of the candidate models within this zone.\footnote{In the working paper version of this paper, we also consider the implications of various types of treatment effect heterogeneity which we impose into the placebo zone data \cite{kettlewell_siminski_2020}.}

\subsection{Candidate models}
The first-stage relationship between DOB and holding a `new' learner's permit following the 2000 reform (0 to 50 MSDH) is shown in Figure \ref{fgr-fs-84} (see Appendix Figure C3 for the 2007 reform, and Figure C4 for scatter plots showing the reduced form for both reforms). Both panels draw on the same underlying data, differing only in the bin-size used in the plots. Panel A uses a `small' bin-size of 2 days, while Panel B uses a `large' bin size of 30 days. 

\begin{figure}[ht!]
	\centering
	\caption{First-stage relationship between DOB and 50 MSDH treatment}\label{fgr-fs-84}
	\begin{tabular}{cc}
	\includegraphics[width=.5\textwidth]{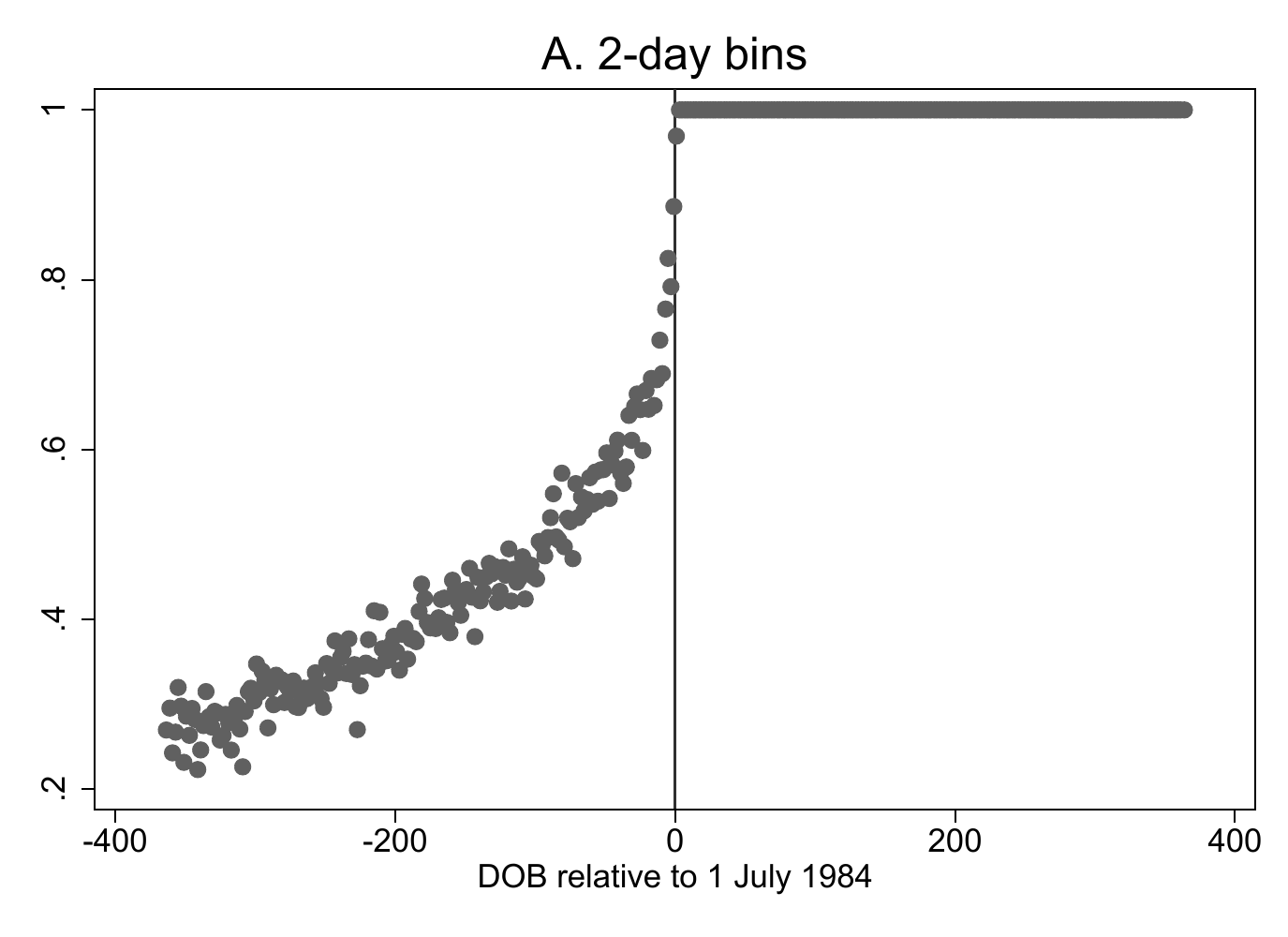} &
	\includegraphics[width=.5\textwidth]{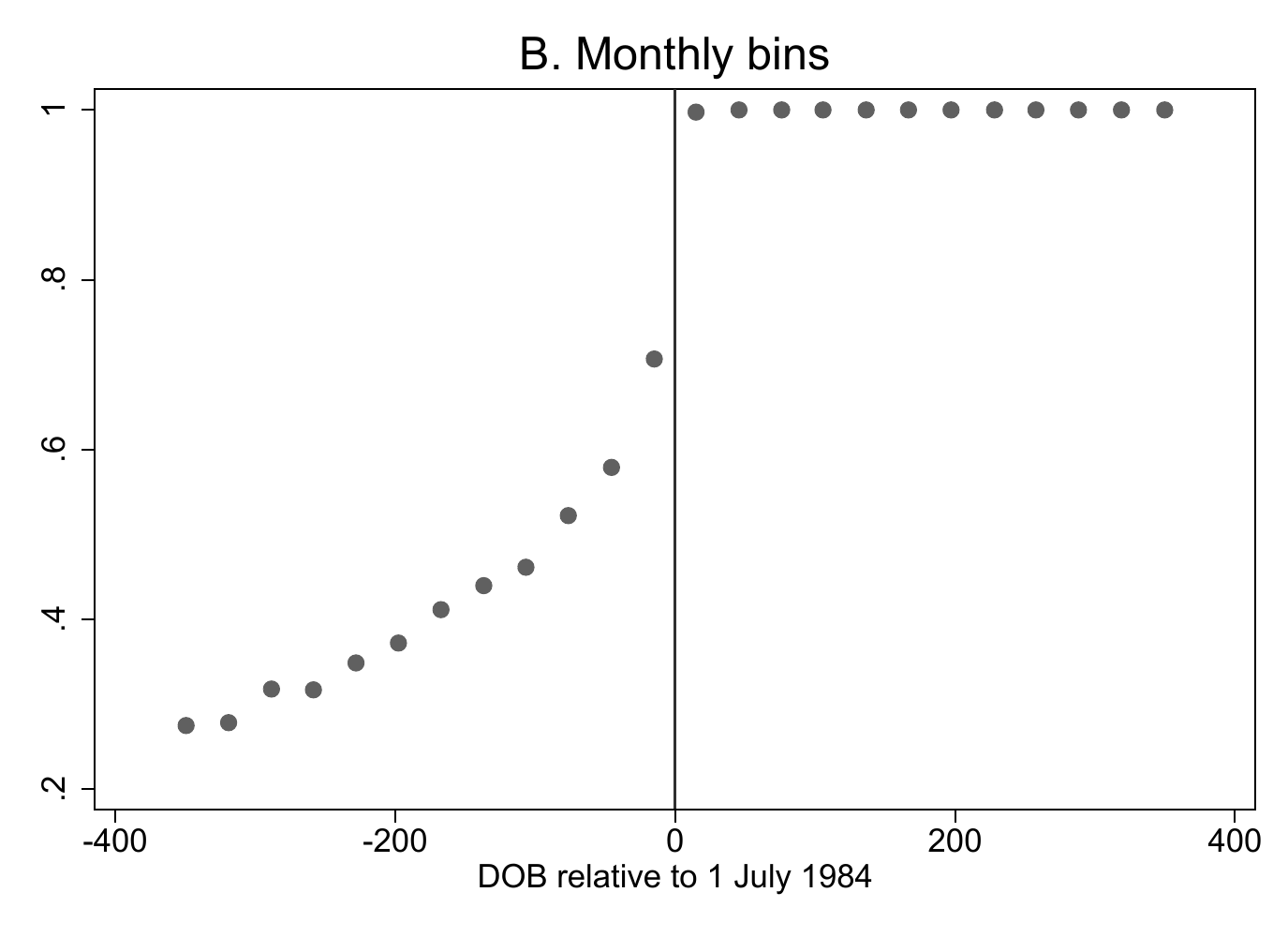} \\
	\end{tabular}	
\begin{tablenotes}[para,flushleft]
	\footnotesize
	\item Notes: Both panels shows the mean value of the `treatment' variable, by DOB. Treatment is defined as obtaining a first learner's permit on or after 1 July 2000, thereby subject to different MSDH requirements. The only difference between panels is the size of the DOB `bins'. 	
\end{tablenotes}
\end{figure}

Figure \ref{fgr-fs-84} shows complete compliance to the right of 1 July 1984.\footnote{While there appears to be very minor non-compliance, this is due to imprecision around DOB, as described in Section B.4. We drop observations where we are uncertain about treatment status in our regression analysis.} It was not possible for anyone born on this date or after to hold an `old' learner's permit due to administrative rules. The pattern on the left side is more complicated. Both panels show a monotonic upward, non-linear pattern. Panel B suggests the presence of a discontinuity at the threshold. In contrast, Panel A suggests no discontinuity, but a kink, caused by a very steep rise on the left side of the threshold.

This figure illustrates that many different estimators could potentially be used to estimate the effect of the reform. Candidate estimators could exploit the apparent kink, or the approximate discontinuity, or both. Or, they could instead employ a between-cohort-IV strategy. Each approach could be implemented using various alternate functional form assumptions (i.e. orders of polynomial, which need not be the same on each side of the threshold). Finally, one can choose between many bandwidths.

We first consider a total of 4,634 alternate candidate specifications, each with symmetrical bandwidth around the threshold. This consists of 14 different models, estimated using each possible bandwidth in the range of 35 to 365 days. In principle, we could consider larger bandwidths as well. This is prevented by practical considerations in our application. People born before 1 July 1983 were eligible for driver's licenses which differed in other important ways. Therefore we need an estimator which does not use data on people born before that date, hence making 365 days the largest feasible bandwidth.

Denoting outcome (i.e. MVA 1-year indicator) for person $i$ by $Y_i$, DOB by $X_i$ (centred at zero around 1 July 1984), treatment (obtained learner's permit after policy change) by $T_i$ and an indicator for DOB $\geq$ 1 July 1984 (1991) by $D_i$, the first 11 candidate models are fuzzy RDD, RPJKD and RKD estimators. Each of these can be treated as instrumental variable models, with the structural equation given by Eq. \ref{eq-structural} and first-stage given by Eq. \ref{eq-first-stage}.  Full details on the estimation equations are in Table C2.\footnote{We only consider a uniform kernel in our application, although it would be straightforward to vary the kernel along with other modelling dimensions. In practice, kernel choice typically has little influence on the estimates \cite{lee_lemieux_2010}.}

\begin{equation}\label{eq-structural}
Y_i = \alpha + \beta T_i + f(X_i,D_i) + e_i
\end{equation}

\begin{equation}\label{eq-first-stage}
T_i = \pi_0 + f(X_i,D_i) + g(X_i,D_i) + \epsilon_i
\end{equation}

Model 1 is a conventional (fully-interacted) linear RDD. Model 2 is an RDD model with a linear fit on the right side of the threshold, and a quadratic on the left. This is motivated by the first-stage relationship in Figure \ref{fgr-fs-84}, characterized by a clearly nonlinear relationship on the left, and perfect linearity on the right. We refer to this as a `mixed polynomial' specification. Model 3 is a conventional (fully-interacted) quadratic RDD.	

The next four candidate models exploit both the discontinuity and the kink for identification. These are RPJKD estimators of the following form. Model 4 is a conventional (fully-interacted) linear RPJKD. Model 5 is a quadratic RPJKD, in which the quadratic term is not interacted with the threshold indicator. Model 6 is an RPJKD model with a linear fit on the right side of the threshold, and a quadratic on the left. Model 7 is a fully-interacted quadratic RPJKD. 

Four more candidate models adopt conventional regression kink designs. Model 8 is a conventional (fully-interacted) linear RKD. Model 9 is a quadratic RKD, in which the quadratic term is not interacted with the threshold indicator. Model 10 is an RKD model with a linear fit on the right side of the threshold, and a quadratic on the left. Model 11 is a fully-interacted quadratic RKD.

The remaining three candidate models are month-of-birth cohort-IV models, which exploit between-cohort variation in the probability of `treatment'. Denoting month-of-birth fixed effects by $\theta_m$, for these models, the first-stage becomes:

\begin{equation}\label{eq-first-stage-cohort-IV}
T_i = \pi_0 + f(X_i) + g(X_i,\theta_m) + \epsilon_i
\end{equation}

Model 12 assumes a linear secular relationship between DOB and the outcome variable. Model 13 assumes a quadratic secular relationship between DOB and the outcome variable. Model 14 assumes a cubic secular relationship between DOB and the outcome variable.

\subsection{The placebo zone}
\begin{figure}[ht!]
	\centering
	\caption{The placebo zone}\label{fgr-placebo-zone}
	\begin{tabular}{cc}
		\includegraphics[width=.5\textwidth]{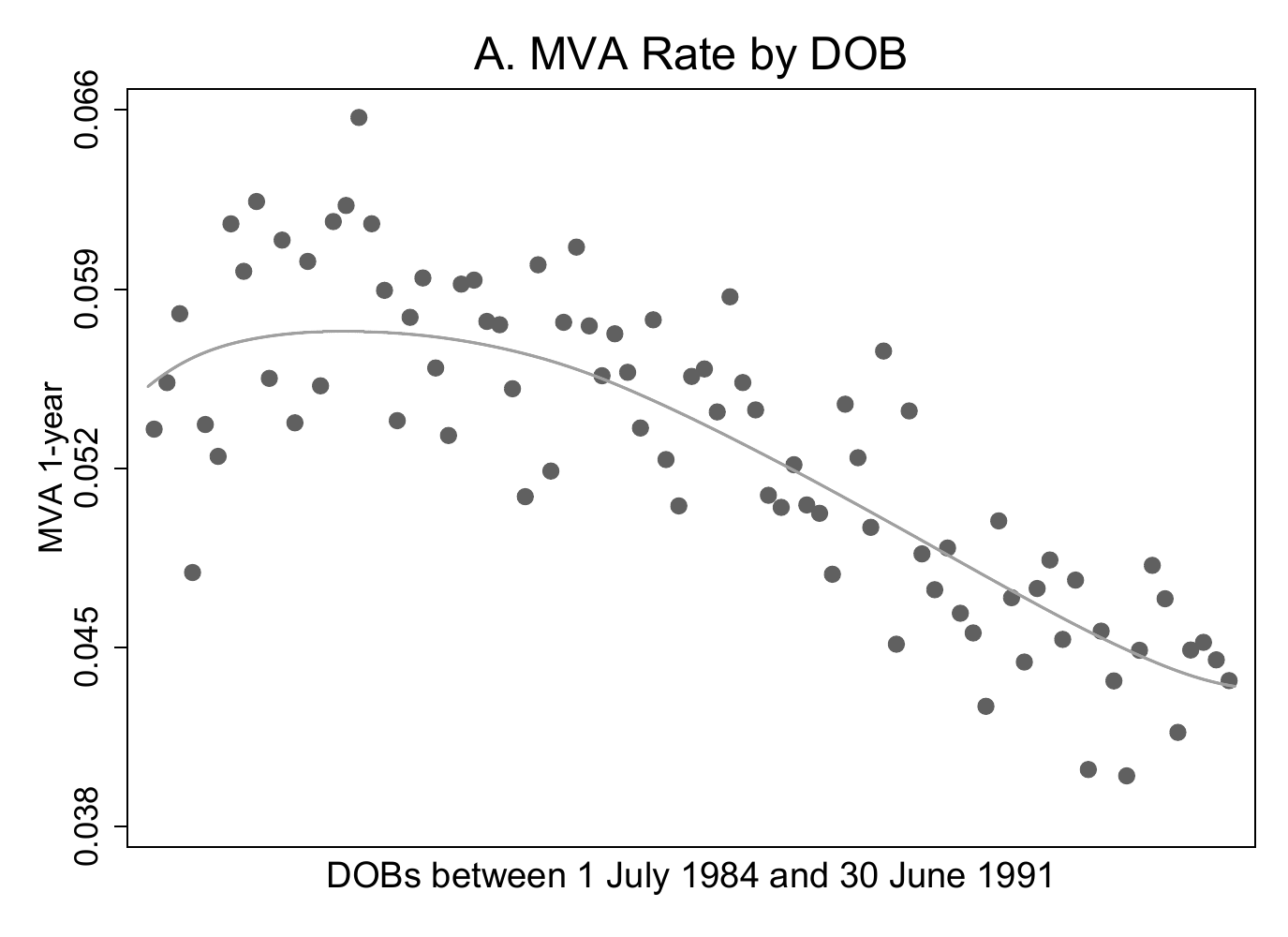} &
		\includegraphics[width=.5\textwidth]{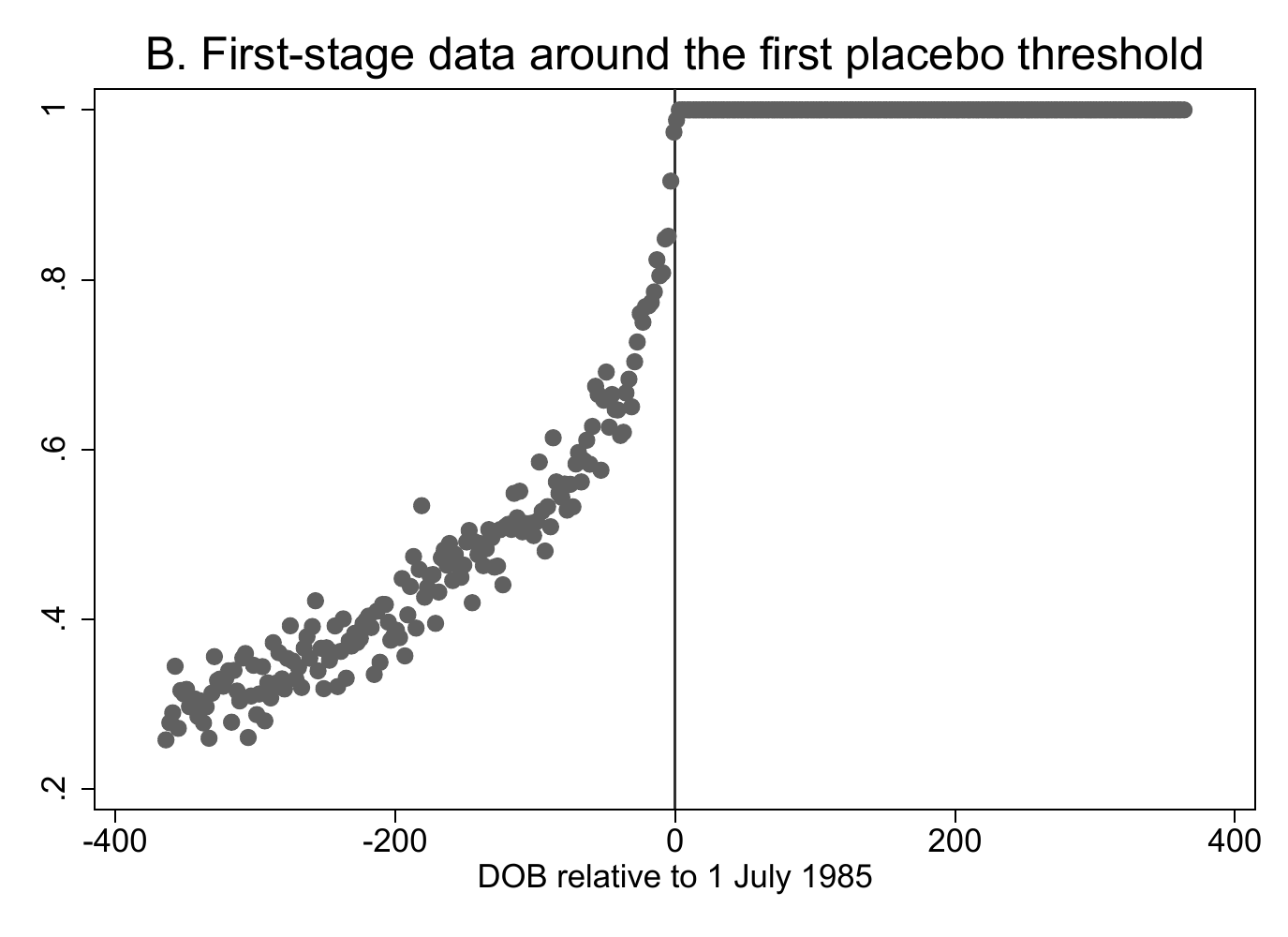} \\
	\end{tabular}	
	\begin{tablenotes}[para,flushleft]
		\footnotesize
		\item Notes: Panel A shows the proportion of people who crashed within one year of receiving a provisional drivers license (the main outcome variable in the analysis) by DOB within the placebo zone. The plot uses 30-day bins of DOB. Panel B is based on Figure \ref{fgr-fs-84} Panel A. Here, however, the range of DOB is shifted by one year, and so is the definition of `treatment', which is a function of date received first Ls. 	
	\end{tablenotes}
\end{figure}

The placebo zone is the set of DOBs between 1 July 1984 and 30 June 1991, inclusive. There were no apparent major licencing policy changes which were likely to have affected MVAs in a way that depends on DOB within this zone. Figure \ref{fgr-placebo-zone} Panel A shows the MVA rate by month of birth within this zone (in 30 day bins), with a lowess fit. Generally, the pattern is relatively smooth, with a slight downward trend, apart from perhaps the first 5 months. 

Within this zone, we create placebo treatments in a way that mimics the true treatment selection process. For example, in the first placebo, persons are deemed treated if they obtained their license on or after 1 July 2001. The first-stage relationship between DOB and this placebo treatment is shown (in 2-day bins) in Figure \ref{fgr-placebo-zone} Panel B, with a 365 day bandwidth around the DOB threshold of 1 July 1985. This relationship closely resembles the true treatment profile around the 1 July 1984 DOB, which we show in Figure \ref{fgr-fs-84}. Similar patterns are found for the other placebo DOB thresholds in this zone.

After collapsing to DOB-level (and weighting by cell-size), we estimate the placebo treatment effect (which we know to be zero and constant across entities) using each of the 4,634 candidate models.\footnote{The results are almost identical when uncollapsed microdata are used instead, but estimation is much faster with collapsed data.}  We repeat this for all 1,826 placebo treatment thresholds, and summarize the performance of each candidate model.

\subsection{Model performance in the placebo zone}
Table \ref{tbl-main-model-perf} summarizes the performance of each candidate model. It would not be practical to report on the performance of all 4,634 candidates. Instead we show only the results for the bandwidth which yields the lowest root mean squared error (RMSE) for each model type. The first clear feature of this table is that for every model considered, large bandwidths (365 days in all but one case) yield the smallest RMSEs, compared with smaller bandwidths. Secondly, most models have appropriate coverage rates.

\begin{table}[ht!]
	\caption{Candidate model performance in the placebo zone}\label{tbl-main-model-perf}
	\begin{tabular}{@{}llllll@{}}
		\toprule
		Model & Description                  & RMSE   & Optimal BW & Coverage & Bias    \\ \midrule
		1     & RDD - linear      & 0.0083 & 365        & 0.962    & -0.0004 \\
		2     & RDD - mixed polynomial       & 0.0199 & 365        & 0.921    & 0.0014  \\
		3     & RDD - quadratic   & 0.0230 & 365        & 0.927    & 0.0015  \\
		\textbf{4}     & \textbf{RPJKD - linear}    & 0.0060 & 365        & 0.936    & 0.0010  \\
		5     & RPJKD - quadratic            & 0.0073 & 365        & 0.980    & -0.0005 \\
		\textbf{6}     & \textbf{RPJKD - mixed polynomial}     & 0.0052 & 365        & 0.992    & 0.0000  \\
		7     & RPJKD - interacted quadratic & 0.0132 & 365        & 0.938    & 0.0005  \\
		8     & RKD - linear      & 0.0096 & 355        & 0.910    & 0.0028  \\
		9     & RKD – quadratic              & 0.0179 & 365        & 0.953    & 0.0019  \\
		\textbf{10}    & \textbf{RKD - mixed polynomial}       & 0.0057 & 365        & 0.984    & 0.0002  \\
		11    & RKD - interacted quadratic   & 0.0177 & 365        & 0.950    & 0.0019  \\
		\textbf{12}    & \textbf{birth cohort-IV - linear}     & 0.0051 & 365        & 0.946    & 0.0006  \\
		13    & birth cohort-IV - quadratic  & 0.0070 & 365        & 0.987    & -0.0007 \\
		14    & birth cohort-IV - cubic      & 0.0124 & 365        & 0.937    & 0.0001  \\
		WA    & Inv-MSE weighted average     & 0.0055 & 365        & n.d.     & 0.0003  \\
		C1    & RDD conventional             & 0.0340 & 117        & 0.966    & 0.0012  \\
		C2    & RDD bias corrected           & 0.0443 & 117/184        & 0.939    & 0.0015  \\
		C3    & RKD conventional             & 0.0346 & 136        & 0.997    & 0.0014  \\
		C4    & RKD bias corrected           & 0.0415 & 136/202        & 0.999    & 0.0019  \\ \bottomrule
	\end{tabular}
	\begin{tablenotes}[para,flushleft]
	\footnotesize
	\item Notes: This table summarizes the performance of each candidate model within the placebo zone. The key statistic is the RMSE of estimated treatment effects. There are 1,826 treatment effect estimates for every model, one for each placebo-zone threshold. The true treatment effect is known to be zero throughout the placebo zone, so zero is the target parameter for every estimator. With the exception of WA and C1-C4, every candidate model is trialled repeatedly with symmetric bandwidths ranging from 30 to 365 days. For each model, results from the bandwidth which yields the lowest RMSE are shown. In addition to the 14 main models, the model labelled WA is an estimator which (for each placebo-zone repetition) uses the inverse-MSE-weighted average of the estimates from the 14 main models, using each of those model's respective optimal bandwidth. Unlike the other models, those labelled C1-C4 use a CCT bandwidth selection procedure and default settings in Stata's -rdrobust- command.
\end{tablenotes}
\end{table}

In our application, four models stand out with the lowest RMSE. The best performing model (RMSE = 0.0051) is `Model 12' -- the month-of-birth cohort-IV model with a linear trend. This is closely followed by `Model 6' (RMSE = 0.0052) -- the RPJKD with mixed polynomial fit (quadratic on the left and linear on the right). Next are the RKD with mixed-polynomials (RMSE = 0.0057) and the linear RPJKD (RMSE = 0.0060). All four have similarly good coverage (at least 93.6\%), and small average bias (0.001 at most).

We also consider a model-averaging approach. It is defined as the weighted average of the estimates from the 14 candidate models (each with full 365 day bandwidth). The weights are set to the inverse of the MSE of each candidate model. The  performance of this weighted average estimator is also shown in Table \ref{tbl-main-model-perf}. Whilst its performance is good, its RMSE is higher than Models 12 and 6. The coverage of this estimator is not shown as its variance has not been derived.

The final four rows of Table \ref{tbl-main-model-perf} summarize the performance of four estimators proposed by CCT, and implemented using Stata’s -rdrobust- command. These are conventional, and bias-corrected estimates using RDD and RKD, respectively.\footnote{The results shown are for models estimated on collapsed microdata. As with the other estimators considered, the results with collapsed (DOB) data are very similar.} The `optimal' bandwidths for these estimators are determined within rdrobust, rather than the placebo zone procedure that we adopt for the other estimators.\footnote{More precisely, the CCT bandwidths shown in Table \ref{tbl-main-model-perf} are the average of bandwidths selected by rdrobust through the placebo zone.} As seen in the table, these bandwidths are much smaller than the others. The key result, however, is that the performance of these estimators, as measured by RMSE, is worse than any of the other candidate models, and an order of magnitude worse than the best performing candidate models. This is consistent with the findings of Card et al. (2017)'s RKD Monte Carlo simulations and our results in Section \ref{sec-sims}.

\subsection{Incorporating asymmetrical bandwidths}
In every model tested on the placebo zone thus far, we have followed conventional practice and imposed the same bandwidth on the left and right sides of the threshold. Here we explore whether model performance can be improved by allowing for asymmetric bandwidths.

In particular, we have so far capped the bandwidth at 365 days on each of the thresholds. This is motivated by practical constraints in our application. Any more than 365 days to the left of the 1 July 1984 threshold would take us into territory where other important policy changes were implemented in a way that relates systematically with DOB. But we do not have the same issue on the right side of the threshold. Similarly, for the 1991 threshold, we have no constraints in the left side, though data constraints prevent us from considering bandwidths greater than 365 days on the right.\footnote{The constraint is due to the fact that drivers who obtain their P1 license after age 25 are dropped from the sample because they are not required to meet the MSDH requirement (see Appendix B). We cannot impose this constraint consistently on the RHS of the 2007 reform because the end-date for our license data mean we do not always observe whether people got their P1 license by age 25.} 

We now repeat the placebo-zone model selection procedure for all 14 candidate models using two similar procedures.

\begin{enumerate}
	\item We fix the bandwidth to 365 days on the left, whilst allowing the bandwidth to vary between 365 days and 730 days on the right. This will be informative for model selection in our analysis of the 2000 reform. The number of placebo thresholds in this exercise is 1,461, due to the need to include a larger maximum bandwidth. 
	\item We fix the bandwidth to 365 days on the right, whilst allowing the bandwidth to vary between 365 days and 730 days on the left. This will be informative for model selection in our analysis of the 2007 reform. The number of placebo thresholds is 1,461. 
\end{enumerate}

The results for Version 1 of this exercise are summarized in Table \ref{tbl-asymm-model-perf}. It shows that performance is improved considerably for every model by allowing larger bandwidths on the right. In some cases, RMSE is reduced by more than 50\%. The optimal right-side bandwidth varies considerably, from 550 up to the 730 day limit. Model 6 is the best performing model, with an optimal RHS bandwidth of 550 days. This is the best performing estimator amongst all candidates for estimating the effect of the 2000 reform. The weighted-average estimator, shown in the lowest row, does just as well as Model 6. Models 12, 4 and 10 continue to perform well.

\begin{table}[ht!]
		\caption{Candidate model performance in the placebo zone V1: Asymmetric bandwidths}\label{tbl-asymm-model-perf}
	\begin{tabular}{lllp{1.9cm}ll}
		\toprule
		Model & Description                  & RMSE    & Optimal RHS BW & Coverage & Bias    \\ \midrule
		1     & RDD - linear      & 0.0069 & 550            & 0.958    & -0.0005 \\
		2     & RDD - mixed polynomial       & 0.0186 & 660            & 0.942    & 0.0016  \\
		3     & RDD - quadratic   & 0.0203 & 710            & 0.930    & 0.0013  \\
		\textbf{4 }    & \textbf{RPJKD - linear}    & 0.0046 & 670            & 0.910    & 0.0022  \\
		5     & RPJKD - quadratic            & 0.0059 & 710            & 0.985    & 0.0001  \\
		\textbf{6}     &\textbf{RPJKD - mixed polynomial}     & 0.0039 & 550            & 0.996    & 0.0005  \\
		7     & RPJKD - interacted quadratic & 0.0056 & 720            & 0.993    & 0.0005  \\
		8     & RKD - linear      & 0.0059 & 730            & 0.879    & 0.0031  \\
		9     & RKD - quadratic              & 0.0166 & 730            & 0.910    & 0.0070  \\
		\textbf{10}    & \textbf{RKD - mixed polynomial}       & 0.0043 & 730            & 1.000    & 0.0006  \\
		11    & RKD - interacted quadratic   & 0.0067 & 730            & 0.997    & 0.0008  \\
		\textbf{12}    & \textbf{birth cohort-IV - linear}     & 0.0042 & 670            & 0.912    & 0.0020  \\
		13    & birth cohort-IV - quadratic  & 0.0056 & 720            & 0.998    & -0.0003 \\
		14    & birth cohort-IV - cubic      & 0.0060 & 700            & 0.990    & -0.0004 \\
		WA    & Inv-MSE weighted average     & 0.0039 & As   above     & n.d      & 0.0010  \\
	    \bottomrule
	\end{tabular}
\begin{tablenotes}[para,flushleft]
	\footnotesize
	\item Notes: The results in this table are from a similar procedure to what is detailed in the Table \ref{tbl-main-model-perf} notes. The only difference is the set of bandwidths considered. The left side bandwidth is fixed at 365 days, the right side bandwidths considered range from 365 days to 730 days. As in Table \ref{tbl-main-model-perf}, results are shown for the bandwidths which yield the lowest RMSE for each model. There are 1,461 treatment effect estimates for every model, one for each placebo-zone threshold. The smaller number of repetitions is a result of the larger maximum bandwidth considered.
\end{tablenotes}
\end{table}

The results for Version 2 of this exercise are summarized in Table \ref{tbl-asymm2-model-perf}. They are similar to those of the previous exercise -- Models 12, 10 and 6 continue to perform well. Optimal bandwidths vary, but are generally considerably larger than the baseline exercise. Model 12 has the lowest RMSE, with an optimal LHS bandwidth of 560 days. This is the single best performing specification amongst all candidates for estimating the effect of the 2007 reform.

\begin{table}[ht!]
	\caption{Candidate model performance in the placebo zone V2: Asymmetric bandwidths}\label{tbl-asymm2-model-perf}
	\begin{tabular}{lllp{1.9cm}ll}
		\toprule
		Model       & Description                          & RMSE    & Optimal LHS BW & Coverage & Bias    \\ \midrule
		1     & RDD - linear        & 0.0049 & 660            & 0.986    & -0.0009 \\
		2     & RDD - mixed polynomial         & 0.0095 & 730            & 0.979    & 0.0020  \\
		3     & RDD - quadratic     & 0.0122 & 730            & 0.977    & 0.0019  \\
		4     & RPJKD - linear      & 0.0047 & 690            & 0.987    & -0.0005 \\
		5     & RPJKD - quadratic              & 0.0041 & 610            & 0.999    & -0.0009 \\
		\textbf{6}     & \textbf{RPJKD - mixed polynomial}       & 0.0040 & 560            & 0.999    & -0.0004 \\
		7     & RPJKD - interacted   quadratic & 0.0122 & 730            & 0.969    & -0.0001 \\
		8     & RKD - linear        & 0.0051 & 370            & 1.000    & -0.0002 \\
		9     & RKD - quadratic                & 0.0054 & 600            & 0.990    & -0.0012 \\
		\textbf{10}    & \textbf{RKD - mixed polynomial}         & 0.0037 & 550            & 0.997    & -0.0005 \\
		11    & RKD - interacted quadratic     & 0.0186 & 370            & 0.942    & 0.0024  \\
		\textbf{12}    & \textbf{birth cohort-IV - linear}       & 0.0036 & 560            & 1.000    & -0.0004 \\
		\textbf{13}    & \textbf{birth cohort-IV - quadratic}    & 0.0037 & 610            & 1.000    & -0.0009 \\
		14    & birth cohort-IV - cubic        & 0.0061 & 730            & 0.997    & 0.0007  \\
		WA    & Inv-MSE weighted average       & 0.0038 & As   above     & n.d      & -0.0005 \\
		\bottomrule
	\end{tabular}
\begin{tablenotes}[para,flushleft]
	\footnotesize
	\item Notes: The results in this table are from a similar procedure to what is detailed in the Table \ref{tbl-main-model-perf} notes. The only difference is the set of bandwidths considered. The right side bandwidth is fixed at 365 days, the left side bandwidths considered range from 365 days to 730 days. As in Table \ref{tbl-main-model-perf}, results are shown for the bandwidths which yield the lowest RMSE for each model. There are 1,461 treatment effect estimates for every model, one for each placebo-zone threshold. The smaller number of repetitions is a result of the larger maximum bandwidth considered.
\end{tablenotes}
\end{table}

\section{Using placebo zone estimates for inference}\label{sec-inference}

Our main interest in this paper is estimation. However the placebo zone may also be useful for inference. As already shown, the placebo zone can be used to assess coverage of confidence intervals stemming from standard approaches. In this section, we propose some new approaches to randomization inference. 

In our main application, the placebo zone consists of 1,826 overlapping data windows, corresponding to 1,826 separate placebo estimates for each (symmetric) estimator. One can use the distribution of these estimates for alternative approaches to inference -- randomization inference, in the spirit of \citeA{ganong_jager_2018}. We discuss two alternative inference approaches. 

\subsection{Approach 1: Fully non-parametric}
Randomization inference considers the estimate of interest alongside the distribution of placebo estimates. Consider an estimate which lies outside of the range of the placebo estimates. If these 1,826 placebo estimates were independent, as per \citeA{ganong_jager_2018}, one would conclude that the two-sided p-value $< 2/1826 = 0.0011$. For an estimate lying inside the range of placebo estimates, $p = 2*\min(i/1826,(1826-i)/1826)$, where $i$ is the rank of the estimate alongside the 1,826 placebo estimates.

However, in our application, the 1,826 placebo estimates are not independent. Indeed they are strongly serially correlated. This is almost certainly the case in other applications as well, given the rolling data window used for consecutive placebo estimates. In our Model 12, the serial correlation of placebo estimates = 0.9895. This equates to an effective sample size (ESS) of just 10 independent observations.\footnote{This calculation draws on Eq. 5 in \citeA{zwiers_storch_1995}.} A more appropriate two-sided p-value for estimates lying outside the placebo zone is $p < 2/ESS$. In our case, for Model 12, $p < 2/10 = 0.2$.

\subsection{Approach 2: Parametric Randomization Inference}
A much more powerful approach to randomization inference is available if one invokes an assumption that the placebo estimates are drawn from a normal distribution. Under this approach, the variance of the placebo estimates is informative of the variance of the treatment effect estimate. This approach to randomization inference facilitates not only p-values but also leads to straightforward calculation of standard errors, t-statistics and confidence intervals. In effect, the variance of placebo estimates is assumed to equal the variance of the treatment effect estimator.

As in the first approach above, it is important to  account for serial correlation of the placebo estimates. Such serial correlation reduces confidence in the estimated variance of the population distribution. This can be easily accounted for with a degrees of freedom adjustment. The test statistic is hence assumed to follow a t-distribution, with degrees of freedom equal to the ESS of placebo estimates minus 1.

This is our preferred approach to randomization inference. Our simulations suggest that it performs well in comparison to alternate approaches such as CCT and IK. See Section \ref{sub-sec-coverage} and Table C1.

In an earlier version of this paper we discuss how one can also draw on the placebo estimates for bias correction, both for estimation and inference \cite[Section 4.5.2]{kettlewell_siminski_2020}. 

\subsection{Randomization inference when the placebo zone is not contiguous}\label{sub-sec-randinf}
In our own application, the placebo zone is contiguous. But in other applications it may not be, particularly if placebo data are available on `both sides' of the real threshold. In such cases, it is not obvious how to determine the `effective sample size', since the estimates on either side of the `gap' may be correlated, but less so than estimates from immediately adjacent thresholds. One approach is to bound the effective sample size. The lower bound essentially ignores this discontinuity in serial correlation, and derives the ESS using a weighted average of the autocorrelations within each contiguous segment. The upper bound treats estimates between each segment as distinct and so the total ESS is the sum of the ESS in each segment.

In our application, this approach would produce tight bounds. To illustrate, if our placebo zone was not contiguous but instead consisted of two equally sized zones on either side of the treatment threshold, the lower bound for the ESS for our best symmetric bandwidth estimator would be 10 and the upper bound would be 11.

\section{Estimated effects of minimum supervised driving hours}\label{sec-results}
The main estimation results are presented in Table \ref{tbl-main-results}. Panel A shows the estimated effects of the 2000 reform and Panel B shows the estimated effects of the 2007 reform. Each panel shows results from five separate estimators -- one in each column.

\begin{table}[ht!]
	\caption{Estimated effects of minimum supervised driving hours}\label{tbl-main-results}
	\begin{tabular}{p{2.1cm}p{2.1cm}p{2.2cm}p{2.1cm}p{2.1cm}p{2.0cm}}
		\toprule
		& Best   estimator & Best sym. cohort-IV & Best sym. RPJKD & Best sym. RKD & Best sym. RDD \\
		\midrule
		& (1)              & (2)                                    & (3)                              & (4)                            & (5)                            \\
		& \multicolumn{5}{c}{A: 2000   Reform (0 $\rightarrow$ 50 hours)}                                                                                                            \\
		\cmidrule(lr){2-6}
		MVA 1-year & -0.0144***	& -0.0132*** & -0.0147*** & -0.0144** & -0.0168***            \\
		SE           & 0.0041    & 0.0049 & 0.0050 & 0.0058 & 0.0058   \\
		p-value      & 0.0005    & 0.0073 & 0.0032 & 0.0129 & 0.0038   \\
		alt. SE & 0.0039 &	0.0047 & 0.0049 & 0.0054 & 0.0080 \\
		alt. p-value & 0.0096 & 0.0203 & 0.0157 & 0.0381 & 0.0552  \\
		Model        & 6         & 12     & 6      & 10     & 1        \\
		BW           & 365 / 550 & 365    & 365    & 365    & 365      \\
		&                  &                                        &                                  &                                &                                \\
		& \multicolumn{5}{c}{B: 2007   Reform (50 $\rightarrow$ 120 hours)}                                                                                                          \\
		\cmidrule(lr){2-6}
		MVA 1-year & 0.0021	& 0.0003 & 0.0006 & -0.0024 & -0.0007                        \\
		SE           & 0.0030    & 0.0033 & 0.0033 & 0.0046 & 0.0035 \\
		p-value      & 0.4790    & 0.9259 & 0.8477 & 0.6069 & 0.8422 \\
		alt. SE & 0.0042 & 0.0047 & 0.0049 & 0.0054 & 0.0080 \\
		alt. p-value & 0.6310 & 0.9496 & 0.9003 & 0.6774 & 0.9328 \\
		Model        & 12        & 12     & 6      & 10     & 1      \\
		BW           & 560 / 365 & 365    & 365    & 365    & 365    \\
		\bottomrule                           
	\end{tabular}
	\begin{tablenotes}[para,flushleft]
	\footnotesize
	\item Notes: This table shows the main estimated effects of the actual policy changes in our main application. Asymptotic standard errors are clustered at the DOB level. Alternate standard errors and p-values use the randomization inference procedure described in Section \ref{sub-sec-randinf}. * $p<0.1$, ** $p<0.05$, *** $p<0.01$.     
	\end{tablenotes}
\end{table}

Column (1) shows results from the `best' estimator. This is the estimator with the lowest RMSE of all candidate models evaluated on the placebo zone. For the 2000 reform, this is `Model 6', with a bandwidth of 365 days on the left and 550 on the right. For the 2007 reform, this is `Model 12' with a bandwidth of 560 days on the left and 365 days on the right.\footnote{Recall that we are constrained to a maximum bandwidth of 365 days on the left of the threshold for the 2000 reform, and a maximum bandwidth of 365 days on the right of the threshold for the 2007 reform. Model 6 is a RPJKD model with quadratic polynomials to the left and linear polynomial to the right of the threshold in each stage. Model 12 is a month-of-birth cohort-IV model, controlling for a linear secular relationship between DOB and the outcome variable.}

These `best' estimates suggest that the first reform had a strong impact on reducing MVAs, while the second reform did not. The first reform is estimated to have reduced the crash rate by -0.014, a reduction of 21\% relative to the predicted value at the threshold for the untreated. The conventional p-value associated with this estimate is 0.0005. We have no reason to be sceptical about the validity of this p-value, since this estimator was found to have good coverage in the placebo zone trial, as well as an estimated bias that is close to zero. Nevertheless, we also show alternate p-values, based on the distribution of placebo estimates, as discussed in the previous section. This p-value is larger (0.0096), though still strongly significant. The alternate p-value is larger, even though the alternate standard error is smaller, due to the small number of degrees of freedom in this approach to inference.\footnote{Just 6 degrees of freedom are used for this estimate. This is equal to the `effective sample size' of placebo estimates calculated in the placebo zone minus 1, taking into account the very strong serial correlation of those estimates (0.9915).} For the 2007 reform, the alternate p-value is also slightly higher than the conventional p-value, but remains very far from any conventional threshold of statistical significance.

Column (2) shows results from the best symmetric estimator -- which is Model 12 with a bandwidth of 365 days on each side. For the 2000 reform, all of the key parameters from this model are similar to those in Column 1. The standard error and both p-values are all slightly larger, but qualitatively the same as in Column (1). The estimate for the 2007 reform is close to zero.

Columns (3), (4) and (5) show the results from the best symmetric RPJKD, RKD and RDD estimators, respectively. In each case, the maximum feasible bandwidth (365 days) is used, consistent with the outcomes of the placebo zone trials. Again, the qualitative conclusions are the same, with strongly significant negative effects of the 2000 reform, and approximately zero for the second reform.

Appendix D delves deeper into the effects of the 2000 reform. We find that: (i) delaying of obtaining a license is at most only a small factor in the treatment effects that we have estimated; (ii) there is no MVA reduction in the following year (12-24 months) after obtaining a license; (iii) the results carry over to only more serious MVAs involving injury; and (iv) our main estimates are similar for males and females.

In Appendix E we undertake a back-of-the envelope cost-benefit analysis using our main estimates. Our estimates imply an average social gain of \$2,300 per person due to the 50 MSDH reform. If we take the conservative view that on average people would complete 20 hours supervised in the absence of the reform, then this would constitute a net social improvement provided that supervisors' and learners' combined cost of obtaining hours is less than \$46 per hour. Since we find no evidence the 120 MSDH reform improved safety, we cannot rule out nil social benefits for that reform.

\section{Other selected applications of the placebo zone approach}\label{sec-further-apps}

In this section we re-evaluate evidence for two `well-known' applications suited to our method; Head Start and MLDA. These applications feature relatively large placebo zones, which makes them suitable for our method. 

\subsection{Head Start}\label{sub-sec-head-start}

Since \citeA{ludwig_miller_2007}'s RDD analysis of the Head Start program, the data from this study have been used widely for illustrative purposes in the RDD methodological literature, including papers by \shortciteA{calonico_cattaneo_titiunik_2014}, \shortciteA{cattaneo_titiunik_vazquez_2017}, \shortciteA{ganong_jager_2018} and \shortciteA{calonico_et_al_2019}.

The unit of analysis is the county. Treatment is eligibility for technical assistance to develop Head Start funding applications. Eligibility is tied to a sharp, arbitrary cut-off in the county-level poverty rate at 59.198 percentage points. The outcome variable is child mortality from Head Start-relevant causes. 

Earlier papers have used a range of methodological approaches to estimate the same discontinuity. \citeA{ludwig_miller_2007} prefer local-linear regressions with a triangular kernel. Citing a lack of consensus on bandwidth selection, they show results using bandwidths of 9, 18 and 36 percentage points, as well as from regular linear and quadratic specifications. \shortciteA{calonico_et_al_2019} use the CCT bandwidth-selection algorithm, which yields bandwidths of 6.81 and 6.98, varying by the use of covariates.

We consider 10 separate estimators, each with a range of alternate bandwidths. These were chosen to examine questions of functional form (linear versus quadratic) kernel (uniform versus triangular), weights (unweighted or population-weighted), and covariates (include or exclude). \textit{A priori}, weighted estimates are likely to be more precise, since residual variance is likely inversely proportional to population size, and population varies greatly (Mean = 38,964; Standard Deviation = 117,460), ranging from 224 to 2,664,438. We also show four estimates using CCT models. Models 11 and 12 are unweighted conventional and bias-corrected estimates with CCT bandwidths. Models 13 and 14 are corresponding weighted estimates. 

We face two challenges for adopting our approach in this context. The first is a relatively small range of the forcing variable within the placebo zone. The placebo zone has a range of 47 percentage points (spanning 15.2 to 59.198 percentage points). When models with relatively large bandwidths are trialled, the effective sample size of the resulting placebo estimates is small. The second challenge is a considerably larger density (about 2.1 times larger) in the placebo zone than in the treatment zone \shortcite<see>[Figure A1]{cattaneo_titiunik_vazquez_2017}. The results of trials within such a high density zone may not be relevant for choosing models to adopt in a low density zone. We address both of these challenges by splitting the placebo zone sample into two independent groups.\footnote{We split the placebo zone observations into two groups because the placebo zone density is 2.1 times greater than the treatment zone density. This approach can be generalized for other contexts where the density is uneven. Practitioners may split the placebo zone into $g$ groups, where $g$ = round(placebo zone density / treatment zone density). It is not clear however if our approach is useful for situations where the treatment zone density is markedly greater than the placebo zone density.} We randomly allocated each county into one of these groups.\footnote{When these random allocations are repeated, the results are generally very similar. The RMSEs for the unweighted specifications are most sensitive to these repetitions, but they seem to always exceed the RMSEs for corresponding weighted specifications, usually by a large factor.} This solves the second challenge, since the resulting density is very similar to that of the treatment zone. It also helps with the first challenge, since the effective sample size of placebo estimates is approximately doubled.

The results of these placebo zone trials are shown in Table \ref{tbl-head-start-candidate}. For every model considered, the optimal bandwidths are either the maximum (15 percentage points), or close to it. This is considerably larger than the bandwidths in \shortciteA{calonico_et_al_2019}.\footnote{The bandwidths in \shortciteA{calonico_et_al_2019} are in turn similar to the average CCT-selected bandwidth within the placebo zone, which are shown in the last four rows of Table \ref{tbl-head-start-candidate}.} Since we are unable to test larger bandwidths, these should be seen as lower bounds for each optimal bandwidth. The results suggest that for this application, the population weights are very helpful -- reducing the RMSE by around 28\% in the linear model. The table also shows that covariates do not help, in fact they increase RMSE slightly. This is perhaps unsurprising, since the set of covariates is not rich and does not account for much residual variation.\footnote{The covariates are: percentage of black and urban population, levels and percentages of population in three age groups (children aged 3 to 5, children aged 14 to 17, and adults older than 25) as well as total population. We do not include `total population' as a covariate whenever we use it as a weight instead.} The results suggest that models with a triangular kernel do worse than a regular rectangular kernel, and that a linear polynomial is preferred to higher orders. The CCT estimators (which are characterized by small bandwidths) perform poorly, but not to the same extent as they do in our main application.

\begin{table}[ht!]
	\caption{Head Start candidate model performance in the placebo zone}\label{tbl-head-start-candidate}
	\begin{tabular}{p{0.8cm}p{5.5cm}lp{1.2cm}p{1.2cm}p{1.3cm}p{1.1cm}}
		\toprule
		Model & Description                              & RMSE   & Optimal LHS BW & Optimal RHS BW & Coverage & Bias    \\
		\midrule
		1     & RDD - linear                              & 0.7763 & 15.0           & 15.0           & 0.972    & -0.020             \\
		\textbf{2}     & \textbf{RDD - linear, weighted}                    & 0.5616 & 15.0           & 15.0           & 0.844    & -0.054          \\
		3     & RDD - linear, with covariates             & 0.7838 & 15.0           & 15.0           & 0.972    & -0.019        \\
		4     & RDD - linear, weighted, with covariates   & 0.5622 & 15.0           & 15.0           & 0.876    & -0.055           \\
		5     & RDD - linear, triangular kernel           & 1.0072 & 15.0           & 15.0           & 1.000    & 0.043             \\
		6     & RDD - linear, weighted, triangular kernel & 0.6198 & 15.0           & 15.0           & 1.000    & -0.036             \\
		7     & RDD - quadratic                           & 1.4649 & 15.0           & 15.0           & 0.890    & 0.147             \\
		8     & RDD - quadratic, weighted                 & 0.7751 & 14.6           & 14.6           & 0.918    & 0.017             \\
		9     & RDD - cubic                               & 1.7051 & 14.6           & 15.0           & 0.968    & 0.254              \\
		10    & RDD - cubic, weighted                     & 0.7842 & 15.0           & 15.0           & 0.982    & 0.057              \\
		C1    & RDD conventional                          & 1.9513 & 4.3            & 4.3            & 0.954    & 0.078              \\
		C2    & RDD bias corrected                        & 2.3386 & 4.3/6.9        & 4.3/6.9        & 0.961    & 0.052              \\
		C3    & RDD conventional - weighted               & 1.0278 & 5.1            & 5.1            & 0.972    & 0.089             \\
		C4    & RDD bias corrected - weighted             & 1.2098 & 5.1/8.0        & 5.1/8.0        & 0.968    & 0.084   \\
		\bottomrule         
	\end{tabular}
	\begin{tablenotes}[para,flushleft]
	\footnotesize
	\item Notes: The results in this table are from a similar procedure to what is detailed in the Table \ref{tbl-main-model-perf} notes. However, these are for the Head Start application. The set of models considered is different, for reasons discussed in the text. The bandwidths considered ranged from 3 to 15 percentage points (in 0.2 percentage point increments) and was allowed to be asymmetric. There are 282 treatment effect estimates for every model, one for each placebo-zone threshold.    
	\end{tablenotes}
\end{table}

The best performing estimator is the weighted linear RDD, with no controls, and with full bandwidth. The estimated discontinuity using this estimator in the treatment zone is shown in Column (1) of Table \ref{tbl-hs-mlda-results}. The estimate is statistically significant, consistent with \citeA{ludwig_miller_2007} and with \shortciteA{calonico_et_al_2019}. But the estimate is also considerably smaller than that of \shortciteA{calonico_et_al_2019}. The randomization inference procedure generates a similar standard error, but a larger p-value. This is primarily because the effective sample size of placebo estimates is small (7).

\afterpage{
\begin{table}[ht!]
	\caption{Head Start and MLDA RDD estimates}\label{tbl-hs-mlda-results}
	\begin{tabular}{lp{2.5cm}p{2.5cm}p{2.5cm}p{2.5cm}}
		\toprule
		& Head Start &   \multicolumn{3}{c}{MLDA}                                       \\
		\cmidrule(lr){2-2} \cmidrule(lr){3-5}
		& Mortality    & Ever Drinks & Drinks Regularly & Proportion of Days Drinks     \\
		& (1)          & (2)         & (3)              & (4)                           \\
		\midrule
		&              &             &                  &                              \\
		Estimated Effect    & -1.323**    & 0.1799***   & 0.2306*** &	0.0726***                    \\
		SE                  & 0.5372       & 0.0285	& 0.0237 &	0.0083   \\
		p-value             & 0.0140      & 0.0000      & 0.0000           & 0.0000                      \\
		alternate SE & 0.5600 & 0.0214 & 0.0283 & 0.0110 \\
		alternate   p-value & 0.0561 & 0.0138 & 0.0148 & 0.0028   \\
		Model               & 2            & 1           & 3                & 1                            \\
		BW                  & 15.00        & 4.68 &	4.93 &	4.68  \\
		\bottomrule
	\end{tabular}
	\begin{tablenotes}[para,flushleft]
	\footnotesize
	\item Notes: This table shows the main estimated effects for the Head Start and MLDA applications, using the best-performing model (lowest RMSE) from the respective placebo-zone trials reported in Tables \ref{tbl-head-start-candidate} and \ref{tbl-mlda-candidate}. Asymptotic standard errors are clustered at unique values of the running variable. Alternate standard errors and p-values use the randomization inference procedure described in Section \ref{sub-sec-randinf}. * $p<0.1$, ** $p<0.05$, *** $p<0.01$.    
\end{tablenotes}
\end{table}
}

\subsection{Minimum legal drinking age and drinking behavior}\label{sub-sec-mlda}
We now illustrate our approach with discontinuities in drinking behaviour at the MLDA. The MLDA context is one of the best known applications of RDD, beginning with \citeA{carpenter_dobkin_2009}. It is featured in econometric textbook treatments of RDD, such as \citeA{angrist_pischke_2015}.

We draw on data from \shortciteA{lindo_siminski_yerokhin_2016}'s analysis for the Australian state of New South Wales. We use the same three self-reported drinking outcomes as Lindo et al.: `Ever drinks', `Drinks regularly' and `Proportion of Days Drinks'. And we use the same data: waves 1-11 of the HILDA survey.

Following \citeA{carpenter_dobkin_2009}, Lindo et al. show estimates from linear specifications with bandwidths up to two years of age. These are centred around the 18th birthday MLDA threshold. Here, we consider the performance of a range of specifications -- linear and quadratic, with and without weights, as well as CCT estimators. We consider a much wider bandwidth range, from three months to five years on the right side (in 90 day increments), with the left side capped at three years. The 3-year cap on the left reflects the limit of data availability in the treatment zone, since all respondents were aged 15 years and over.

Table \ref{tbl-mlda-candidate} shows results from the placebo zone trials, for which the placebo zone consists of 18-30 year old respondents.\footnote{The results are qualitatively similar when a longer placebo zone is used (eg. 18-40 years, or 18-50 years) or if the maximum bandwidth is changed (e.g. 10 years, or 20 years). These are available on request.} In many respects, the results are consistent across outcome variables used, and indeed consistent with the earlier applications we have shown: (i) long bandwidths are optimal for each estimator -- much larger than those selected by CCT’s procedure; (ii) linear RDD yields the lowest RMSEs; (iii) the CCT estimator does poorly, with or without bias adjustment. 

\begin{table}[ht!]
	\caption{MLDA candidate model performance in the placebo zone}\label{tbl-mlda-candidate}
	\begin{tabular}{llllll}
		\toprule
		Model & Description                       & RMSE   & Optimal RHS BW & Coverage & Bias      \\
		\midrule
		\multicolumn{6}{c}{\underline{A: Ever Drinks}}                                                                                       \\
		\textbf{1}     & \textbf{RDD - linear}             & 0.0212 &	4.68 &	0.957 &	-0.003      \\
		2     & RDD - quadratic          & 0.0434 &	3.70 &	0.914 &	0.006        \\
		3     & RDD - weighted linear    & 0.0233 &	4.68 &	0.900 &	-0.005     \\
		4     & RDD - weighted quadratic & 0.0379 &	4.19 &	0.919 &	0.007    \\
		C1    & RDD conventional                    & 0.0680	& 0.99 &	0.838 &	0.003      \\
		C2    & RDD bias corrected                  & 0.0776	& 1.58 &	0.843 &	0.002    \\
		\multicolumn{6}{c}{\underline{B: Drinks Regularly}}                                                                                       \\
		1     & RDD - linear             & 0.0302 &	4.93 &	0.995 &	0.012      \\
		2     & RDD - quadratic          & 0.0551 &	2.71 &	0.981 &	0.009       \\
		\textbf{3}     & \textbf{RDD - weighted linear}    & 0.0278 &	4.93 &	1.000 &	0.006      \\
		4     & RDD - weighted quadratic & 0.0531 &	4.93 &	0.948 &	0.013       \\
		C1    & RDD conventional                    & 0.0744	& 1.02 &	0.952 &	0.006         \\
		C2    & RDD bias corrected                  & 0.0861	& 1.59 &	0.957 &	0.005        \\
		\multicolumn{6}{c}{\underline{C: Proportion of Days Drinks}}                                                                               \\
		\textbf{1}     & \textbf{RDD - linear}             & 0.0117 &	4.68 &	1.000 &	0.005      \\
		2     & RDD - quadratic          & 0.0261 &	3.21 &	0.938 &	0.003        \\
		3     & RDD - weighted linear    & 0.0119 &	4.93 &	1.000 &	0.002      \\
		4     & RDD - weighted quadratic & 0.0257 &	4.93 &	0.890 &	0.007    \\
		C1    & RDD conventional                    & 0.0372 &	1.02 &	0.890 &	0.002      \\
		C2    & RDD bias corrected                  & 0.0430 &	1.61 &	0.886 &	0.001    \\
		\bottomrule 
	\end{tabular}
	\begin{tablenotes}[para,flushleft]
	\footnotesize
	\item Notes: The results in this table are from a similar procedure to what is detailed in the Table \ref{tbl-main-model-perf} notes. However, these are for the MLDA application. The set of models considered is different, for reasons discussed in the text. For each of the three outcome variables, the bandwidth is allowed to vary from three months to 12 years on the right side, with the left side capped at three years. There are 210 treatment effect estimates for every model, one for each placebo-zone threshold.    
\end{tablenotes}
\end{table}

Table \ref{tbl-hs-mlda-results} shows the estimated discontinuities at the MLDA, using the placebo-zone-optimal models we have identified. These are each linear RDD models, with a bandwidth of three years on the left, and between 4.68 and 4.93 years on the right, as per Table \ref{tbl-mlda-candidate}. These results are directly comparable to those in Lindo et al.’s Figure 3. Each of the point estimates is similar to Lindo et al.’s 2-year bandwidth estimates.

\section{Conclusion and practical considerations}\label{sec-discussion}
Regression Discontinuity Design and related estimators are amongst the most important tools of empirical economics. When using such estimators, however, applied researchers are typically faced with choosing between hundreds or thousands of candidate specifications. The large number of candidates is due to the numerous dimensions by which these estimators can vary -- bandwidth, functional form, kernel, covariates are some of these dimensions, and these need not be the same on either side of the threshold. Various guidelines have been developed for model selection, but these generally only address one of these dimensions, whilst keeping others constant. In practice, contemporary applied work in leading economics journals still relies more on robustness testing than on model selection algorithms. Many such papers provide no explicit justification for model specification.

We have outlined a new approach for model selection which allows the performance of all candidate models to be assessed. The approach is conceptually straightforward. Each candidate model is assessed on its performance in estimating treatment effects in a placebo zone of the running variable – where the true effect is known to be zero. The RMSE of the resulting placebo estimates is the summary statistic by which each estimator is judged.

We have proven that this new approach is asymptotically optimal under restrictive conditions. More importantly, our simulations show that the approach also performs well under very different conditions, including with DGPs based on well-known applications. Our approach has potential to be useful for model-selection in a wide range of applications. We have demonstrated its use with three such applications within the paper. Researchers can implement the approach using our Stata command -pzms-. We recommend that researchers consider our approach whenever the available range of the running variable is relatively wide. One can compare the likely performance of our approach to other approaches by conducting a simulation exercise based on the data from any given application. This is quick and easy to do using our companion Stata command -pzms\_sim-, which implements simulations following the steps we discuss in Section \ref{sub-sec-sims-real-degps}. However the approach should not be seen as a completely automated procedure for unproblematically choosing an objectively best specification. In this section we discuss some complications and suggestions for using the approach judiciously.

\subsection{Applying the Approach for Fuzzy RDD and RKD}
Our method is relatively straightforward to apply for testing a set of candidate sharp-RDD models. For designs that rely on a first stage (e.g., fuzzy-RDD and RKD), a first stage relationship between the treatment variable and running variable may not exist in the placebo zone.\footnote{Our own application is unusual since the placebo treatments have a natural definition -- as a function of date obtained learners license.} In these settings, one option is to use our approach to choose an estimator for the reduced-form discontinuity (or kink) only -- that is, the discontinuity (or kink) of the outcome variable at the threshold. Such an approach may be useful in cases where it can be reasonably assumed that the first-stage discontinuity (or kink) estimate is relatively insensitive to the specification used. This is often the case in practice -- see for example \citeA{Abdulkadiroglu_etal_2014}.\footnote{This is often the case in practice, sometimes because the first-stage results from policy rules that function in a nearly deterministic way. Of the fuzzy-RDD papers we identified in Table \ref{tbl-recent-rd}, the majority arguably fall in this category in our view.}   

\subsection{How to set the maximum bandwidth for the placebo zone tests?}
In any given application of our proposed method, the analyst must choose a maximum bandwidth for the set of candidate models. This choice will depend on the specific constraints of the application. In principle, one would like to consider all possible bandwidths, but this is not practical. If the chosen maximum bandwidth is too large, the number of thresholds within the placebo zone will be too small for the procedure to be informative about model performance.\footnote{Larger bandwidths are also likely to yield higher serial correlation in placebo estimates.} 

When natural constraints do not occur, we suggest that researchers undertake some data examination before choosing a maximum bandwidth. A key consideration, for a given model type, is the relationship between treatment effect estimates and bandwidth. In Figure C7 we plot this for our applications in Section \ref{sec-further-apps}. For Head Start, the estimated treatment effect changes markedly up to a bandwidth of around 15, which indicates worth in setting a maximum at least equal to this value. Beyond 15, the effect size is more stable and (arguably) the discrepancy between estimates with bandwidths between 15 and 30 are not economically important. Since our estimator selects 15 as the preferred bandwidth when we set this as the maximum, there is little value in choosing a higher bandwidth and this seems like a sensible choice (if the solution had been interior, we would suggest increasing the maximum and reassessing). For MLDA outcomes, the estimates are fairly stable after five years.\footnote{We also suggest researchers consider the ESS. The ESS of the placebo estimates for the Head Start applications is quite small (5), due to the relatively small placebo zone. A larger maximum bandwidth would decrease the ESS even further. Importantly, however, even with this small ESS, our monte carlo simulations suggest that our procedure still outperforms popular alternative approaches (Table \ref{tbl-sim-stylized-dgps}, Panel E). The small ESS perhaps has greater implications for our alternate inference procedure, by which the estimate is only marginally significant.}        

\subsection{Allowing for heterogeneous treatment effects}
Our approach is perhaps most useful for model selection within (rather than between) a class of estimators. For example, consider the large set of candidate RDD estimators for a given application. Our approach assesses performance of such models with different bandwidths and different polynomial orders. Each of those candidate estimators has the same target parameter, and so comparing performance is relatively unproblematic.

Comparing performance between classes of models is more problematic, because they often estimate different parameters. Fuzzy-RDD models estimate LATEs, while RKDs estimate MTEs, RPJKDs estimate a weighted average of a LATE and a MTE (under additional assumptions of local MTE stability), while cohort-IV estimates a weighted average of a different set of LATEs. Our approach can be used to compare performance between such models. But this can only be done unproblematically if one is willing to assume that selection into treatment is unrelated to potential gains from that treatment. In our own application, this may be a reasonable assumption. It is less reasonable in many other applications.

More generally, researchers adopting our approach should carefully consider the implications of potential treatment effect heterogeneity. To be clear, placebo treatment effects in the raw data are precisely zero. This implies that model performance is assessed in a constant-treatment-effect context. This may be informative for model selection in more general contexts. But a more nuanced approach is to explore the implications for model performance if treatment effect heterogeneity is imposed into the placebo zone. In \citeA{kettlewell_siminski_2020} (Section 4.6) we demonstrate how one might go about this using our own MSDH application as an example.

\bibliographystyle{apacite}
\bibliography{references}            

\end{document}

\appendix
\begin{center}
	\section*{*Online Appendices*}
\end{center}

\setcounter{table}{0}
\renewcommand{\thetable}{A\arabic{table}}
\setcounter{figure}{0}
\renewcommand{\thefigure}{A\arabic{figure}}
\setcounter{equation}{0}
\renewcommand{\theequation}{A\arabic{equation}}

\section{Proof of Asymptotic Optimality}\label{sec-proof}

The proof below is based on the following key insight. To establish asymptotic optimality, it is sufficient to show that for each candidate specification, the MSE of the threshold estimate is the same as the MSE of each relevant placebo estimate. That is, to show that:

\begin{equation}\label{eq-theory1}
\textrm{MSE}(\hat{\tau}) = \textrm{MSE}(\hat{\tau}_k), \quad \textrm{for all } |k|\geq b,
\end{equation}

where $X=k$ is the location of each placebo threshold, and $b$ is the bandwidth used for this estimator.\footnote{The condition $|k|\geq b$ ensures that the discontinuity itself does not influence any of the placebo estimates.} $\textrm{MSE}(\hat{\tau}_k)=E(\hat{\tau}_k-\tau_k)^2=E(\hat{\tau}_k)^2$ since $\tau_k=0$. If equation (\ref{eq-theory1}) holds, this implies that: $E(\hat{\tau}_k)^2=$ $\textrm{MSE}(\hat{\tau})$ for all $|k|\geq b$, and hence $E(\hat{\tau}_k)^2=$ is independent of $k$.

For each candidate specification, we observe ${\hat{\tau}}_k$ for a finite set of $k$. Assume $k\in(b,\ b+1,\ b+2,\ldots,\ b+m)$. The mean of the $m+1$ observed squared-placebo estimates is $\frac{1}{m+1}\sum_{k=b}^{b+m}{{\hat{\tau}}_k}^2$. Since $E{({\hat{\tau}}_k)}^2$ is the same for all $k$, as $m$ gets large, $\underset{m\rightarrow\infty}{lim}{\frac{1}{m+1}\sum_{k=b}^{b+m}{{\hat{\tau}}_k}^2}=\textrm{MSE}(\hat{\tau})$. Therefore, as the placebo zone becomes large, the MSE of each candidate estimator is revealed as a simple function of the corresponding observed placebo estimates. And the optimal specification (the one with the lowest MSE of the threshold estimate) is also the specification which yields the lowest mean squared placebo estimates.

We will show that equation (\ref{eq-theory1}) holds under the following assumptions:
\begin{enumerate}
	\item $X$ is uniformly distributed
	\item $Y_i(0)$ and $Y_i(1)$ are homoscedastic and have the same variance: \\
	$Var(Y_i(0) | X) = Var(Y_i(0)) = Var(Y_i(1) | X) = Var(Y_i(1))$
	\item 	The Conditional Expectations of $Y_i(0)$ and $Y_i(1)$ with respect to $X$ are continuous, smooth, and have zero fourth derivatives: \\
	$\frac{d^4}{dX^4}E[Y_i(0)|X] = 0$    and     $\frac{d^4}{dX^4}E[Y_i(1)|X] = 0$.
	\item 	The Average Treatment Effect has a zero second derivative with respect to X: \\
	$\frac{d^2}{dX^2} E[Y_i(1) - Y_i(0) | X] = 0$
\end{enumerate}

Consider a set of local linear candidate estimators, each with a distinct, symmetric bandwidth ($b$).\footnote{All of the results below also hold under the same assumptions if the set of candidate estimators is expanded to include higher-order polynomials. The required assumptions are more restrictive if the set of candidate estimators includes local-linear estimators with asymmetric bandwidth (which requires a zero 3rd derivative in place of assumption 3), or zero-order polynomial estimators (which requires a zero 2nd derivative in place of assumption 3).} Let ${\hat{\tau}}_b$ denote the estimated treatment effect from a local linear estimator with bandwidth $b$. There is a finite sample of $n_b$ observations within this bandwidth, on each side of the threshold. The estimated treatment effect is ${\hat{\tau}}_b={\hat{\alpha}}_{2b}-{\hat{\alpha}}_{1b}$, where ${\hat{\alpha}}_{2b}$ and ${\hat{\alpha}}_{1b}$ can be estimated using two independent linear regressions, one on each side of the threshold, using data within the appropriate bandwidth. Dropping the $i$ subscript for simplicity, these regression equations are:

\begin{equation}\label{eq-theory2}
y=\alpha_1+\beta_1x+\varepsilon,\ \ \ \ \ \ \ \ -b<x<0
\end{equation}
\begin{equation}\label{eq-theory3}
y=\alpha_2+\beta_2x+\varepsilon,\ \ \ \ \ \ \ \ 0<x<b
\end{equation}

Similarly, let ${\hat{\tau}}_{bk}$ denote the estimated discontinuity at placebo threshold $X = k$ (where the true discontinuity is zero), using the same local linear specification and bandwidth as above.
${\hat{\tau}}_{kb}={\hat{\alpha}}_{k2b}-{\hat{\alpha}}_{k1b}$, where ${\hat{\alpha}}_{k1b}$ and ${\hat{\alpha}}_{k2b}$ are the estimates from the regressions below:

\begin{equation}\label{eq-theory4}
y=\alpha_{k1}+\beta_1\left(x-k\right)+\varepsilon,\ \ \ \ \ \ \ \ \left(k-b\right)<x<k
\end{equation}
\begin{equation}\label{eq-theory5}
y=\alpha_{k2}+\beta_2\left(x-k\right)+\varepsilon,\ \ \ \ \ \ \ \ k<x<(k+b)	
\end{equation}

The MSE of any treatment effect estimator is
\begin{equation}\label{eq-theory6}
\textrm{MSE}\left(\hat{\tau}\right)={E(\hat{\tau}-\tau)}^2=Var\left(\hat{\tau}\right)+{Bias\left(\hat{\tau}\right)}^2
\end{equation}

For each local linear estimator, $\textrm{MSE}\left({\hat{\tau}}_b\right)$ is a function of the variance of the estimated constant from both regressions in equations (\ref{eq-theory2}) and (\ref{eq-theory3}), and the bias of the estimator:	

\begin{equation}\label{eq-theory7}
\textrm{MSE}\left({\hat{\tau}}_b\right)=E{({\hat{\alpha}}_{2b}-{\hat{\alpha}}_{1b}-\tau)}^2=Var\left({\hat{\alpha}}_{1b}\right)+Var\left({\hat{\alpha}}_{2b}\right)+{Bias\left({\hat{\tau}}_b\right)}^2
\end{equation}

Assumptions (3) and (4) imply that the true DGP takes the following form:

\begin{equation}\label{eq-theory8}
y=\alpha+(\tau+\vartheta x)\textbf{1}(x>0)+\theta_1x+\theta_2x^2+\theta_3x^3+\varepsilon,	
\end{equation}

In other words, the assumptions imply that the CEF can be globally cubic, with a potential discontinuity ($\tau$) and kink ($\vartheta$) at the threshold.\footnote{$\theta_1$, $\theta_2$ and $\theta_3$ can be any real numbers. The inclusion of any other term to equation (\ref{eq-theory8}) would violate assumption (3). As we show in this section, the bias of local-linear RDD estimators is a function of the third derivative of the DGP’s CEF. If this derivative is not a constant (implied by assumption 3), the bias of placebo estimates does not equal the bias of the threshold estimate, and so the proof does not hold.} Following standard characterisations of omitted variable bias (see for example \citeA{Wooldridge2010} pp. 66-67 and \citeA{Greene2003} pp. 148-149), it can be shown that:

\begin{equation}\label{eq-theory9}
E({\hat{\alpha}}_{1b}|x)=\alpha+\theta_2\delta_{L2b}+\theta_3\delta_{L3b}
\end{equation}

Where $\delta_{L2b}$ is the constant from the linear projection of $x^2$ on $x$, and $\delta_{L3b}$ is the constant from the linear projection of $x^3$ on $x$, both with $-b<x<0$. Similarly, for ${\hat{\alpha}}_{2b}$, using data on the RHS $(0<x<b)$:

\begin{equation}\label{eq-theory10}
E({\hat{\alpha}}_{2b}|x)=(\alpha+\tau)+\theta_2\delta_{R2b}+\theta_3\delta_{R3b}
\end{equation}

Therefore,

\begin{equation}\label{eq-theory11}
E\left({\hat{\tau}}_b|x\right)=E({\hat{\alpha}}_{2b}|x)-E({\hat{\alpha}}_{1b}|x)=\left(\alpha+\tau\right)+\theta_2\delta_{R2b}+\theta_3\delta_{R3b}-(\alpha+\theta_2\delta_{L2b}+\theta_3\delta_{L3b})
\end{equation} 

However, $\delta_{R2b}=\delta_{L2b}$ and $\delta_{R3b}={-\delta}_{R3b}$.\footnote{Given the assumed uniform distribution of $x$, the constant from a linear projection of $x^2$ on $x$ is the same with the domain of $x$ restricted to $-b<x<0$ as it is with $0<x<b$. The constant from a linear projection of $x^3$ on $x$ also remains the same except with opposite sign, for $-b<x<0$ versus $0<x<b$.} Therefore,

\begin{equation}\label{eq-theory12}
E\left({\hat{\tau}}_b|x\right)=\tau+2\theta_3\delta_{R3b}
\end{equation} 

so that

\begin{equation}\label{eq-theory13}
Bias\left({\hat{\tau}}_b|x\right)=2\theta_3\delta_{R3b}.	
\end{equation} 

This bias is proportional to $\theta_3$, and unrelated to any other parameters of the true DGP. $Bias\left({\hat{\tau}}_b|x\right)$ is hence proportional to the third derivative of the DGP’s CEF.
 
We now show that $Bias\left({\hat{\tau}}_b|x\right) = Bias\left({\hat{\tau}}_{kb}|x\right)$, for $|k|\geq b$.
Substituting $x_k=x-k$, equations (\ref{eq-theory4}) and (\ref{eq-theory5}) are equivalent to

\begin{equation}\label{eq-theory14}
y=\alpha_{k1}+\beta_1x_k+\varepsilon,\ \ \ \ \ \ \ \ -b<x_k<0 
\end{equation} 

\begin{equation}\label{eq-theory15}
y=\alpha_{k2}+\beta_2x_k+\varepsilon, \ \ \ \ \ \ \ \ 0<x_k<b
\end{equation}

and the DGP in equation (\ref{eq-theory8}) can be expressed as

\begin{equation}\label{eq-theory16}
y=\alpha+\tau+(\theta_1+\vartheta)(x_k+k)+\theta_2{(x_k+k)}^2+\theta_3{(x_k+k)}^3+\varepsilon    
\end{equation}

if $k\geq b$. And, similarly,

\begin{equation}\label{eq-theory17}
y=\alpha+\theta_1(x_k+k)+\theta_2{(x_k+k)}^2+\theta_3{(x_k+k)}^3+\varepsilon
\end{equation}

if $k\leq -b$.

Expanding equation (\ref{eq-theory16}) and collecting terms:

\begin{equation}\label{eq-theory18}
y=\pi_0+\pi_1x_k+\pi_2{x_k}^2+\theta_3{x_k}^{3}+\varepsilon
\end{equation}

Where $\pi_0=\alpha+\tau+k\theta_1+k\vartheta+k^2\theta_2+k^3\theta_3,  \pi_1=\theta_1+\vartheta+2k\theta_2+3k^2\theta_3,  \pi_2=\theta_2+3k\theta_3$

And similarly if one expands (\ref{eq-theory17}).

Equations (\ref{eq-theory14}), (\ref{eq-theory15}) and (\ref{eq-theory18}) are equivalent to equations (\ref{eq-theory2}), (\ref{eq-theory3}) and (\ref{eq-theory8}), respectively, with $x>0$, or similarly if $x < 0$, with the threshold at $x_k=0$. As shown above, the bias of the RDD estimate is proportional only to the third derivative of the true DGP’s CEF. The third derivative (${6\theta}_3$) of (\ref{eq-theory18}) is the same as for (\ref{eq-theory8}), and so $Bias\left({\hat{\tau}}_{kb}|x\right)=Bias\left({\hat{\tau}}_b|x\right)$, for $|k|\geq b$, noting again the assumed uniform distribution of $x$.

It is trivial to show that $Var\left({\hat{\tau}}_{kb}\right)=Var\left({\hat{\tau}}_b\right)$, given assumptions (1) and (2). Therefore $\textrm{MSE}\left({\hat{\tau}}_{kb}\right)=\textrm{MSE}\left({\hat{\tau}}_b\right)$, which we have argued above is sufficient for establishing the asymptotic optimality of our procedure under assumptions (1)--(4).

\setcounter{table}{0}
\renewcommand{\thetable}{B\arabic{table}}
\setcounter{figure}{0}
\renewcommand{\thefigure}{B\arabic{figure}}

\pagebreak
\newpage

\section{An application: Minimum supervised driving hours and motor vehicle accidents}\label{sec-policy-background}       

\subsection{Overview}
Globally, motor vehicle accidents (MVAs) are the leading cause of death for children and young adults, with more than 1.3 million people aged 5-29 years dying from MVAs each year \cite{who_2018}. To reduce the fatality rate for young drivers, governments around the world have introduced graduated driver licensing (GDL). GDL limits the exposure of young drivers to risky situations with the goal of better preparing them for unsupervised driving. It typically operates in three stages: a learner stage in which driving is supervised; a provisional stage in which driving is unsupervised but subject to restrictions; and an unrestricted stage. To progress, drivers are required to demonstrate competence by passing written exams and practical driving tests.

During the learner stage drivers usually need to complete a mandatory number of supervised driving hours -- the MSDH requirement. Most U.S. states mandate between 40-60 hours \cite{IIFHS_2020}. In Australia, the three most populous states (New South Wales (NSW), Victoria and Queensland) require 100-120 hours. 

It is generally believed that GDL as a system has reduced MVAs for young drivers \cite{mcknight_peck_2002,foss_2007,shope_2007}; however, there is little evidence on the independent effects of different components of GDL. Typically researchers rank GDL systems by some measure of ‘strictness’ and use state variation in regulatory settings to identify policy effects \shortcite<e.g.>{dee_etal_2005,chen_etal_2006,traynor_2009,trempel_2009,karaca_ridgeway_2010,masten_etal_2011,lyon_etal_2012,steadman_etal_2014}.\footnote{\citeA{moore_morris_2020} identify the causal effect of one common component of GDL -- night-time passenger restriction -- on MVAs in NSW, Australia. Using  variation in MVAs by time-of-day and a difference-in-differences design, they find large reduction effects.} Results consistently show that states with stricter GDL systems experience lower rates of fatalities and MVAs involving injury among teenage drivers. 

We are aware of only one study \cite{gilpin_2019} that attempts to estimate the independent \textit{causal} effect of MSDH on MVAs.\footnote{\citeA{trempel_2009} and \shortciteA{mccartt_etal_2010} estimate models that control for MSDH in U.S. state-level studies but do not control for state fixed-effects or time trends. \shortciteA{obrien_etal_2013} study an increase from 0 to 30 MSDH in Minnesota using a before-after design.} \citeA{gilpin_2019} uses a difference-in-differences design with variation between and within U.S. states and finds going from no MSDH requirement to having some MSDH requirement counter-intuitively increased fatalities overall, but had no effect per licensee.  

\subsection{Policy environment and causal variation}
NSW adopted GDL on 1 July 2000. Prior to this, a licensing system with GDL features operated. Under the pre-July 2000 system, the minimum age for obtaining a learner license was 16 years, there was a minimum six-month learner period and one year provisional license period, and the minimum age for obtaining a provisional license was 17 years. There was no MSDH requirement. The introduction of GDL resulted in two restricted provisional license periods -- provisional 1 (P1) and provisional 2 (P2), which remain in place today. It also resulted in a large increase in MSDH -- from 0 to 50 hours. Importantly, the six-month minimum learner period and 17 years minimum age for obtaining a provisional license remained in place. 

Although the 1 July 2000 MSDH increase was not implemented in isolation, it was implemented in such a way that people born up to one year before 1 July 1984 experienced the same provisional licensing conditions as those born after this date. This is because people born within one year prior to 1 July 1984 turned 16 before the introduction of GDL (meaning they could obtain their learner license before 1 July 2000 and avoid the increase to MSDH) but turned 17 \textit{after} 1 July 2000, meaning they could not avoid the new GDL provisional regulations. Consequently, the GDL experience of people born within one year of 1 July 1984 only differs with regards to the 50 MSDH requirement.

The GDL system was expanded on 1 July 2007. The most significant changes were passenger restrictions for night-time driving for P1 drivers, a zero-tolerance policy for speeding (immediate three-month suspension of license) and an increase in MSDH from 50 to 120 hours (minimum 20 hours night-time driving). There was also an increase to the minimum learner period from six to 12 months. In Table \ref{tbl-gdl} we highlight the main difference between the pre- and post-July 2007 regimes \cite<see>[for a detailed comparison]{bates_2012}. As with the 1 July 2000 policy changes, the 17 years minimum age for obtaining a provisional license meant that people born up to one year prior to 1 July 1991 (meaning they would turn 17 \textit{after} 1 July 2007) could obtain their learner license before 1 July 2007 and avoid the MSDH increase but would be subject to the same provisional regulations as those born after 1 July 1991.\footnote{Because the minimum learner period also increased at the same time as MSDH in 2007, these policy effects may be confounded in our analysis. To separate these effects and isolate the impact of increased driving practice, we consider the impact of the policy change on time spent on the learner license and how our results change when we control for this.}

\begin{table}[ht!]
	\caption{NSW GDL characteristics}\label{tbl-gdl}
	\begin{tabular}{p{4cm} p{5.5cm}p{5.5cm}}
		\toprule
		& 1 July 2000–-30 June 2007 & From 1 July 2007 \\
		\midrule
		MSDH & 50 & 120\textsuperscript{a} \\
		Min. learner age & 16 years & 16 years \\
		Min. learner period & 6 months & 1 year \\
		Min. P1 age & 17 years & 17 years \\
		Min. P1 period & 1 year & 1 year \\
		P1 restrictions & Max speed (90km/h); 4 demerit points\textsuperscript{b}; engine restrictions\textsuperscript{c} & Max speed (90km/h); 4 demerit points; engine restrictions\textsuperscript{c}; night-time passenger restrictions; immediate license suspension for speeding \\
		Blood alcohol limit & 0.02 (0.00)\textsuperscript{d} & 0.00 \\
		\bottomrule
	\end{tabular}
	\begin{tablenotes}[para,flushleft]
		\footnotesize
		\item Notes: \textsuperscript{a}In NSW drivers receive financial penalties and demerit points for driving offences. Drivers who accrue a critical number of demerit points have their license suspended (4 for P1 drivers, 12 for unrestricted license drivers). \textsuperscript{b}Minimum 20 hours at night. In December 2009 new rules were introduced that allowed learners to convert hours with a qualified driving instruction at a ratio 3:1 with regular supervised driving (limited to 10 hours). \textsuperscript{c}Since 11 July 2005 P1 drivers have been prohibited from driving certain high-powered vehicles. \textsuperscript{d}Lowered to this on 3 May 2004. 	
	\end{tablenotes}
\end{table}

Our empirical analysis exploits the fact that people born just before 1 July 1984 (1991) are likely to be statistically similar to those born just after 1 July 1984 (1991) but differ in their MSDH experience. Since people often delay getting their license until sometime after their 16th birthday, there is a positive slope in the probability of treatment on the left-hand-side of the threshold, while everyone is treated on the right-hand-side. 

\subsection{Compliance}
We do not observe learner driving hours so cannot assess compliance directly. However, the limited Australian evidence supports high compliance with MSDH regulations. For example, surveys of newly licensed drivers found 98.2\% complied with NSW's 50 MDSH requirement \shortcite{bates_etal_2010}, while only 12.8\% admitted to rounding up hours and 4\% to including additional hours not undertaken in Queensland \shortcite{scott-parker_etal_2011}. A survey by \shortciteA{bates_etal_2014b} also found strong agreement from parent supervisors about the accuracy of recorded hours.

One reason to expect high compliance in our setting is because learner drivers are required to record all journeys in a log book, with each entry signed off by the supervising driver (not a P1 or P2 driver). If there is evidence of falsification the learner may be barred from taking the practical driving test for up to six weeks and fined (fines also apply to supervising drivers). 

A related question is whether the policy is binding at all. \shortciteA{bates_etal_2010} compared new drivers in NSW to Queensland when NSW had a 50 MSDH requirement and Queensland had none. The average self-reported hours was only slightly higher in NSW (73 compared to 64). However, while 98.8\% of drivers reported completing at least 50 hours in NSW, more than half in Queensland reported doing less than this. A 50 MSDH requirement would have therefore been binding for a significant portion of learners in Queensland. It is important to note that any effects of increased MSDH we estimate will be driven by the subsample of learners who would have completed less than the minimum requirement in the absence of the policy.

\subsection{Data}\label{sec-data}
Our data are individual level administrative records supplied by the NSW Centre for Road Safety (CRS). Driver licencing data come from the universe of licensing history for NSW drivers born from 1 January 1980. For these individuals, we know their age in (completed) weeks at the time they obtain their license. The MVA data are from a separate dataset containing the universe of police reported MVAs from 1 January 1996 to 26 October 2017. MVAs are accidents occurring on NSW roads in which at least one vehicle was towed away or one of the occupants was injured or killed, which by law must be reported to NSW Police (we exclude motorcycle crashes from the analysis). We link the license and MVA datasets using a unique identifier provided by CRS. 

\subsubsection{Main variables}

Our outcome variables are indicators for whether an MVA occurred within certain periods. We focus primarily on the probability a person was the driver in an MVA within one year of obtaining his/her P1 license (during which drivers are typically aged 17-20 years). The one year criterion matches the mandatory time period before a P1 driver can take the test to become a P2 driver, and therefore reflects an expected period of progression in driver safety. We also find little evidence that the MSDH reforms improve driver safety beyond this period. In further analysis we limit attention to MVAs that resulted in injury to a driver or passenger or resulted in fatality. 

Our running variable, date of birth (DOB), is constructed as follows. For each entry a person has in the license dataset (for example, when they renew their license or move to a different license class), we observe that person’s age in weeks on that day. That means that for people with one entry, we know their precise DOB within 6 days. For people with multiple entries we can narrow that window down; for more than 50\% of people we can narrow it down to within three days. We use the midpoint of the minimum and maximum possible DOB as our variable, considering all licensing history data available to us.

\begin{figure}[ht!]
	\centering
	\caption{DOB distribution plots}\label{fgr-dob-density}
	\begin{tabular}{cc}
		\includegraphics[width=.5\textwidth]{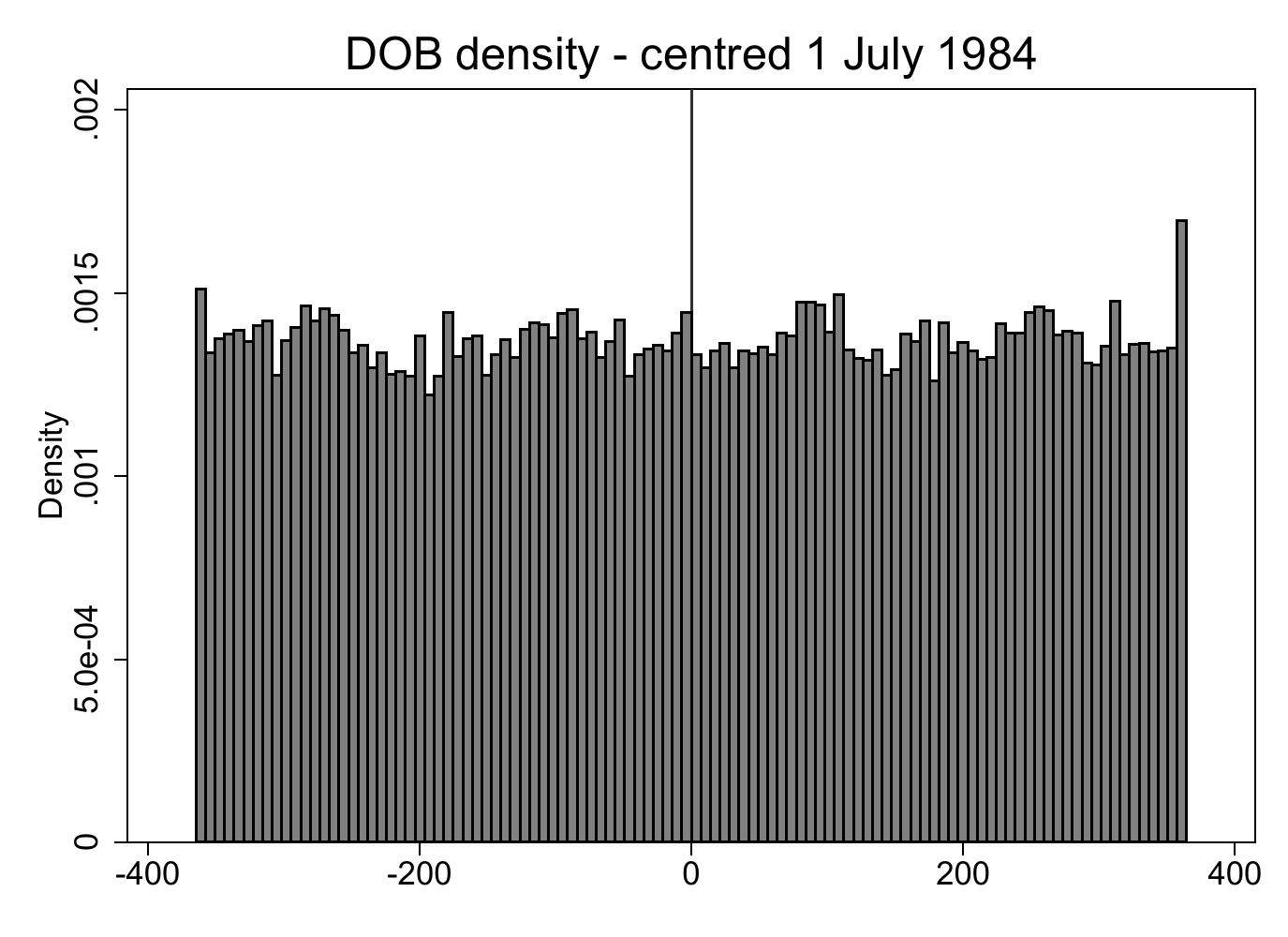} &
		\includegraphics[width=.5\textwidth]{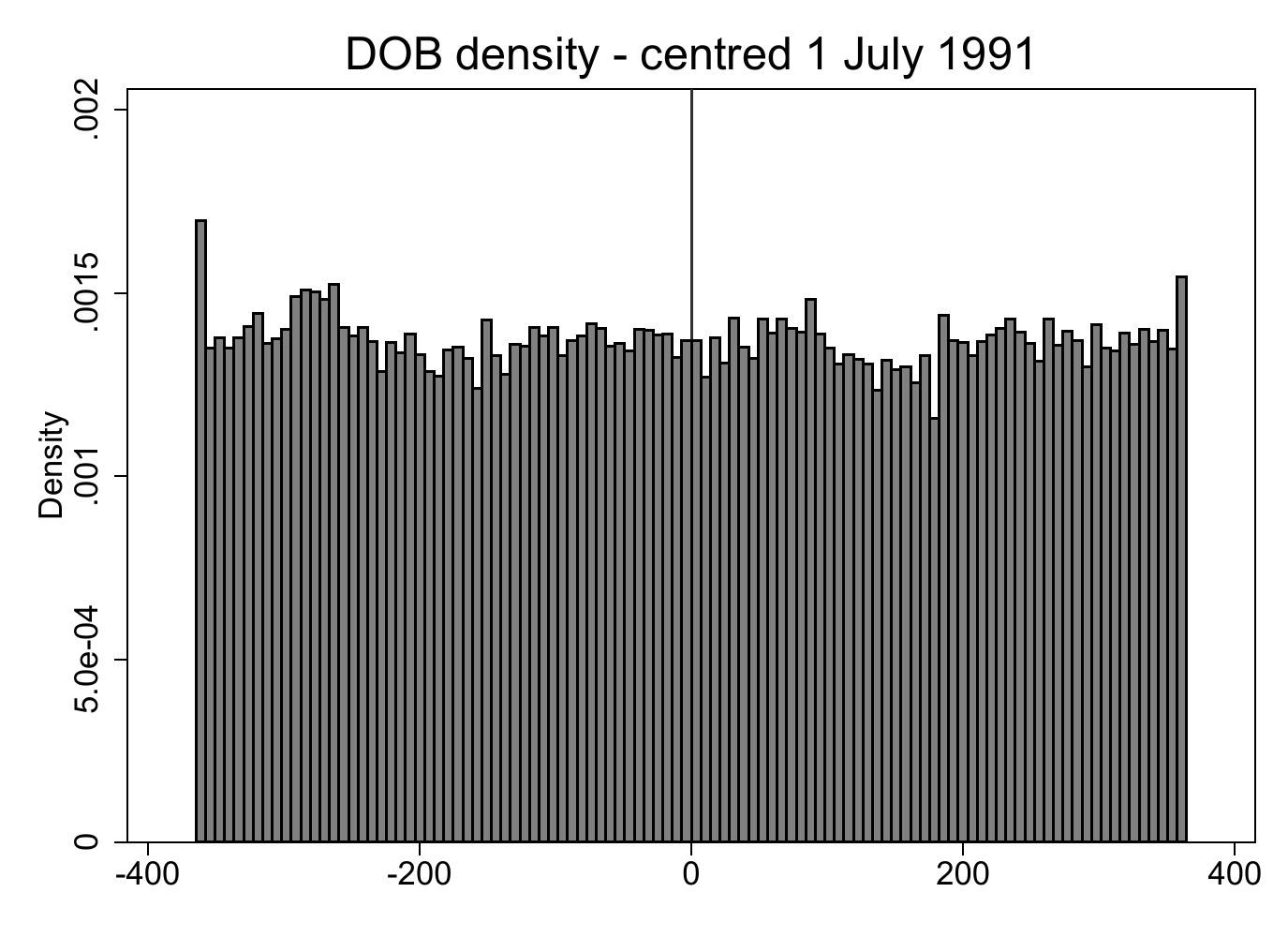} \\
	\end{tabular}
	\begin{tablenotes}
		\footnotesize
		\item $n$ = 154,524 in Panel A. $n$ = 160,301 in Panel B. 7-day bins.   
	\end{tablenotes}
\end{figure}

See Figure \ref{fgr-dob-density} for density plots for DOB. 

\subsection{Matching MVAs to driver license records}

From October 2002 onwards, we can match 99.6\% of MVAs involving drivers who, according to the MVA data are between 17-20 years old and licensed in NSW, to the license data.\footnote{The match rate is almost identical (99.5\%) if we instead look at all people who, based on their age recorded in the MVA data, we can be certain were born after 1980.} Prior to this date the match rate is discontinuously lower by around 20 percentage points between July 2002 and October 2002 and 15 percentage points pre-July 2002 (see Appendix Figure \ref{fgr-match-rates}). CRS cited an improvement in record keeping practices as a reason for the discontinuity but were unable to provide further details. Our analysis in Appendix Figure \ref{fgr-match-rates} indicates that the discontinuities are not limited to any subset of MVAs by characteristics, which would have allowed us to exclude inconsistently recorded MVAs. To address the missing MVAs we therefore inflate the MVA indicators we use as dependent variables for people who are not matched to an MVA (i.e. are recorded as having not had an MVA in our raw data) by a factor equal to the probability they actually did have an MVA given what we know about the rate of non-matched MVAs. 

\begin{figure}[ht!]
	\centering
	\caption{Match rates: License and MVA data}\label{fgr-match-rates}
	\includegraphics[width=1.0\textwidth]{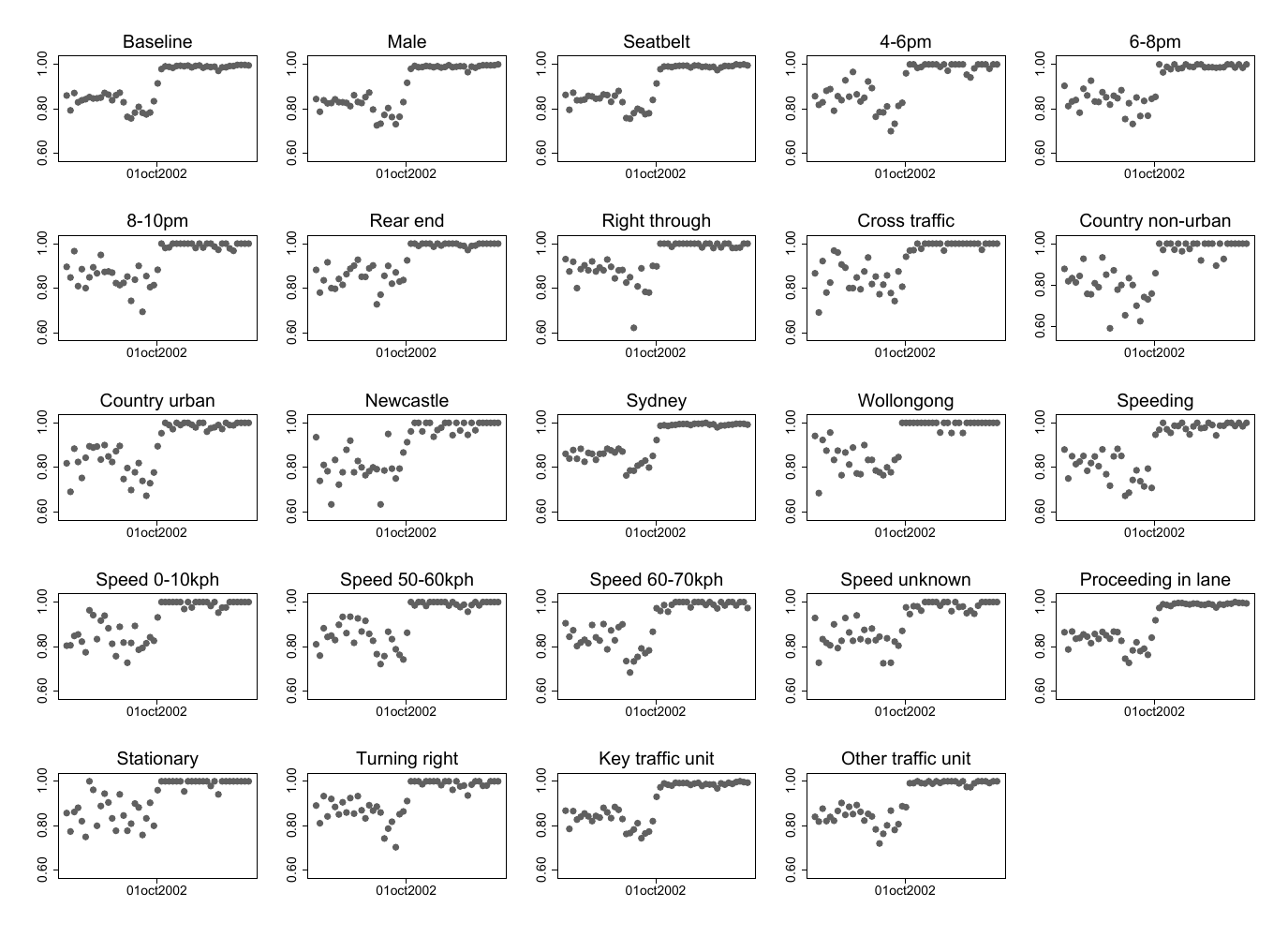}
	\begin{tablenotes}
		\footnotesize
		\item Notes: Each scatter point corresponds to the average percentage of MVAs for 17-20 year old drivers that can be matched to license records for NSW licensed drivers (15 day groupings). Baseline: The full sample of 17-20 year old drivers; Male: Males only; Seatbelt: driver wearing seatbelt; 4-6pm: MVA between 4-6pm; 6-8pm: MVA between 6-8pm; 8-10pm: MVA between 8-10pm; Rear end: MVA reason, rear-ender; MVA reason, right through; Cross traffic: MVA reason, cross traffic; Country non-urban: MVA in country non-urban region; Country urban: MVA in country urban region; Newcastle: MVA in Newcastle region; Sydney: MVA in Sydney region; Wollongong: MVA in Wollongong region; Speeding: speeding involved in MVA; Speed 0-10kph: main vehicle travelling between 0-10kph; Speed 50-60kph: main vehicle travelling between 50-60kph; Speed 60-70kph: main vehicle travelling between 60-70kph; Speed unknown: main vehicle speed unknown; Proceeding in lane: manoeuvre before crash, proceeding in lane; Stationary: manoeuvre before crash, stationary; Turning right: manoeuvre before crash, turning right; Key traffic unit: vehicle was key traffic unit in MVA; Other traffic unit: vehicle was not key traffic unit in MVA.             
	\end{tablenotes}
\end{figure}

Focusing on our main dependent variable (any MVA within 12 months of obtaining P1 license), our preferred approach adjusts the crash probability for a person obtaining their P1 license on day $t$ by:

\begin{equation}\label{eq-adj-factor}
1-\left(1-\left[\sum_{t}^{t+365}\frac{MVA_t}{n_t}\right] \times 0.15\right)^{\min\{{t^*-t,365}\}} \times \left(1-\left[\sum_{t}^{t+365}\frac{MVA_t}{n_t}\right] \times 0.20\right)^{\mathbf{1.}[t>t^*] \times \min\{{t-t^*,130}\}} 
\end{equation}

where $MVA_t$ is total number of matched MVAs involving drivers aged 17-20 years, $n_t$ is the total number of licensed drivers aged 17-20 years (so that $\sum_{t}^{t+365}\frac{MVA_t}{n_t}$ is the unadjusted probability of being involved in an MVA within one year of obtaining a P1 license on day $t$), $t^*$ indicates 1 July 2002 and $\mathbf{1}$.$\left[t>t^\ast\right]$ is an indicator for $t>t^\ast$. We also use Eq. \ref{eq-adj-factor} to adjust indicators for MVAs involving injury since the match rates for these crashes is almost identical to the rate for MVAs overall. For MVA indicators over six-month windows we replace 365 with 183. 

A more sophisticated approach of obtaining the adjustment factor is to replace 0.15 and 0.20, which are approximate rates at which MVAs can be matched to license data in the pre-July 2002 and July-October 2002 periods, with daily estimates for this rate and calculate

\begin{equation}\label{eq-adj-factor-exact} 
1-\sum_{t}^{t+365}\left(1-\left[\sum_{t}^{t+365}\frac{MVA_t}{n_t}\right] \times Match \ rate_t \right).
\end{equation} 

\begin{figure}[ht!]
	\centering
	\caption{Adjustment factors}\label{fgr-adjust-factors}
	\includegraphics[width=0.5\textwidth]{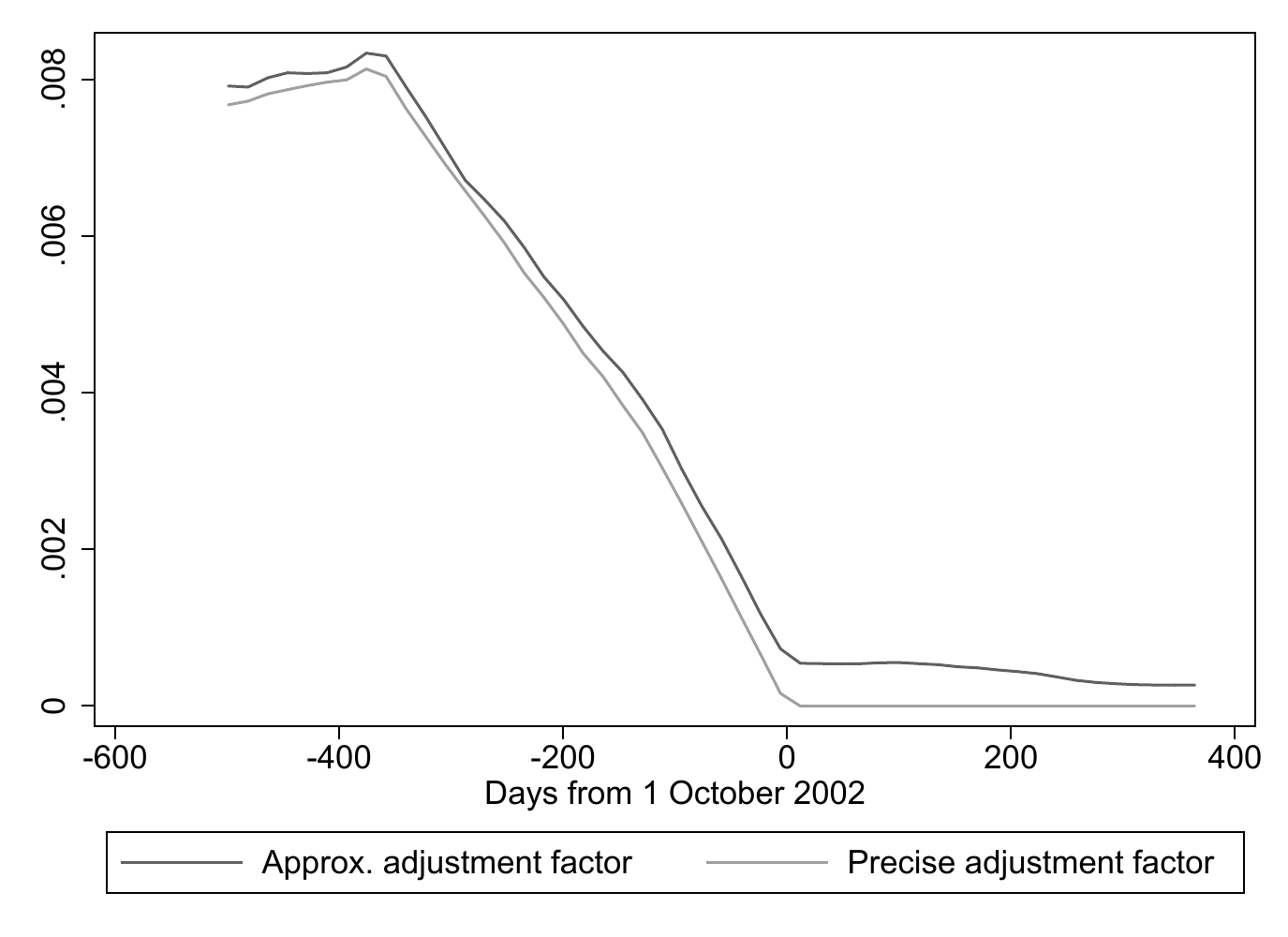}
	\begin{tablenotes}
		\footnotesize
		\item Notes: The y-axis shows by how much the MVA within first year of P1 license probability is adjusted for a person obtaining their P1 license on day $t$ (the x-axis) in order to account for missing MVA data. Formulaic details are in Eq. \ref{eq-adj-factor} (Approx. adjustment factor) and Eq. \ref{eq-adj-factor-exact} (Precise adjustment factor).   
	\end{tablenotes}
\end{figure}

This approach allows for more variation in the match rate; however, it may suffer from rare events bias since there are often few MVAs on a given day (expanding the time unit can solve this but setting a new time unit is arbitrary). In practice, the two approaches give very similar adjustment factors (Appendix Figure \ref{fgr-adjust-factors}) and after confirming the choice had no effect on our main results, we committed to using the simpler and more transparent approximation approach. 

\subsection{Sample restrictions}

Throughout our analysis we always maintain the following sample restrictions:

\begin{itemize}
	\item \textit{Exclusion of people whose license history violates the GDL rules} -- some people in our dataset appear to have licencing histories that violate rules of the GDL system. For example, some people obtain their P1 license before turning 17 years; others are on their learner license for less than the mandatory period. These violations may have a variety of unobserved causes, such as people moving from interstate or data error. Only around 0.3\% of people in our analysis sample violate the GDL rules and we exclude them throughout the analysis. 
	
	\item \textit{Exclusion of people whose eligibility to avoid treatment status is uncertain} -- in our models, people cannot avoid ‘treatment’ if they are born after a certain date. People whose minimum and maximum possible DOB straddles the threshold date are dropped in the models we estimate (see discussion in Section \ref{sec-data}). Recall that for people with multiple license records we can narrow this range. However, since the reforms may have affected licensing behavior (and consequently the accuracy of DOB), we only use the first license record when imposing this restriction.
	
	\item \textit{Exclusion of people who obtained their P1 license after age 25} -- these people are not subject to the MSDH requirements. Only 8.7\% of people in our dataset obtained their P1 license after age 25.
\end{itemize}

\subsection{Additional details on the data}
\begin{itemize}
\item \textit{Change in MVA record keeping 2014} -- in 2014, a policy change meant that NSW Police reported fewer MVAs from this year onwards. This was due to NSW Police no longer being refquired to attend a crash scene and investigate for tow away MVAs where nobody was injured or killed. This policy change is largely inconsequential for us as we only consider the periods July 2000-June 2008 in our analysis, and most people we observe obtain their P1 license more than 12 months before 2014. Moreover, exposure to this period (in terms of days on P1 after 2014) is a smooth function of DOB. We therefore ignore this change. 
	
\item \textit{License suspensions and demotions} -- in our analysis we focus on the first date a person obtained their P1 license and ignore any suspensions (e.g. for speeding or drunk driving), demotions (i.e. being made to redo the learner class due to a serious driving offence) or moves out of NSW after this date. If these events are unrelated to MSDH, which seems reasonable, then our MVA rates will be equally affected by them in the treatment and untreated groups.  While we do not observe suspensions or moves in our data, we note that in 99.1\% of cases the expiry date for learner license matches the date of effect for the first time obtaining a P1 license, which indicates demotions are rare.

\item \textit{Other policy changes between 200 and 2007}  -- Policy changes that may have affected MVAs in our data window are lowering the Blood Alcohol Limit from 0.02 to 0 (3 May 2004), engine restrictions (11 July 2005) and changes to the GDL system that occurred on 1 July 2007 for P1 drivers (see Table \ref{tbl-gdl}), in particular night-time passenger restrictions. However, days exposed to these policies is a smooth function of DOB (see Appendix Figure \ref{fgr-restrictions}) so they are unlikely to have affected MVAs in a way that discontinuously depends on DOB.
\end{itemize}

\subsubsection{Descriptive Statistics}

\begin{figure}[ht!]
	\centering
	\caption{Scatter and fit plots: Main variables by DOB}\label{fgr-main-vars-by-dob}
	\begin{tabular}{cc}
		\includegraphics[width=0.5\textwidth]{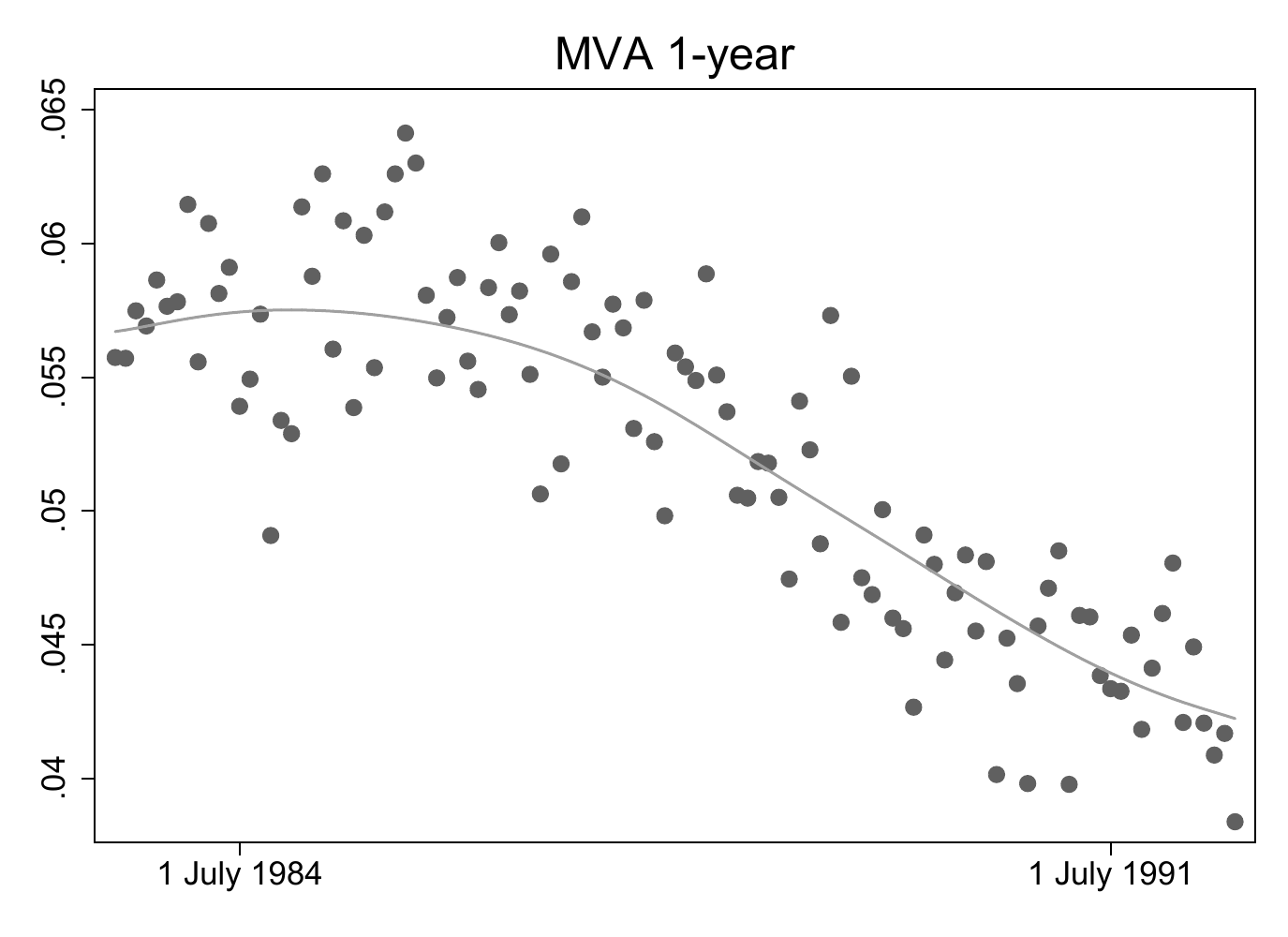} &
		\includegraphics[width=0.5\textwidth]{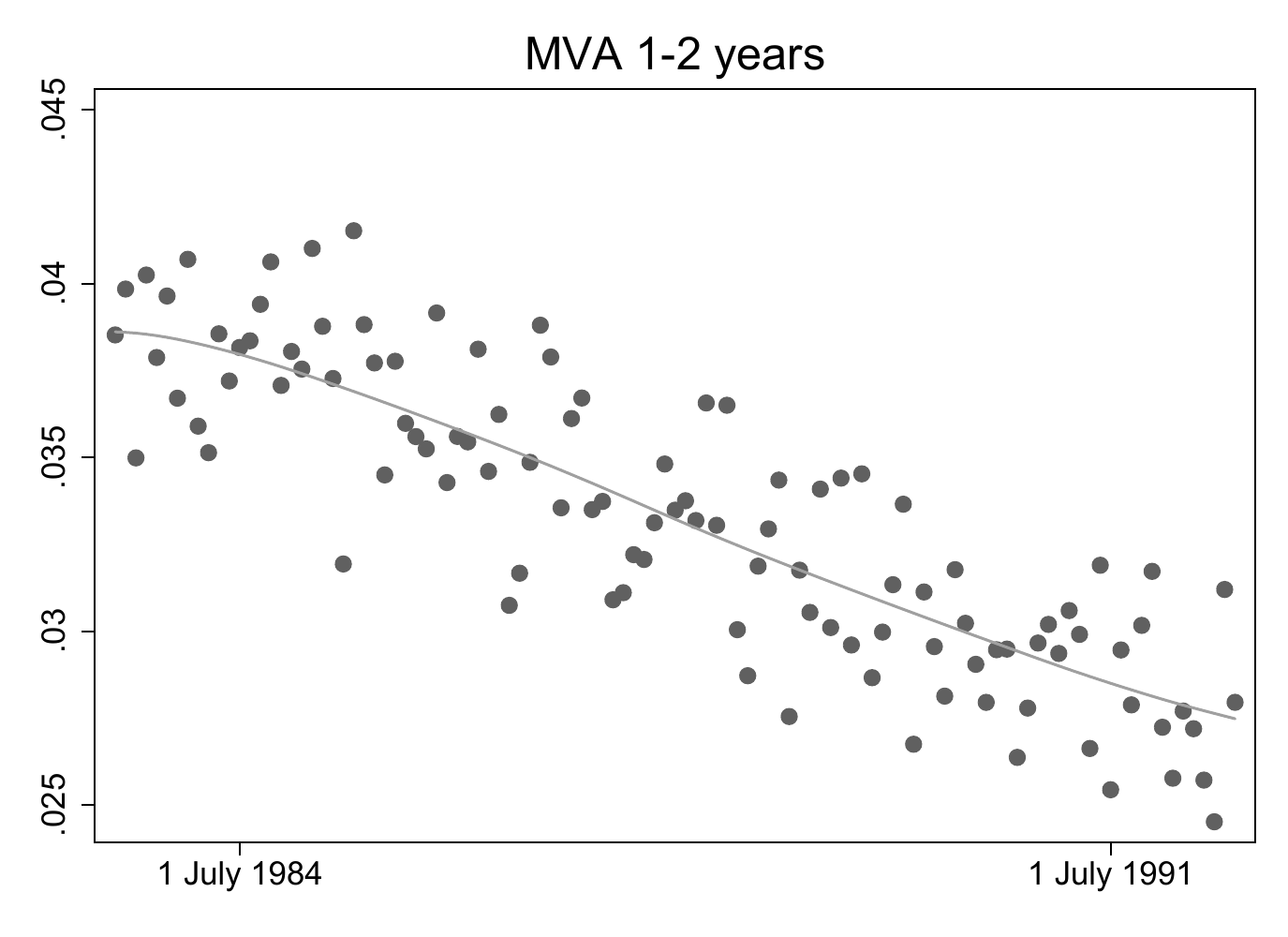} \\
		\includegraphics[width=0.5\textwidth]{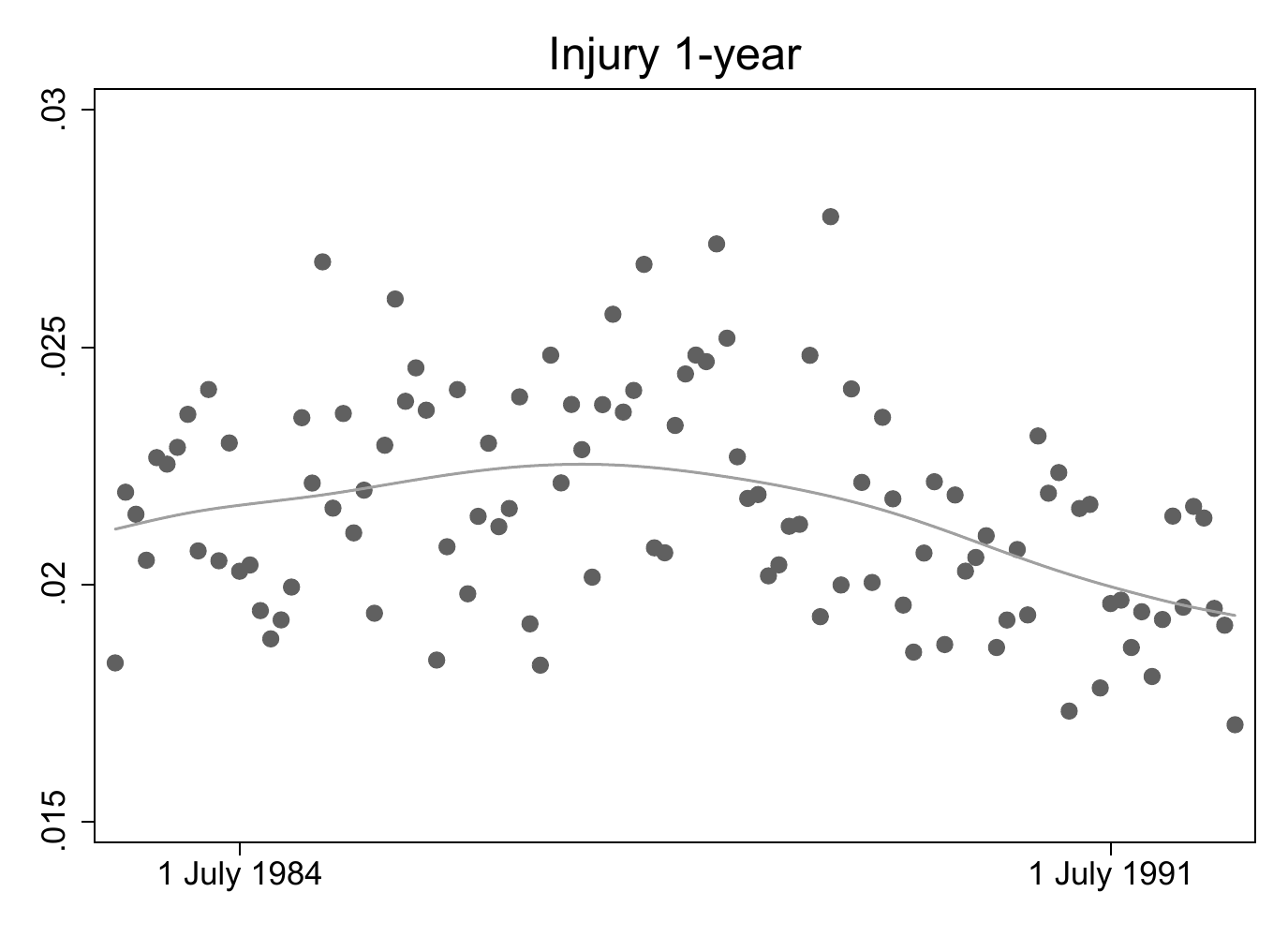} &
		\includegraphics[width=0.5\textwidth]{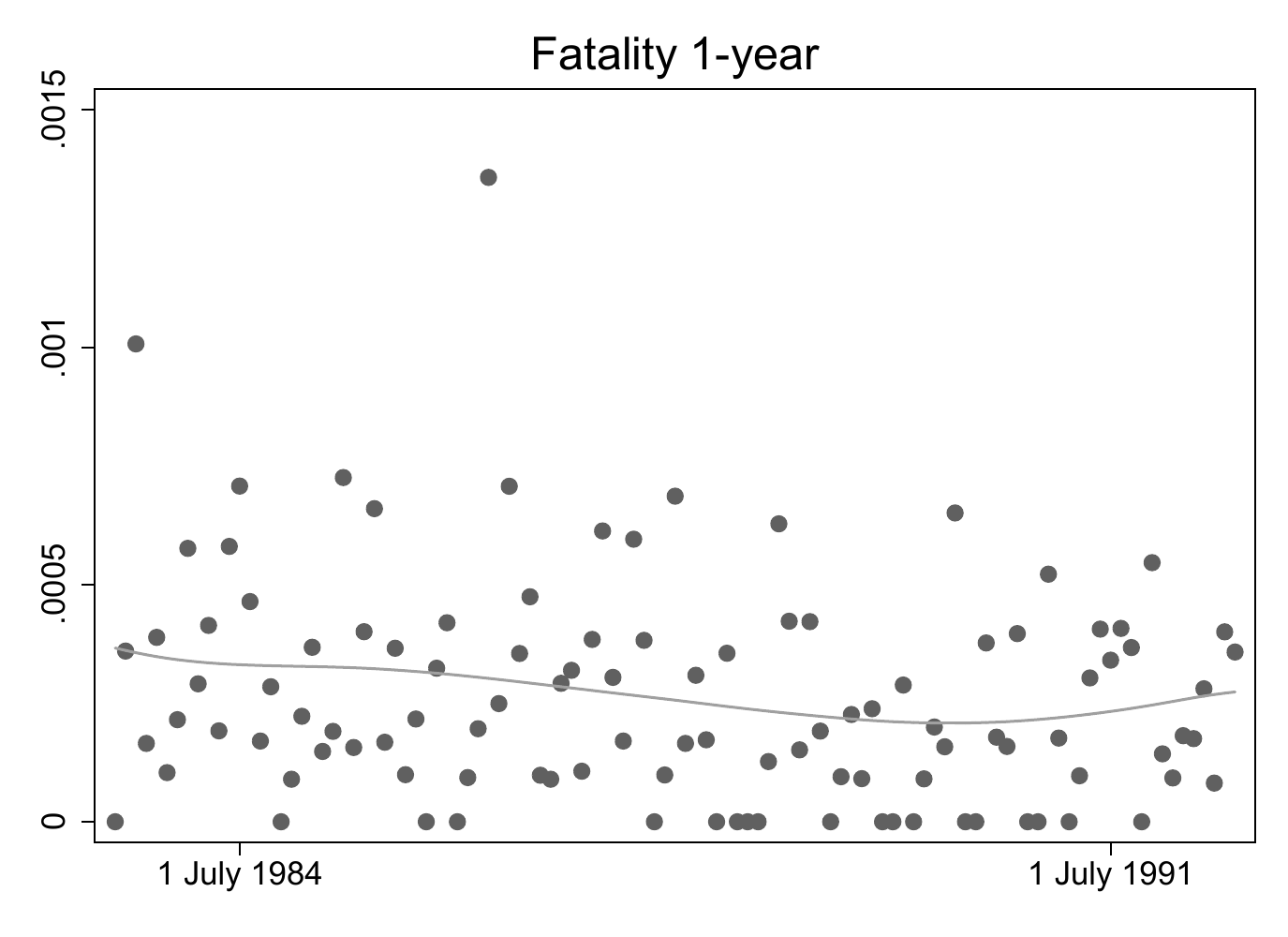} \\
		\includegraphics[width=0.5\textwidth]{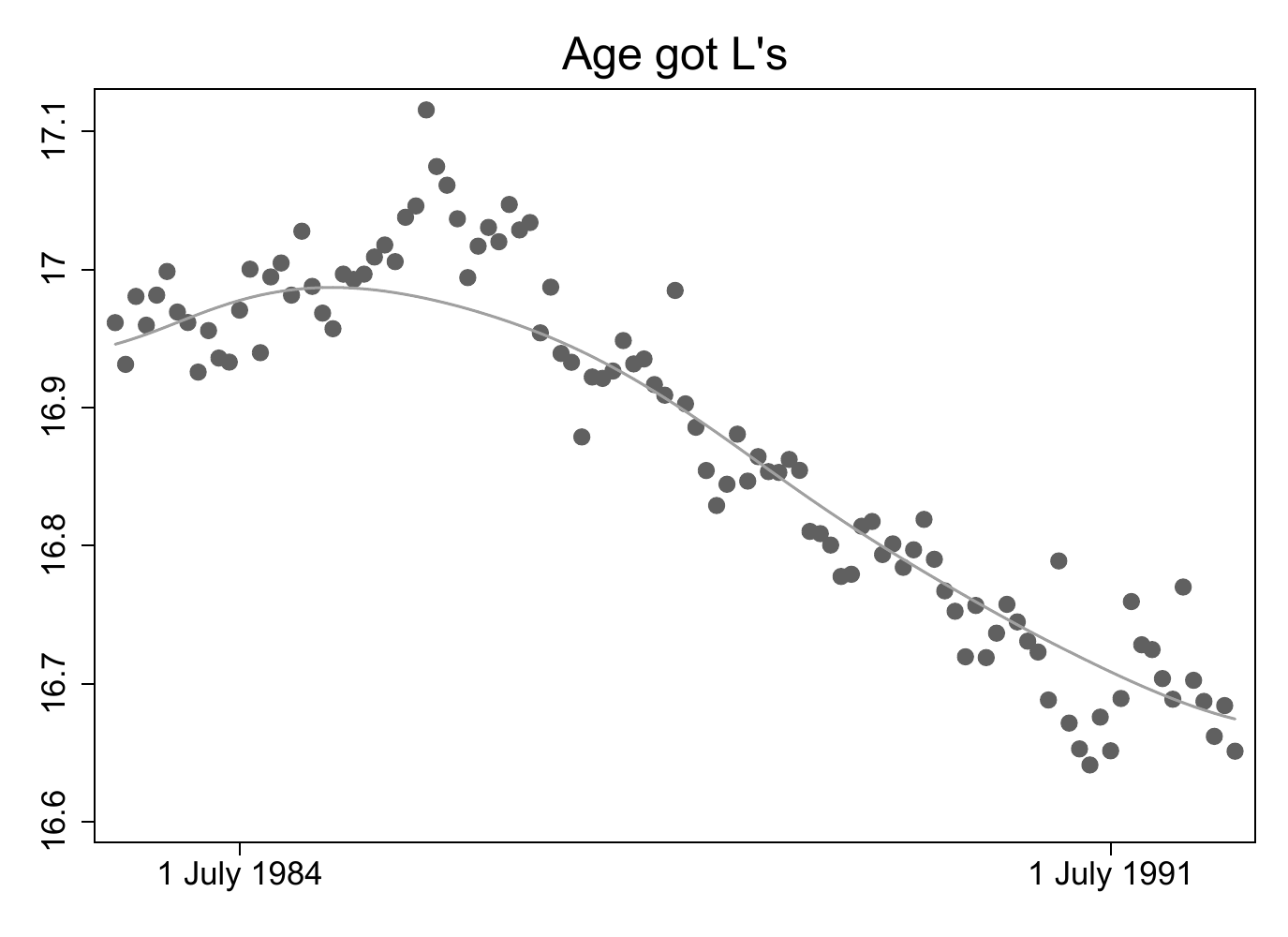} &
		\includegraphics[width=0.5\textwidth]{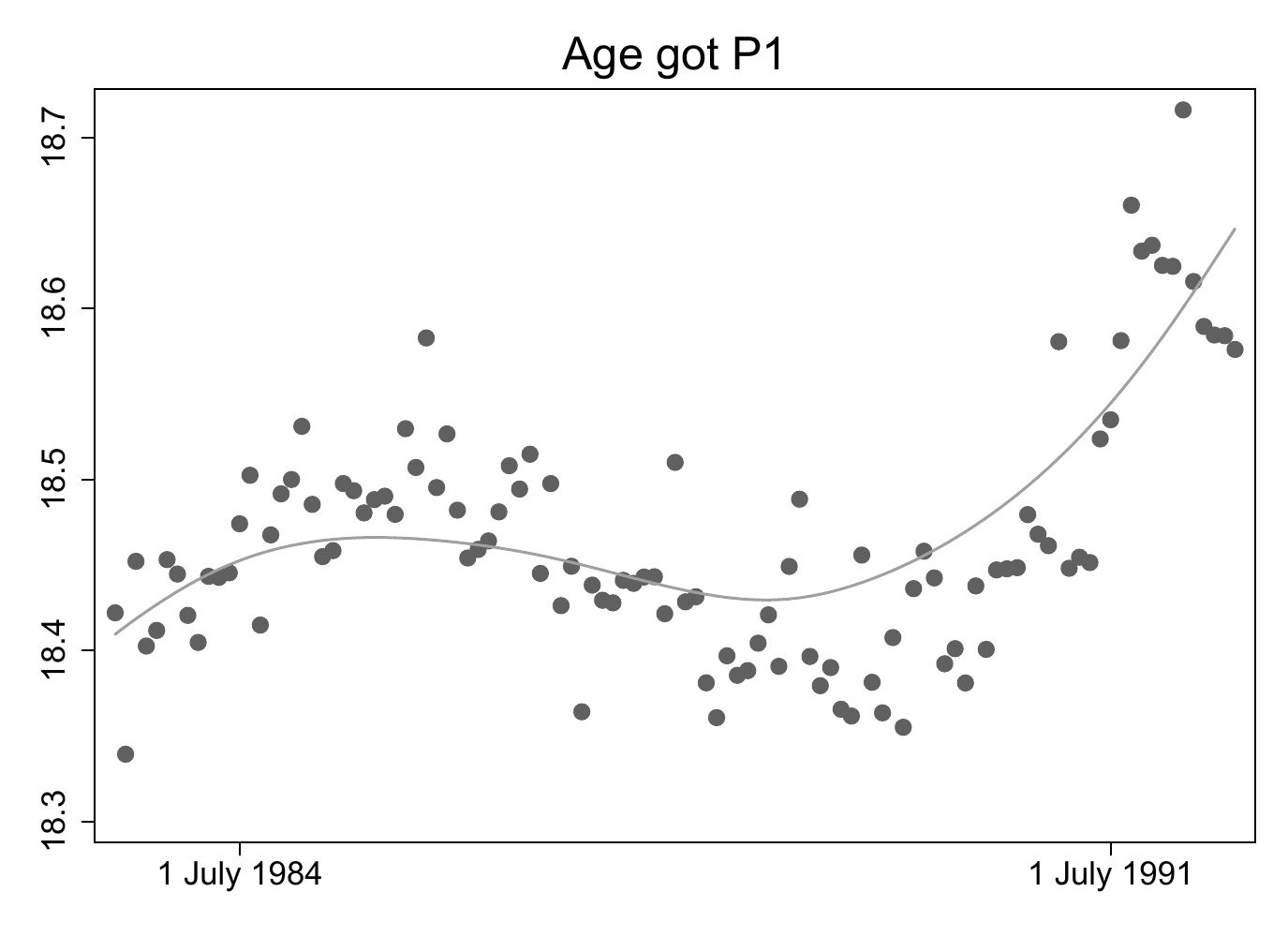} \\
	\end{tabular}
	\begin{tablenotes}
		\footnotesize
		\item Notes: Scatter plots 30-day bin size with lowess trend lines. 
	\end{tablenotes}
\end{figure}

Sample means for the main variables in our study are in Table \ref{tbl-descriptives}. We focus on two birth cohorts, centred $\pm$365 days from the key dates for our two policy reforms. In Figure \ref{fgr-main-vars-by-dob} we plot the variables by DOB for all years.

\begin{table}[ht!]
	\caption{Sample means by birth cohort}\label{tbl-descriptives}
	\begin{tabular}{p{4cm} p{5.5cm}p{5.5cm}}
		\toprule
		& 1 July 1983--30 June 1985 & 1 July 1990--30 June 1992 \\
		\midrule
		MVA 1-year & 0.057 & 0.044 \\
		MVA 1-2 years & 0.038 & 0.028 \\
		Injury 1-year & 0.022 & 0.020 \\
		Fatality 1-year & $<$0.000 & $<$0.000 \\
		Age got L's & 16.970 & 16.700 \\
		Age got P1 & 18.451 & 18.538 \\
		\midrule
		$n$ & 154,524 & 160,301 \\ 
		\bottomrule
	\end{tabular}
	\begin{tablenotes}[para,flushleft]
		\footnotesize
		\item Notes: this table shows sample means of the key variables for observations within each of the two `treatment zones' --. i.e. people born within one year of 1 July 1984, and 1 July 1991, respectively	
	\end{tablenotes}
\end{table}

\begin{figure}[!htb]
	\centering
	\caption{Distributions for age obtaining license}\label{fgr-dist-age-lic}
	\begin{tabular}{cc}
		\includegraphics[width=.5\textwidth]{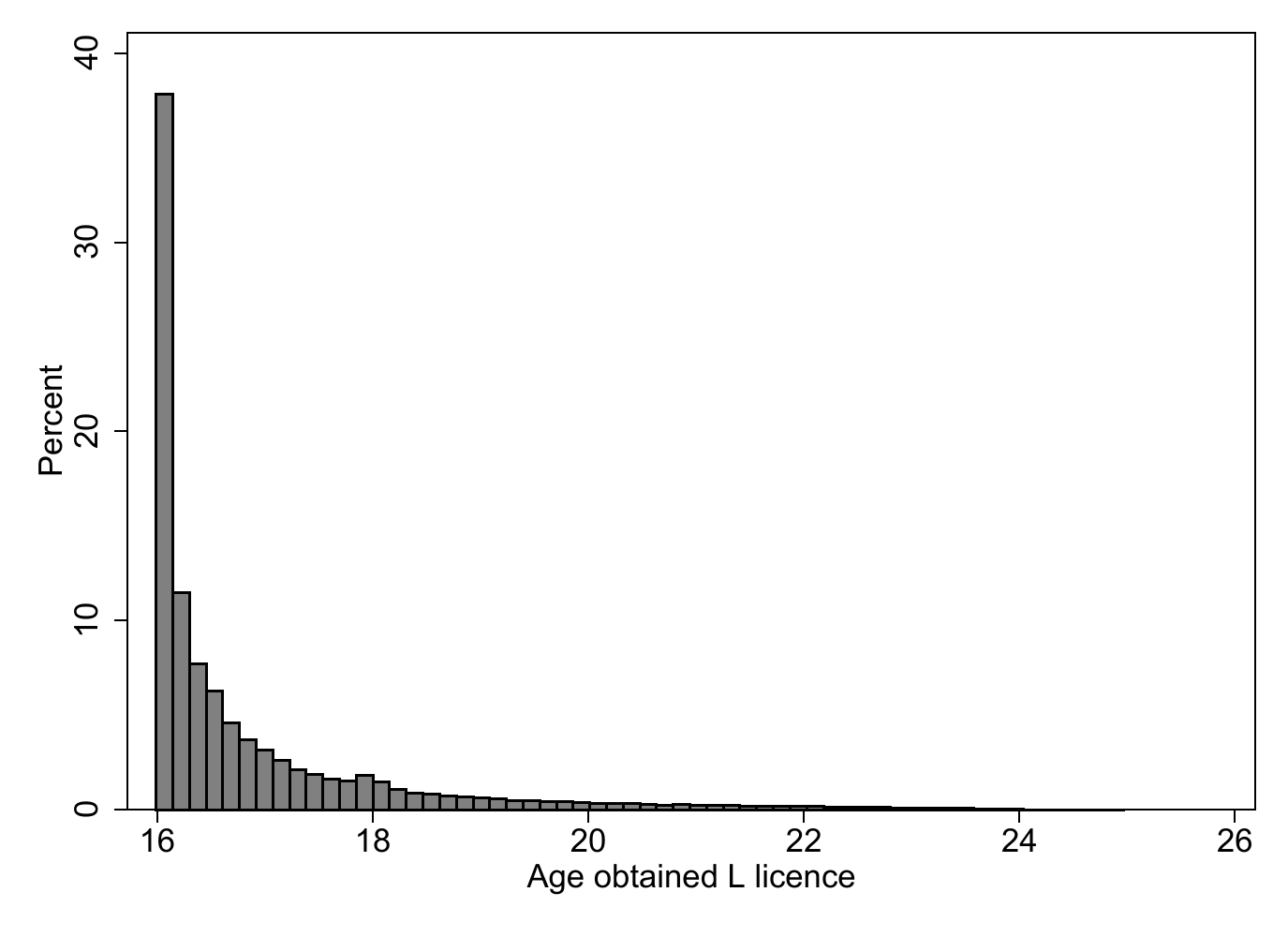} &
		\includegraphics[width=.5\textwidth]{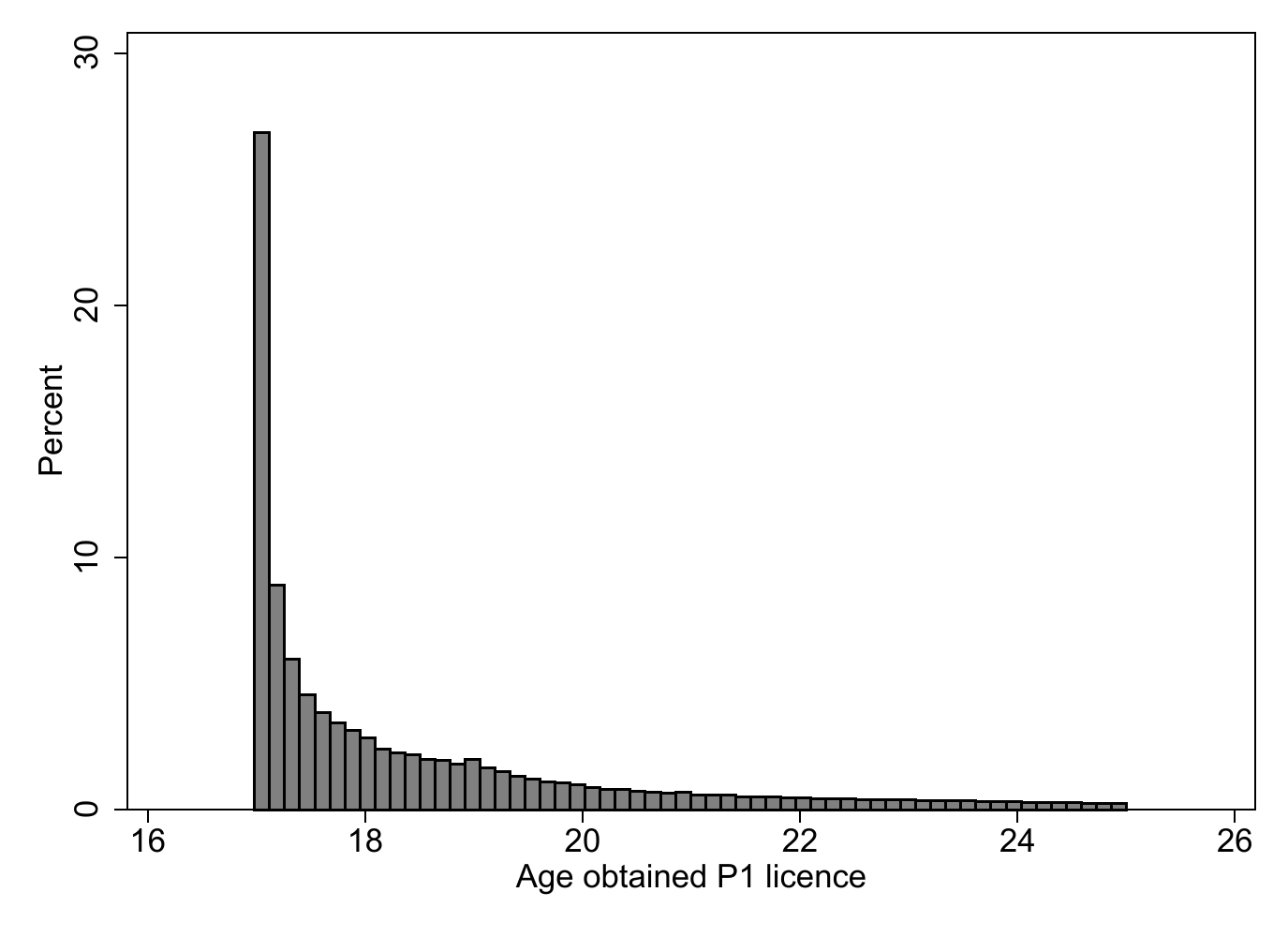} \\
	\end{tabular}
	\begin{tablenotes}
		\footnotesize
		\item Notes: Sample includes all NSW licensees born between 1 July 1983 and 30 June 1992 ($n=704,468$).   
	\end{tablenotes}
\end{figure}

MVA incidences are generally lower for younger birth cohorts, although this is weaker for more serious MVAs that involve injury. For the circa 1 July 1984 cohort, the probability of any MVA within 12 months of obtaining P1 is 5.7\%. This falls to 3.8\% for the next 12 months, consistent with young drivers becoming safer with age and experience. The average age at which people obtain their learner license is 17 years, a full 12 months later than they become eligible. However, the mass of observations are just after the 16th birthday (Figure \ref{fgr-dist-age-lic}). Most people obtain their license shortly after they become eligible. Similarly, there is a large mass who obtain their P1 license shortly after their 17th birthday, while the average is 18.5 years for both birth cohorts.

\setcounter{table}{0}
\renewcommand{\thetable}{C\arabic{table}}
\setcounter{figure}{0}
\renewcommand{\thefigure}{C\arabic{figure}}

\pagebreak
\newpage

\section{Additional tables and figures}

\begin{figure}[ht!]
	\centering
	\caption{Stylized DGPs for Monte Carlo simulations}\label{fgr-stylized-dgps}
	\includegraphics[width=0.55\textwidth]{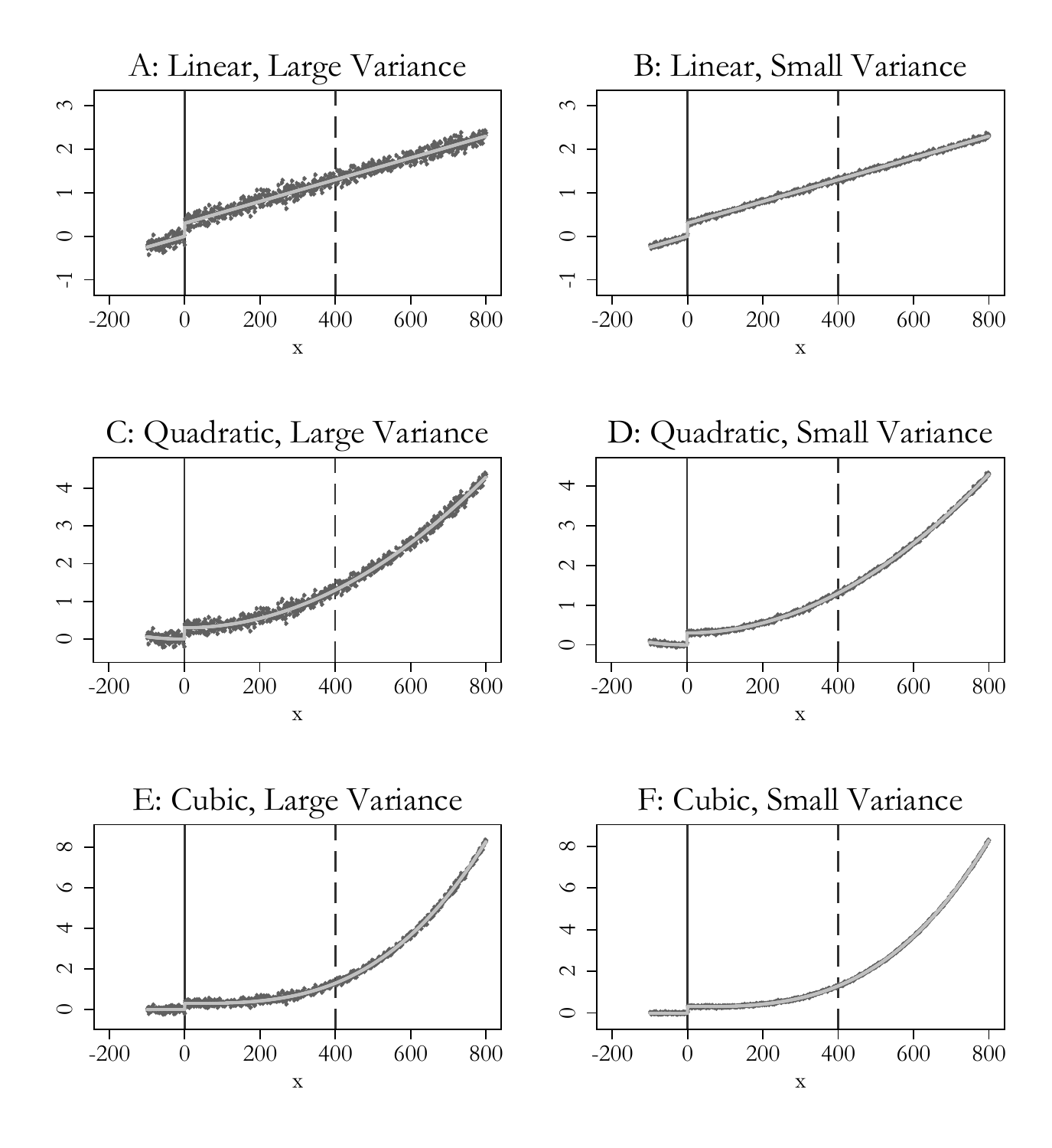}	
	\includegraphics[width=0.55\textwidth]{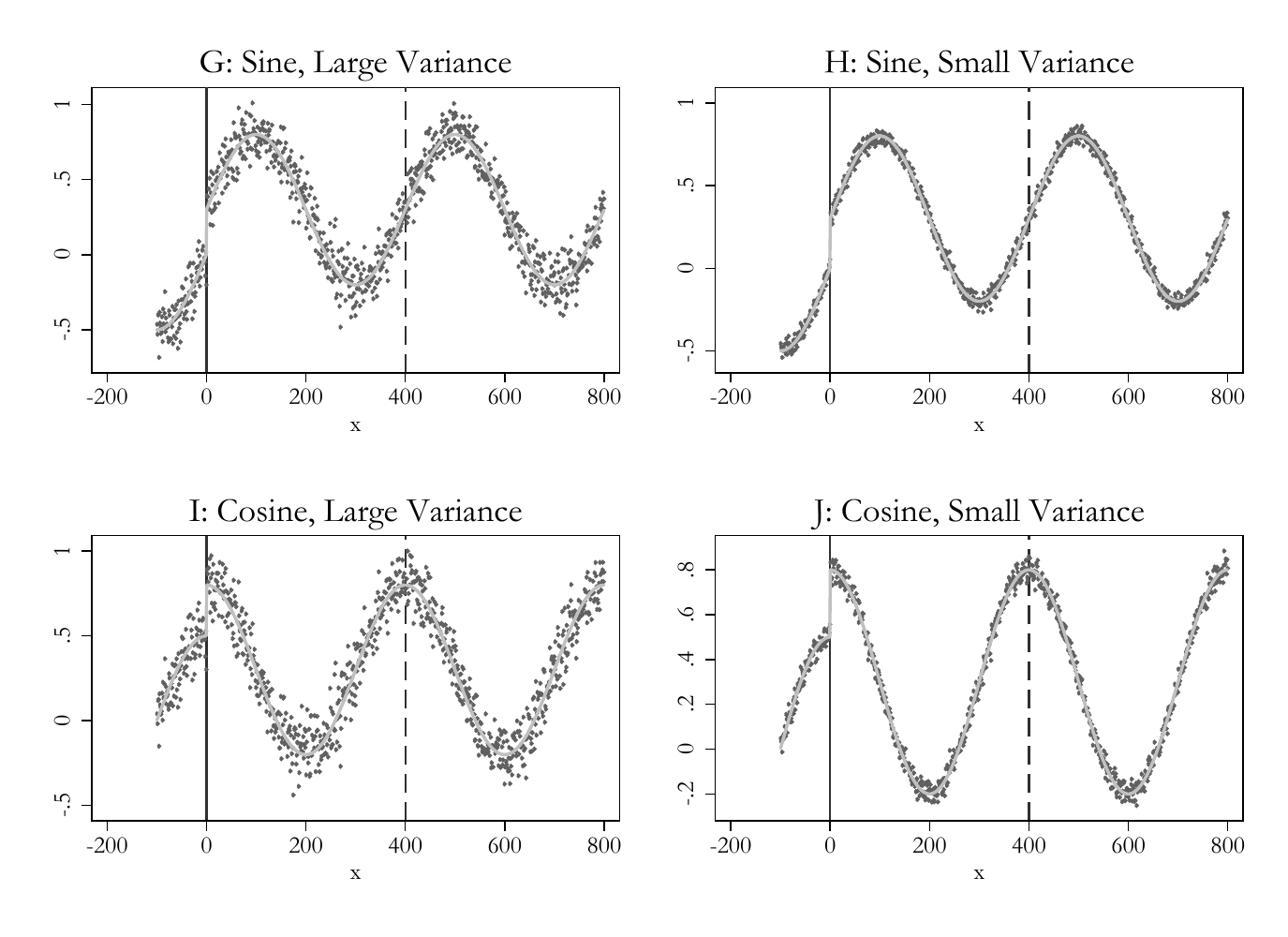}		
	\begin{tablenotes}
		\footnotesize
		\item Notes: This figure shows the `stylized' data generating processes for the Monte Carlo simulations. Each panel shows the C.E.F. used in each iteration, as well as the full distribution of the data in the first iteration. The outcome variable is given by the equation $y = 0.3(x>0) + f(x) + \epsilon$ with $\sigma^2 = 0.1^2$ (representing `large' error variance), or $0.03^2$ (`small' error variance). For panels A and B: $f(x) = x/400$. For panels C and D: $f(x) = (x/400)^2$. For panels E and F: $f(x) = (x/400)^3$. For panel G and H: $f(x) = sin(2\pi x/400)/2$ For panel I and J: $f(x) = cos(2\pi x/400)/2$.  
	\end{tablenotes}
\end{figure}

\begin{figure}[ht!]
	\centering
	\caption{DGPs based on prominent examples}\label{fgr-real-dgps}
	\includegraphics[width=1.0\textwidth]{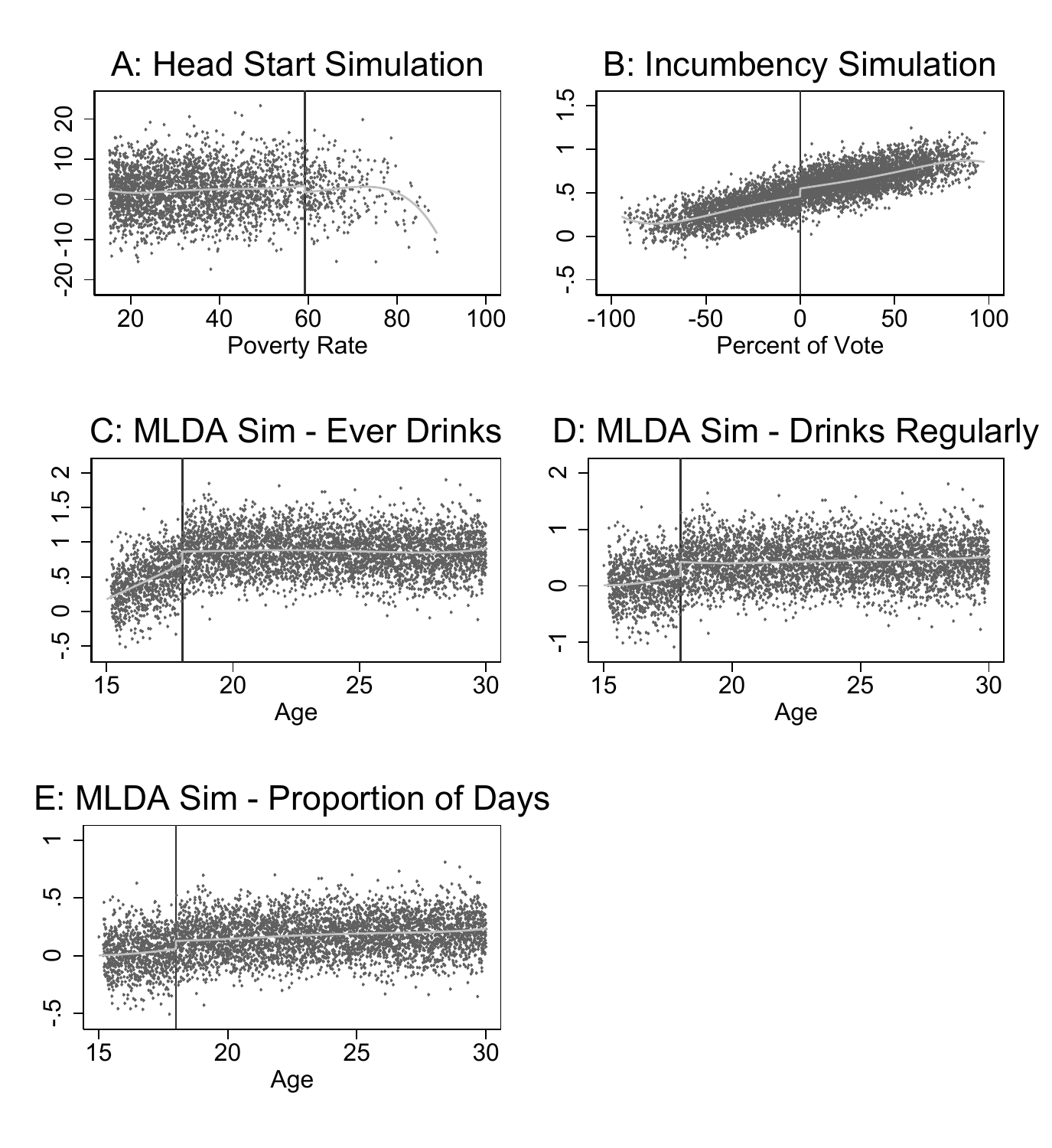}		
	\begin{tablenotes}
		\footnotesize
		\item Notes: The simulated DGPs are constructed by fitting a 5th order global polynomial through the support of the original data, allowing a discontinuity and a kink at the threshold, and then fitting a beta distribution to the same data to summarise the distribution of the running variable. 
	\end{tablenotes}
\end{figure}

\clearpage 
\newpage

\begin{longtable}[ht!]{p{4cm}*{8}{p{1.1cm}}}
	\caption{Coverage rates and confidence interval lengths from Monte Carlo simulations}\label{tbl-sim-coverage} \\
	\toprule
	& \multicolumn{2}{c}{KS}       & \multicolumn{2}{c}{KS alt.} & \multicolumn{2}{c}{CCT}  &  \multicolumn{2}{c}{IK}     \\
	\cmidrule(lr){2-9}
	& cov   & CI & cov   & CI   & cov  & CI & cov & CI \\
	& \multicolumn{8}{c}{A: Linear DGP} \\
	\cmidrule(lr){2-9}
	Baseline DGP                          & 0.952             & 0.094              & 0.961             & 0.110              & 0.928             & 0.184              & 0.937             & 0.160              \\
	Small error variance                  & 0.952             & 0.028              & 0.961             & 0.033              & 0.929             & 0.055              & 0.942             & 0.045              \\
	Small placebo zone                    & 0.942             & 0.107              & 0.901             & 0.124              & 0.912             & 0.245              & 0.934             & 0.170              \\
	Small placebo zone and error variance & 0.942             & 0.032              & 0.901             & 0.037              & 0.916             & 0.074              & 0.945             & 0.048              \\
	& \multicolumn{8}{c}{B: Quadratic DGP} \\
	\cmidrule(lr){2-9}
	Baseline DGP                          & 0.945             & 0.109              & 0.951             & 0.120              & 0.928             & 0.184              & 0.902             & 0.165              \\
	Small error variance                  & 0.955             & 0.035              & 0.948             & 0.037              & 0.927             & 0.055              & 0.759             & 0.048              \\
	Small placebo zone                    & 0.938             & 0.113              & 0.915             & 0.135              & 0.912             & 0.245              & 0.927             & 0.173              \\
	Small placebo zone and error variance & 0.949             & 0.036              & 0.927             & 0.044              & 0.912             & 0.073              & 0.887             & 0.051              \\
	& \multicolumn{8}{c}{C: Cubic DGP} \\
	\cmidrule(lr){2-9}
	Baseline DGP                          & 0.949             & 0.115              & 0.947             & 0.126              & 0.929             & 0.184              & 0.924             & 0.171              \\
	Small error variance                  & 0.926             & 0.037              & 0.933             & 0.039              & 0.924             & 0.055              & 0.917             & 0.056              \\
	Small placebo zone                    & 0.938             & 0.119              & 0.921             & 0.142              & 0.912             & 0.245              & 0.926             & 0.177              \\
	Small placebo zone and error variance & 0.926             & 0.039              & 0.924             & 0.046              & 0.913             & 0.073              & 0.926             & 0.058              \\
	& \multicolumn{8}{c}{D: Sine DGP} \\
	\cmidrule(lr){2-9}
	Baseline DGP                          & 0.889             & 0.151              & 0.940             & 0.173              & 0.931             & 0.189              & 0.900             & 0.220              \\
	Small error variance                  & 0.890             & 0.053              & 0.937             & 0.061              & 0.929             & 0.064              & 0.782             & 0.069              \\
	Small placebo zone                    & 0.886             & 0.157              & 0.895             & 0.174              & 0.911             & 0.246              & 0.910             & 0.226              \\
	Small placebo zone and error variance & 0.878             & 0.055              & 0.908             & 0.061              & 0.918             & 0.078              & 0.819             & 0.072              \\
	& \multicolumn{8}{c}{E: Cosine DGP} \\
	\cmidrule(lr){2-9}
	Baseline DGP                          & 0.942             & 0.153              & 0.954             & 0.173              & 0.927             & 0.185              & 0.936             & 0.186              \\
	Small error variance                  & 0.951             & 0.055              & 0.970             & 0.061              & 0.933             & 0.058              & 0.937             & 0.057              \\
	Small placebo zone                    & 0.921             & 0.156              & 0.908             & 0.169              & 0.913             & 0.245              & 0.939             & 0.203              \\
	Small placebo zone and error variance & 0.937             & 0.056              & 0.938             & 0.061              & 0.914             & 0.073              & 0.941             & 0.063                           \\
	& \multicolumn{8}{c}{F: Head Start DGP} \\
	\cmidrule(lr){2-9}
	Mortality                 & 0.930 & 3.569 & 0.912 & 3.763 & 0.941 & 6.045 & 0.941 & 3.758 \\
	& & & & & & & & \\
	& \multicolumn{8}{c}{G: Political incumbency DGP} \\
	\cmidrule(lr){2-9}
	Wins                      & 0.952 & 0.044 & 0.963 & 0.058 & 0.947 & 0.054 & 0.918 & 0.060 \\
	& & & & & & & & \\	
	& \multicolumn{8}{c}{H: MLDA DGP} \\
	\cmidrule(lr){2-9}
	Ever Drinks               & 0.950 & 0.103 & 0.936 & 0.152 & 0.928 & 0.239 & 0.933 & 0.123 \\
	Drinks Regularly          & 0.946 & 0.130 & 0.919 & 0.197 & 0.928 & 0.304 & 0.932 & 0.149 \\
	Proportion of Days Drinks & 0.949 & 0.059 & 0.937 & 0.090 & 0.928 & 0.138 & 0.927 & 0.066 \\
	\bottomrule   
\end{longtable}
\begin{tablenotes}[para,flushleft]
	\footnotesize
	\item Notes: For the columns `KS', coverage rate and average confidence interval length is based on conventional cluster-robust standard errors. For the columns `KS alt.', they use the approach outlined in 5.3. For the columns `CCT', they are based on the robust confidence intervals discussed in CCT and implemented using the -rdrobust- package for Stata. For the columns 'IK' they use conventional standard errors after choosing a bandwidth using the bwselect (IK) in the rdbwselect\_2014 command. For further details on the simulations see Table 2.    
\end{tablenotes}

\begin{figure}[ht!]
	\centering
	\caption{First-stage relationship between DOB and 120 MSDH treatment}\label{fgr-fs-91}
	\begin{tabular}{cc}
		\includegraphics[width=.5\textwidth]{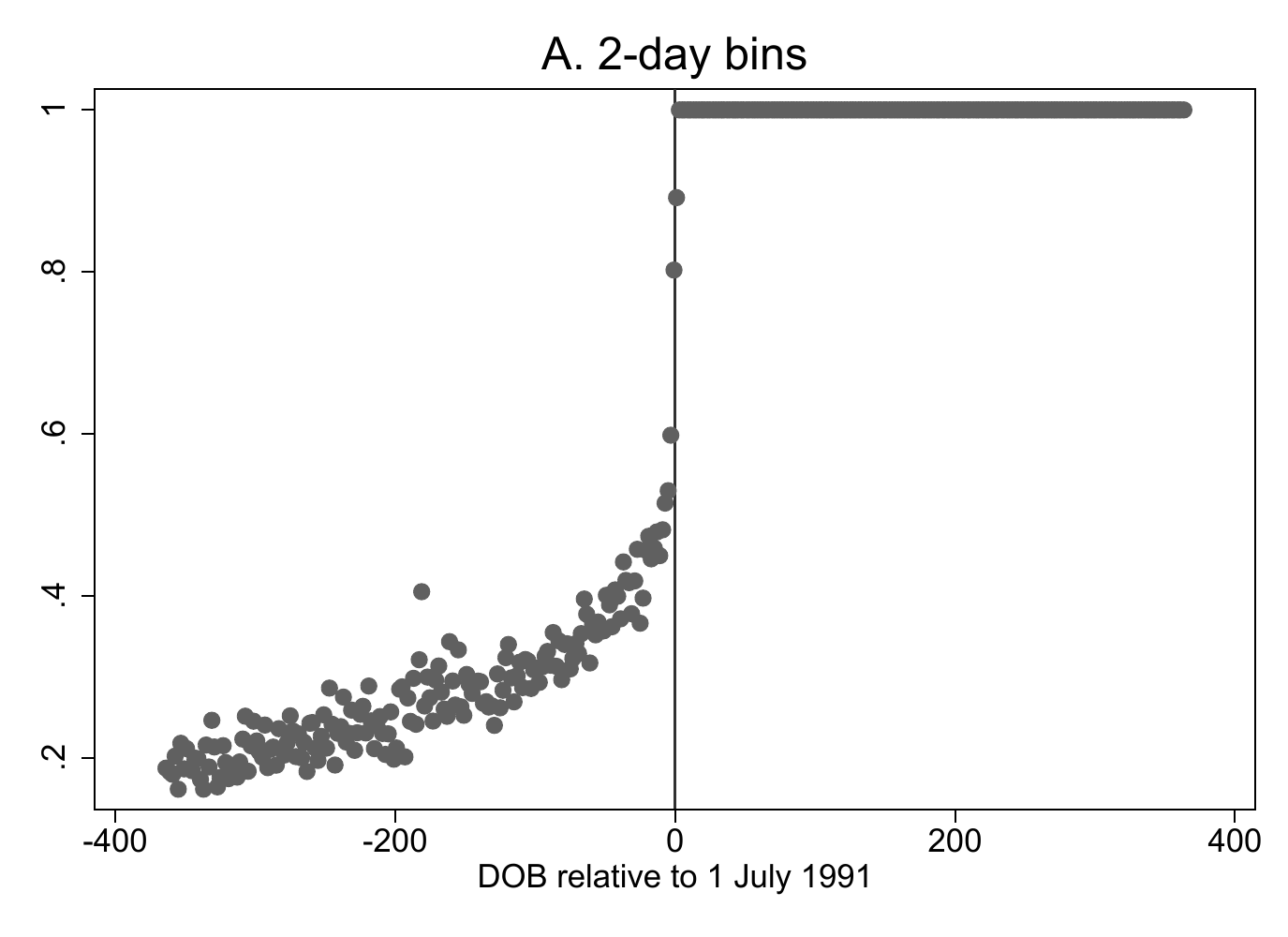} &
		\includegraphics[width=.5\textwidth]{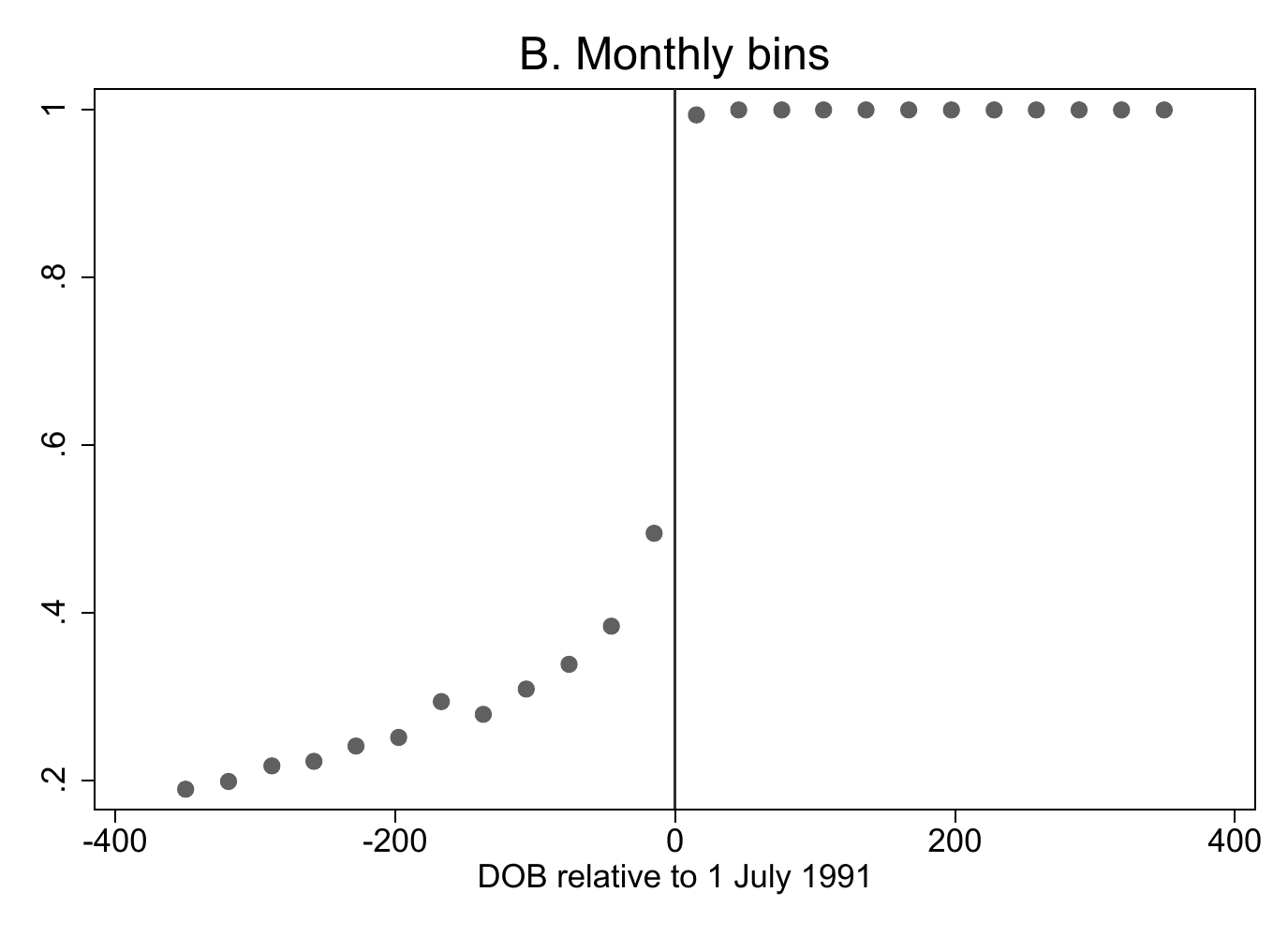} \\
	\end{tabular}
\end{figure}

\begin{figure}[ht!]
	\centering
	\caption{Reduced-Form Relationships between DOB and MVA 1-year}\label{fgr-reduced-form}
	\begin{tabular}{cc}
		\includegraphics[width=.5\textwidth]{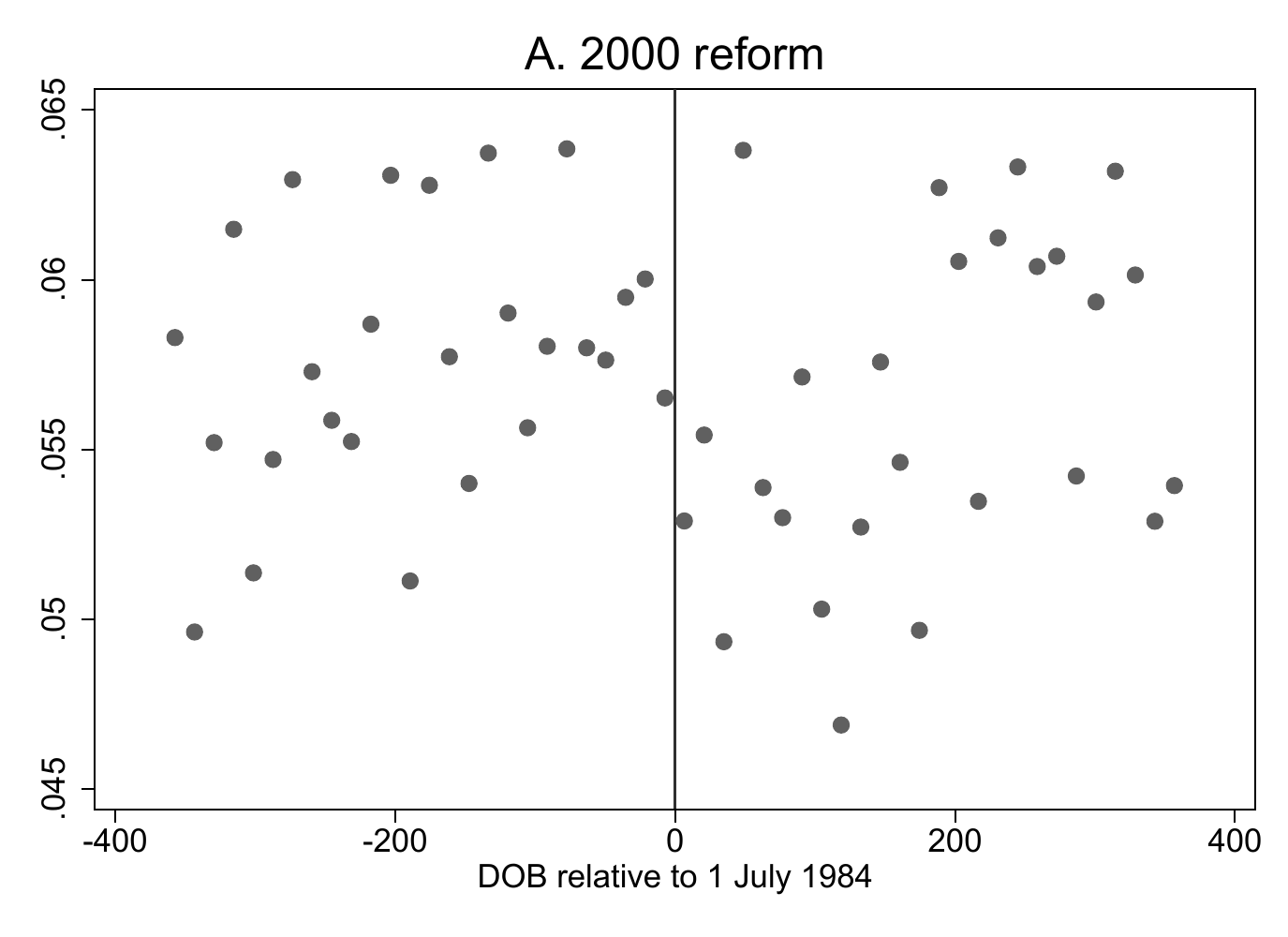} &
		\includegraphics[width=.5\textwidth]{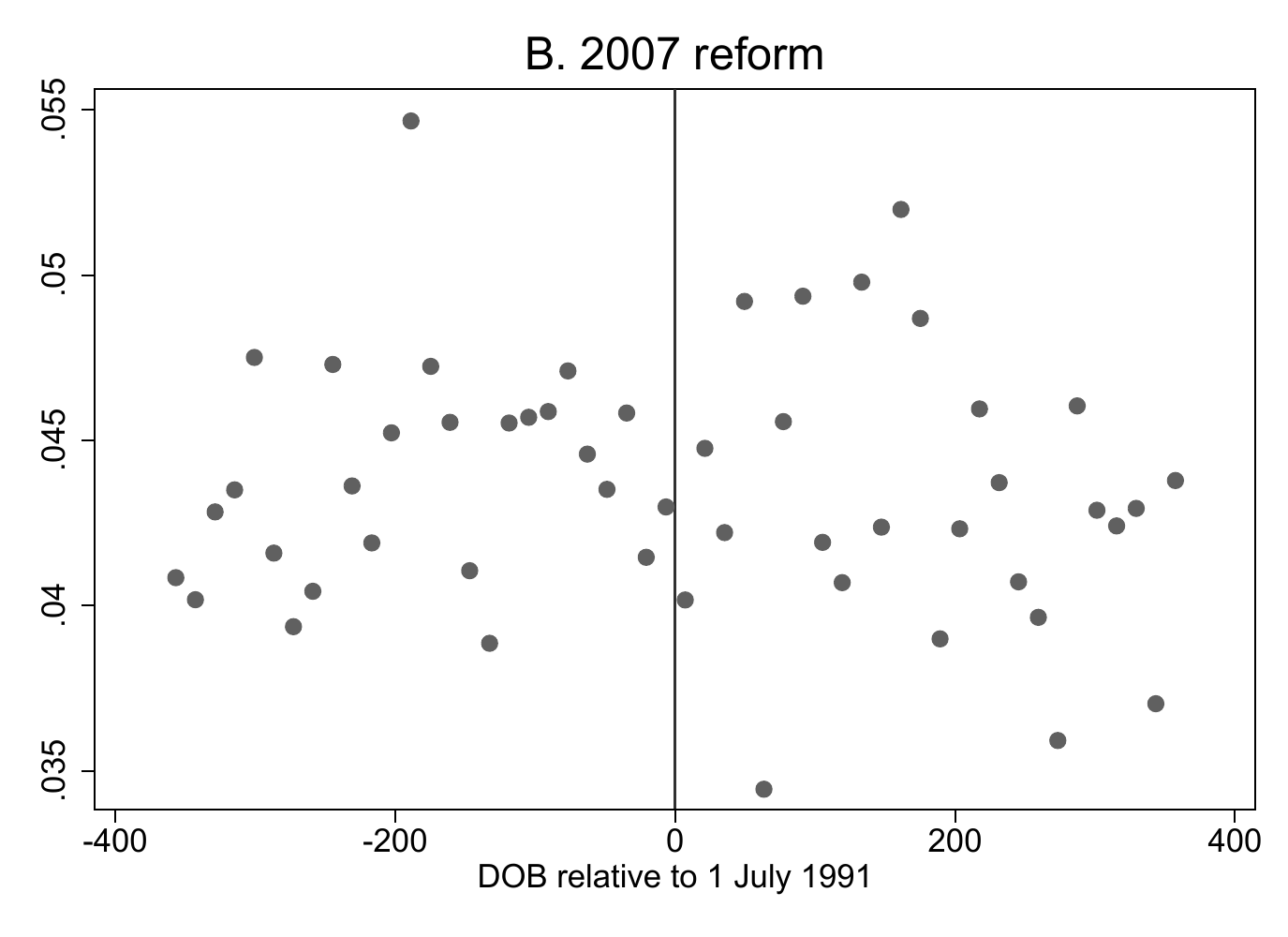} \\
	\end{tabular}
	\begin{tablenotes}
		\footnotesize
		\item Notes: Scatter plots use 14-day bin size.      
	\end{tablenotes}
\end{figure}

\begin{figure}[ht!]
	\centering
	\caption{Exposure to BAC, engine and passenger restrictions by DOB}\label{fgr-restrictions}
	\begin{tabular}{cc}
		\includegraphics[width=.5\textwidth]{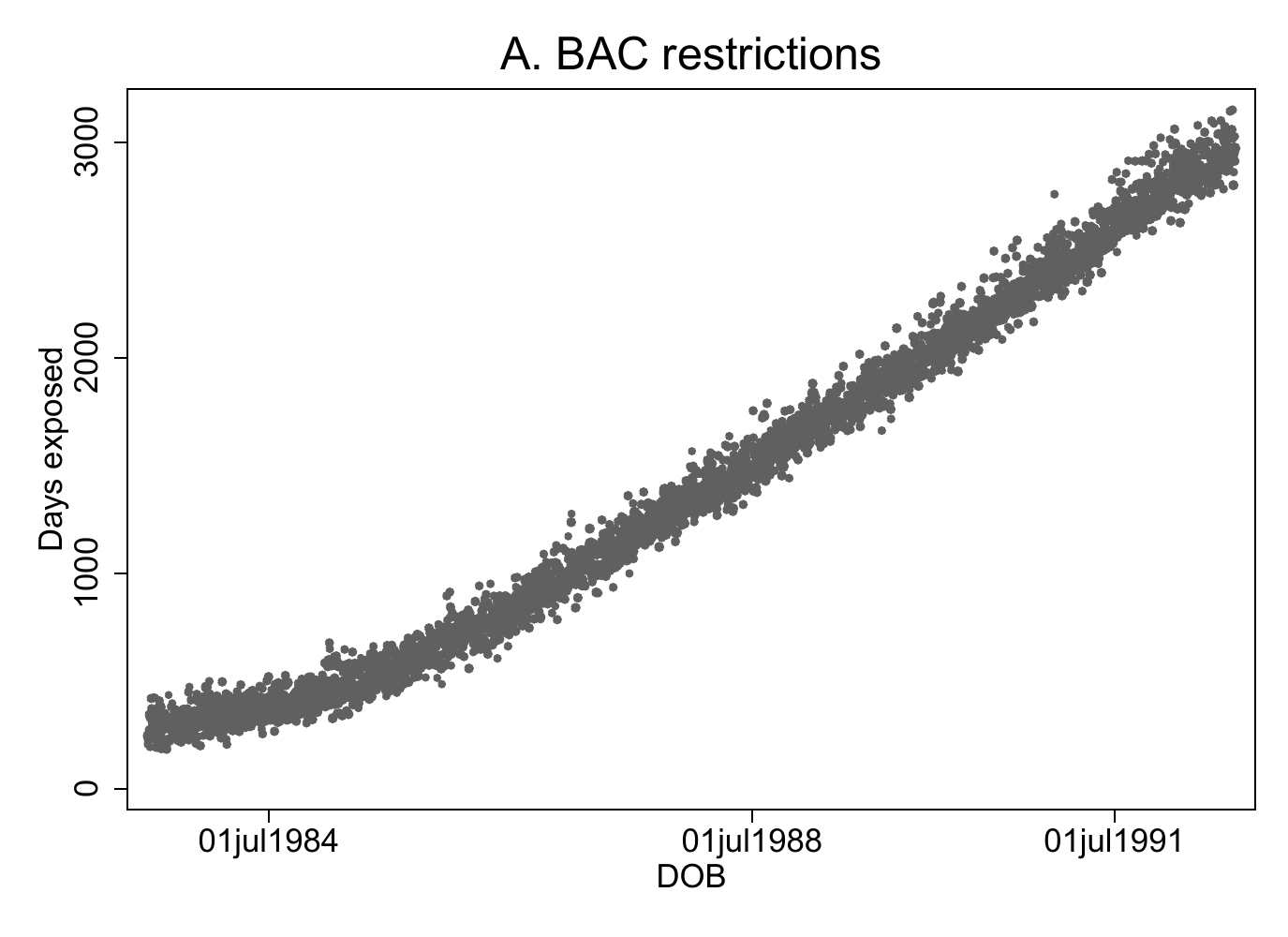} &
		\includegraphics[width=.5\textwidth]{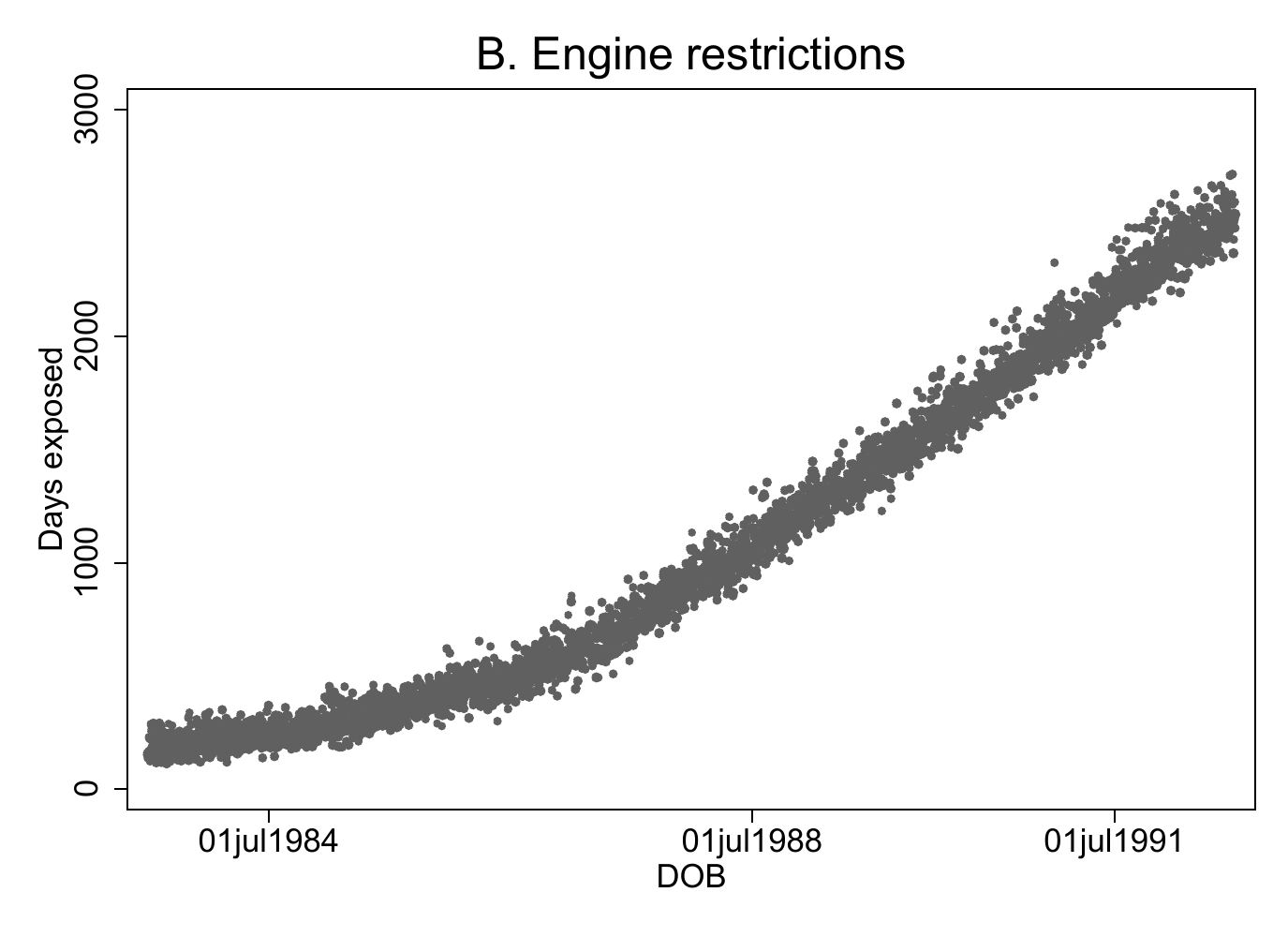} \\
		\includegraphics[width=.5\textwidth]{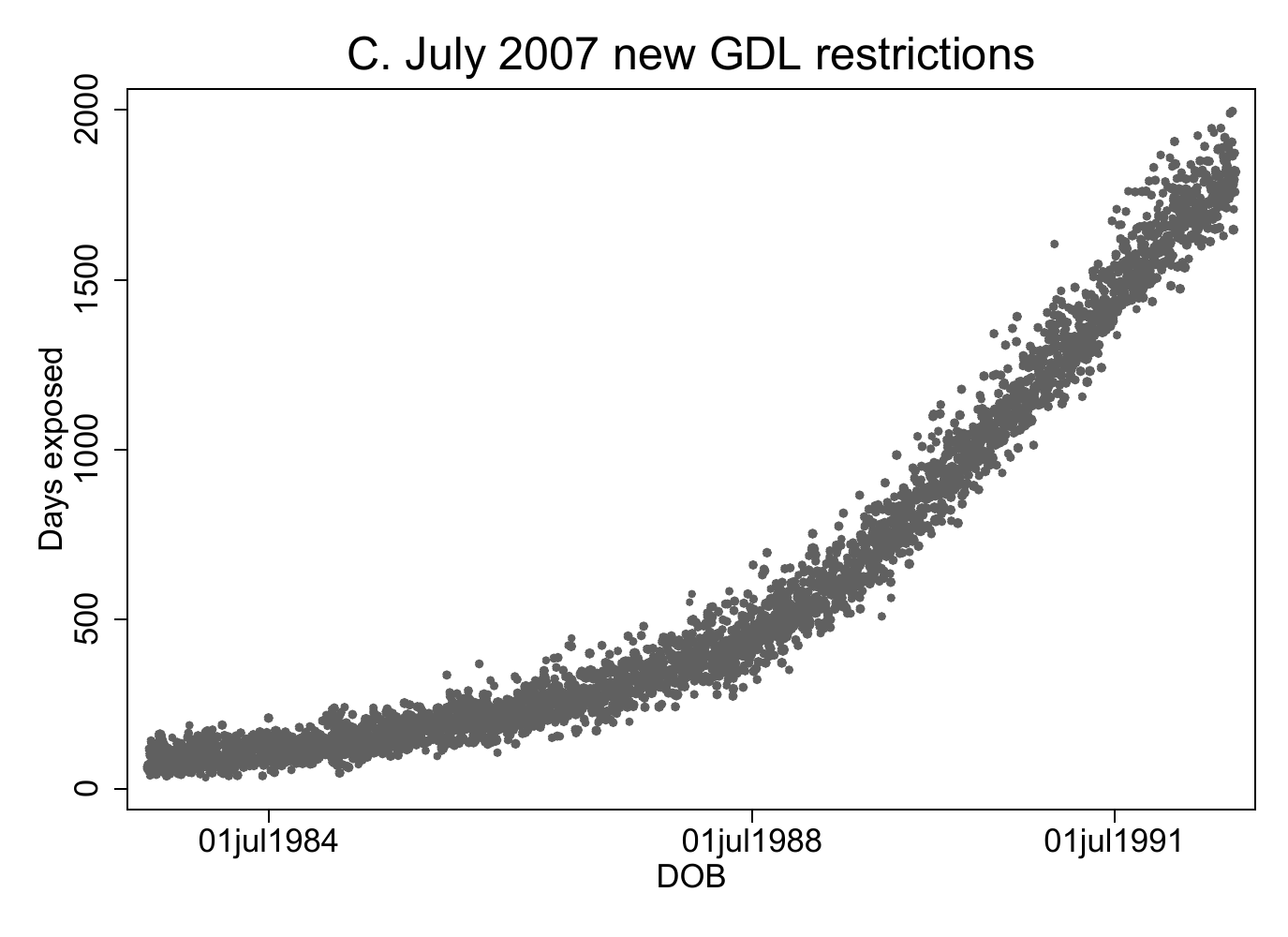} & \\
	\end{tabular}
	\begin{tablenotes}
		\footnotesize
		\item Notes: Day exposed is the maximum of zero and the date their P1 license expired minus the relevant policy change date.
	\end{tablenotes}
\end{figure}

\begin{figure}[ht!]
	\centering
	\caption{Relationship between DOB and age obtained P1 license}\label{fgr-p1-age}
	\begin{tabular}{cc}
		\includegraphics[width=.5\textwidth]{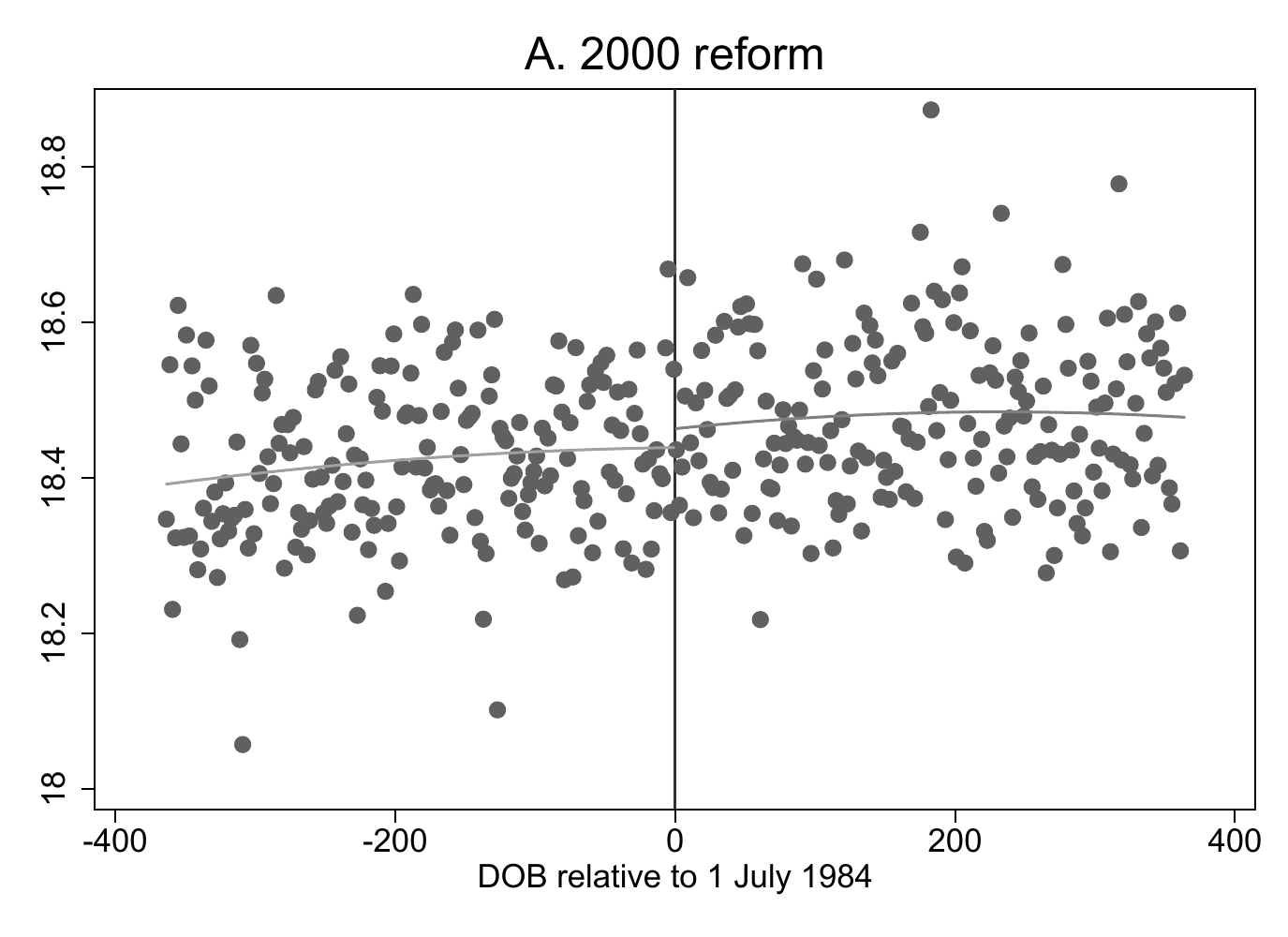} &
		\includegraphics[width=.5\textwidth]{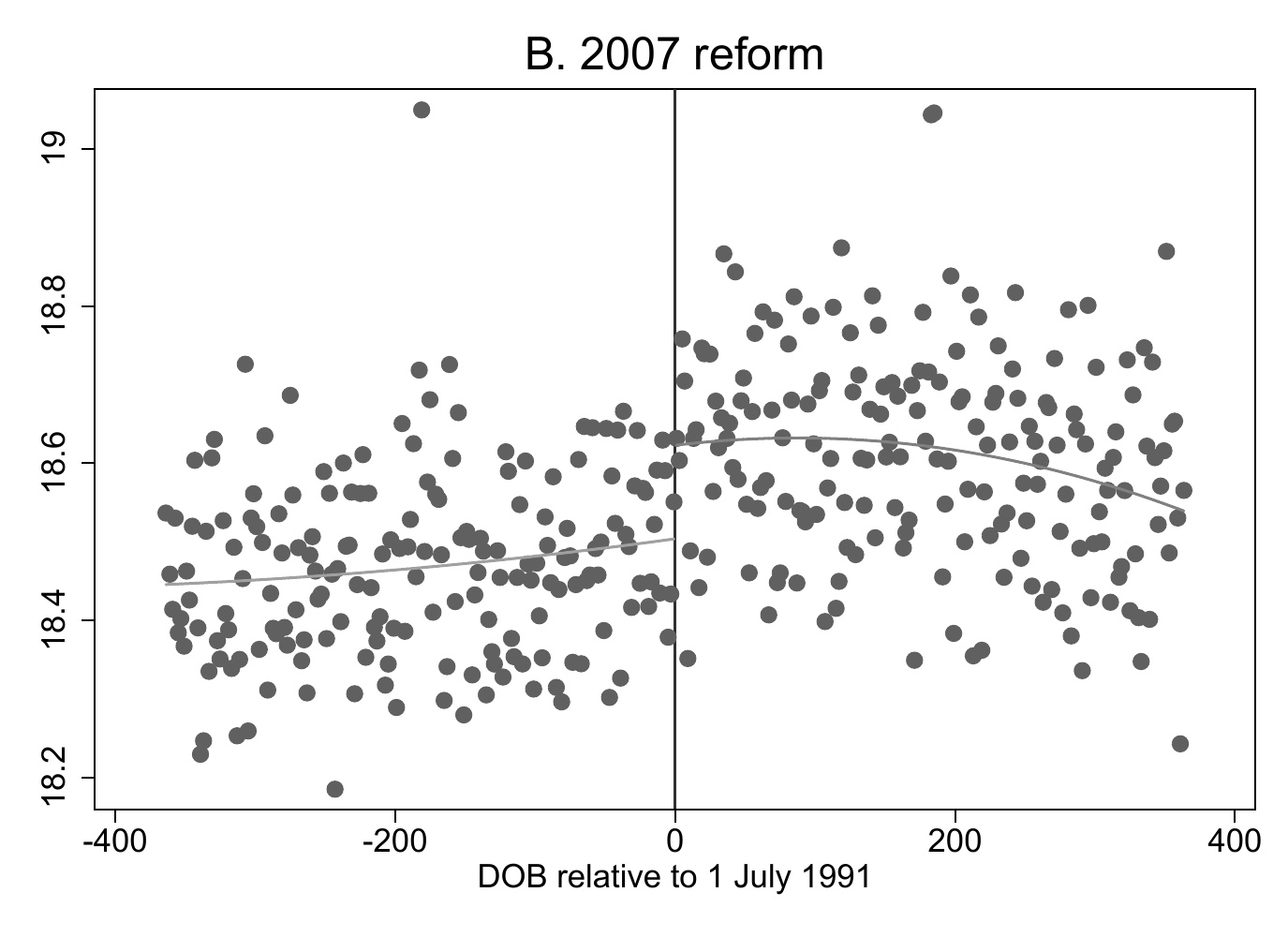} \\
	\end{tabular}
	\begin{tablenotes}
		\footnotesize
		\item Notes: Scatter point correspond to 2-day bins.      
	\end{tablenotes}
\end{figure}

\afterpage{
	\begin{landscape}
		\begin{table}[ht!]
			\caption{Candidate model equations}\label{tbl-equations}
			\begin{tabular}{llll}
				\toprule
				Model & Description & \multicolumn{1}{c}{$f(.)$} & \multicolumn{1}{c}{$g(.)$} \\
				\midrule
				\multicolumn{4}{c}{RDD models} \\
				1 & RDD - linear & $f(.)=\gamma_1 X_i + \gamma_2 X_i D_i$ & $g(.) = \pi_1 D_i$ \\
				2 & RDD - mixed polynomial & $f(.)=\gamma_1 X_i + \gamma_2 X_i D_i + \gamma_3 X_i^2 (1-D_i)$ & $g(.) = \pi_1 D_i$ \\ 
				3 & RDD - quadratic & $f(.)=\gamma_1 X_i + \gamma_2 X_i D_i + \gamma_3 X_i^2 (1-D_i)+ \gamma_4 X_i^2 D_i$ & $g(.) = \pi_1 D_i$ \\ 
				\multicolumn{4}{c}{RPJKD models} \\
				4 & RPJKD - linear & $f(.)=\gamma_1 X_i$ & $g(.) = \pi_1 D_i + \pi_2X_i D_i$ \\ 
				5 & RPJKD - quadratic & $f(.)=\gamma_1 X_i + \gamma_2 X_i^2$ & $g(.) = \pi_1 D_i + \pi_2X_i D_i$ \\ 
				6 & RPJKD - mixed polynomial & $f(.)=\gamma_1 X_i + \gamma_2 X_i^2 (1-D_i)$ & $g(.) = \pi_1 D_i + \pi_2X_i D_i$ \\ 
				7 & RPJKD - interacted quadratic & $f(.)=\gamma_1 X_i + \gamma_2 X_i^2 (1-D_i) + \gamma_3 X_i^2$ & $g(.) = \pi_1 D_i + \pi_2X_i D_i$ \\ 
				\multicolumn{4}{c}{RKD models} \\
				8 & RKD - linear & $f(.)=\gamma_1 X_i$ & $g(.) =\pi_1 X_i D_i$ \\
				9 & RKD - quadratic & $f(.)=\gamma_1 X_i + \gamma_2 X_i^2$ & $g(.) = \pi_1 X_i D_i$ \\
				10 & RKD - mixed polynomial & $f(.)=\gamma_1 X_i + \gamma_2 X_i^2(1-D_i)$ & $g(.) = \pi_1 X_i D_i$ \\ 
				11 & RKD - interacted quadratic & $f(.)=\gamma_1 X_i + \gamma_2 X_i^2 + \gamma_3 X_i^2(1-D_i)$ & $g(.) = \pi_1 X_i D_i$ \\
				\multicolumn{4}{c}{Birth cohort-IV models} \\
				12 & Birth cohort-IV - linear & $f(.)=\gamma_1 X_i$ & $g(.) = \theta_m$ \\
				13 & Birth cohort-IV - quadratic & $f(.)=\gamma_1 X_i + \gamma_2 X_i^2$ & $g(.) = \theta_m$  \\
				14 & Birth cohort-IV - cubic & $f(.)=\gamma_1 X_i + \gamma_2 X_i^2 + \gamma_3 X_i^3$ & $g(.) = \theta_m$  \\
				\bottomrule
			\end{tabular}
			\begin{tablenotes}[para,flushleft]
				\footnotesize
				\item Notes: This table summarizes the functional forms of the models included in the placebo zone trials in our main application. The functions $f(.)$ and $g(.)$ are components of the full specifications shown in equations 3, 4, and for models 12-14, equation 5.	
			\end{tablenotes}
			
		\end{table}
	\end{landscape}
}

\begin{figure}[ht!]
	\centering
	\caption{Relationship between bandwidth and RDD estimates: Head Start and MLDA}\label{fgr-hs-mlda_by_bw}
	\includegraphics[width=1.0\textwidth]{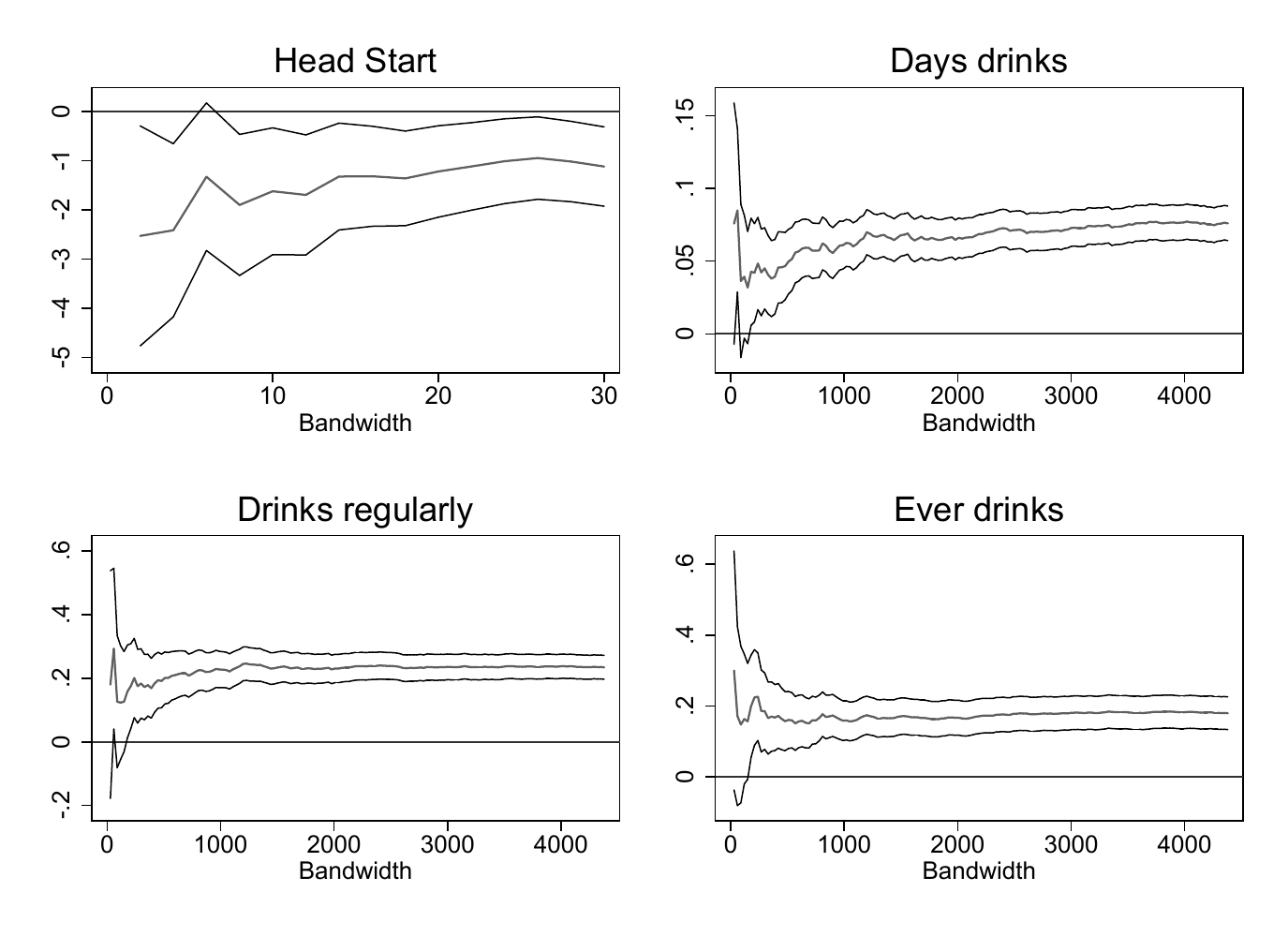} 
	\begin{tablenotes}
		\footnotesize
		\item Notes: This figure shows RDD estimates and confidence intervals for the Head Start and MLDA applications by bandwidth (in intervals of 2 units for Head Start and 30 for MLDA) using local linear regression. For MLDA, the maximum LHS bandwidth is three years; values above this in the figure only refer to the RHS (with the LHS fixed at three years). Asymptotic standard errors used to construct confidence intervals are clustered at unique values of the running variable.    
	\end{tablenotes}
\end{figure}

\clearpage 
\newpage

\begin{table}[ht!]
	\caption{Candidate model performance in the placebo zone: `shifted' first-stage}\label{tbl-altfs-model-perf}
	\begin{tabular}{@{}llllll@{}}
		\toprule
		Model & Description                  & RMSE   & Optimal BW & Coverage & Bias    \\ \midrule
		1     & RDD - linear      & 0.0060 & 365        & 0.962    & -0.0006 \\
		2     & RDD - mixed polynomial       & 0.0114 & 365        & 0.922    & -0.0001 \\
		3     & RDD - quadratic   & 0.0129 & 270        & 0.971    & 0.0006  \\
		\textbf{4 }    & \textbf{RPJKD - linear }   & 0.0053 & 365        & 0.932    & 0.0006  \\
		5     & RPJKD - quadratic            & 0.0057 & 365        & 0.978    & -0.0004 \\
		\textbf{6}     & \textbf{RPJKD - mixed polynomial}     & 0.0046 & 365        & 0.985    & 0.0000  \\
		7     & RPJKD - interacted quadratic & 0.0111 & 365        & 0.939    & 0.0007  \\
		8     & RKD - linear      & 0.0113 & 350        & 0.921    & 0.0033  \\
		9     & RKD - quadratic              & 0.0209 & 365        & 0.964    & 0.0012  \\
		\textbf{10}    & \textbf{RKD - mixed polynomial}       & 0.0052 & 365        & 0.984    & 0.0001  \\
		11    & RKD - interacted quadratic   & 0.0205 & 365        & 0.955    & 0.0013  \\
		\textbf{12}    & \textbf{birth cohort-IV - linear}     & 0.0048 & 365        & 0.953    & 0.0004  \\
		13    & birth cohort-IV - quadratic  & 0.0054 & 365        & 0.981    & -0.0005 \\
		14    & birth cohort-IV - cubic      & 0.0102 & 365        & 0.932    & 0.0004  \\
		WA    & Inv-MSE weighted average     & 0.0050 & 365        &   n.d.       & 0.0001   \\ \bottomrule
	\end{tabular}
	\begin{tablenotes}[para,flushleft]
		\footnotesize
		\item Notes: The results in this table are from a similar procedure to what is detailed in the Table 3 notes. In this case, the first-stage relationship from the treatment zone is imposed into (each repetition of) the placebo zone, after collapsing the data to DOB level. This procedure is explained in the text.
	\end{tablenotes}
\end{table}

\setcounter{table}{0}
\renewcommand{\thetable}{D\arabic{table}}
\setcounter{figure}{0}
\renewcommand{\thefigure}{D\arabic{figure}}
\setcounter{equation}{0}
\renewcommand{\theequation}{D\arabic{equation}}

\section{Further results for the 2000 MSDH reform}\label{append-2000-further}
Table \ref{tbl-further-results} delves deeper into the effects of the 2000 reform. Corresponding results for the 2007 reform are generally precise zeros and are available on request. The structure of this table is the same as the previous table, and the same five estimators are used throughout.\footnote{We use the same set of estimators across each of the outcome variables (and sub-populations) here. This approach has the advantage of transparency and internal consistency, which helps to interpret the drivers of the main estimates. An alternative approach is to choose a different set of preferred estimators (using the placebo zone approach) for each outcome variable and sub-population.}

\begin{table}[ht!]
	\caption{Further results for the 2000 reform to driving hours}\label{tbl-further-results}
	\begin{tabular}{p{3.3cm} *{5}{p{2.0cm}}}
		\toprule
		& Best   estimator & Best sym. cohort-IV & Best sym. RPJKD & Best sym. RKD & Best sym. RDD \\ 
		\midrule
		& (1)            & (2)                                  & (3)                            & (4)                          & (5)                          \\
		& \multicolumn{5}{c}{A: Age of Obtaining Provisional License (Mechanism)}                                                                            \\
		\cmidrule(lr){2-6}
		MVA 1-year                  & -0.0144***     & -0.0132***                           & -0.0147***                     & -0.0144**                    & -0.0168***                   \\
		SE                          & 0.0041         & 0.0049                               & 0.0050                         & 0.0058                       & 0.0058                       \\
		controlling for age got P1s & -0.0144***     & -0.0118**                            & -0.0133**                      & -0.0131**                    & -0.0155**                    \\
		SE                          & 0.0042         & 0.0052                               & 0.0052                         & 0.0061                       & 0.0060                       \\
		&                &                                      &                                &                              &                              \\
		& \multicolumn{5}{c}{B:   Timing of Treatment Effect}                                                                                                  \\
		\cmidrule(lr){2-6}
		MVA 6 months                & -0.0094***     & -0.0074**                            & -0.0078**                      & -0.0072                      & -0.0093**                    \\
		SE                          & 0.0031         & 0.0035                               & 0.0036                         & 0.0044                       & 0.0044                       \\
		MVA 6-12 months             & -0.0048*       & -0.0059*                             & -0.0070**                      & -0.0069*                     & -0.0077*                     \\
		SE                          & 0.0028         & 0.0034                               & 0.0034                         & 0.0040                       & 0.0040                       \\
		MVA 1-2 years               & 0.0035         & 0.0026                               & 0.0027                         & 0.0019                       & 0.0033                       \\
		SE                          & 0.0036         & 0.0045                               & 0.0046                         & 0.0053                       & 0.0053                       \\
		&                &                                      &                                &                              &                              \\
		& \multicolumn{5}{c}{C:   Serious MVAs}                                                                                                             \\
		\cmidrule(lr){2-6}
		Injury                      & -0.0084***     & -0.0093***                           & -0.0100***                     & -0.0102***                   & -0.0110***                   \\
		SE                          & 0.0026         & 0.0032                               & 0.0032                         & 0.0038                       & 0.0036                       \\
		Fatality                    & -0.0002        & -0.0001                              & -0.0002                        & -0.0002                      & -0.0002                      \\
		SE                          & 0.0003         & 0.0004                               & 0.0004                         & 0.0005                       & 0.0005                       \\
		&                &                                      &                                &                              &                              \\
		& \multicolumn{5}{c}{D: Heterogeneity   by Sex}                                                                                                        \\
		\cmidrule(lr){2-6}
		MVA 1-year males            & -0.0132**      & -0.0146**                            & -0.0139*                       & -0.0114                      & -0.0163*                     \\
		SE                          & 0.0059         & 0.0072                               & 0.0073                         & 0.0086                       & 0.0085                       \\
		MVA 1-year females          & -0.0164***     & -0.0111*                             & -0.0159**                      & -0.0181**                    & -0.0177**                    \\
		SE                          & 0.0056         & 0.0066                               & 0.0068                         & 0.0080                       & 0.0083                       \\ \bottomrule
	\end{tabular}
	\begin{tablenotes}[para,flushleft]
		\footnotesize
		\item Notes: Asymptotic standard errors are clustered at the DOB level. * $p<0.1$, ** $p<0.05$, *** $p<0.01$.     
	\end{tablenotes}
\end{table}

One possible explanation for the 2000 reform reducing MVA is delayed timing of obtaining a provisional license, which would support the idea that maturity rather than improved driving skill lowered MVAs. Appendix Figure \ref{fgr-p1-age} suggests a possible small delay effect. Panel A considers the extent to which this explains the main treatment effect. The first rows show the original estimates, while the next rows show estimates from the same models, but controlling for a quadratic of age (in days) of obtaining a provisional license. The estimated effects are generally slightly smaller when these controls are included. In the `best' estimator, the treatment effect estimate is actually unchanged, while in the other models, this reduction is no more than 11\%. Thus we conclude that delaying of obtaining a license is at most only a small factor in the treatment effects that we have estimated.\footnote{Moreover, for the 2007 reform we observe a much stronger delay effect, yet our treatment effect estimates indicate no effect on MVAs.}

Panel B shows results which consider the timing of the treatment effects. As may be expected, the majority (65\% in the `best' model) of the treatment effect is confined to the first 6 months after obtaining a provisional license. The effect in the 6-12 month period is also at least marginally significant across the estimators, and its magnitude is not small. The effect in the following year (12-24 months after obtaining a license) is not statistically significant in any column.

Panel C shows results for serious MVAs. It shows strongly significant negative effects for the subset of MVAs in which one or more people were injured. The effect size (-0.0084 in the preferred model) is large (-30\% relative to the predicted value at the threshold for the untreated). The estimate is larger when the other estimators are used. The effects for fatalities are not statistically significant, which reflects a lack of statistical power stemming from a relatively small number of fatalities.

Panel D shows results by sex. The preferred estimator suggests that the effects are similar by sex, as do the results of the other estimators.

\setcounter{table}{0}
\renewcommand{\thetable}{E\arabic{table}}
\setcounter{figure}{0}
\renewcommand{\thefigure}{E\arabic{figure}}
\setcounter{equation}{0}
\renewcommand{\theequation}{E\arabic{equation}}

\section{Cost-benefit analysis}\label{append-cba}
Our cost-benefit analysis estimates the social benefits from reducing the probability of an MVA in the first 12 months of unsupervised driving. Since we find no evidence of reduced risk beyond this period, we assume those benefits are zero.

We proceed by first estimating the reduction in the rate of MVAs at the policy threshold for each MVA type (non-injury, injury and fatality) using our `best' model for each reform (see Table 6). Since our point estimates for fatalities are imprecise owing to low frequencies, we assume fatality risk decreased by the same percentage as injury risk in our baseline calculations. We also show how the estimates change if instead we assume no change in the fatality risk.

We multiply the MVA rates by the total social costs associated with each type of MVA under the assumption that no one is treated ($T=0$) and under the assumption that everyone is treated ($T=1$). The difference between these estimates is the total social benefit per person. Social benefits for non-injury MVAs are taken from BITRE (2009). Specifically, we use the average repair cost for MVAs. We therefore ignore other costs associated with these MVAs such as towing costs, time lost and administrative fees associated with, for example, insurance claims. On the other hand, the BITRE value includes all MVAs, even severe MVAs resulting in injury or fatality, and as such the repair costs may be overstated. Moreover, non-injury crashes comprise a negligible proportion of total social benefits so doubling or tripling this value has little material effect on the estimates. 

Social benefit estimates for crashes involving injury and fatality are taken from NRMA (2017). They are estimated using the willingness-to-pay method, which uses hypothetical scenario analysis to infer preferences. A strength of this approach is that it should capture all the information that goes into people's individual preferences. A drawback is that people may be unsure of their preferences, particularly if they have never experienced an MVA. This approach also ignores externalities, and may be subject to hypothetical bias. 

Table \ref{tbl-cba-2000} steps through our calculations for the 2000 (0 to 50 MSDH) reform. We have set out this table in such a way that it is easy to substitute our chosen social benefits parameters for other parameters, if desired.

\begin{table}[ht!]
	\caption{CBA -- 2000 reform (0 to 50 MSDH)}\label{tbl-cba-2000}
	\begin{tabular}{lcccc}
		\toprule
		& Non-injury & Injury & Fatality & Alt. fatality \\
		\cmidrule(lr){2-5}
		Policy effect & -0.0058238 & -0.008405 & -0.0002131 & -0.000101968 \\
		Prediction $T=0$ & 0.039571 & 0.0279324 & 0.000480 & 0.000480 \\
		Prediction $T=1$ & 0.0337472 & 0.0195274 & 0.0002669 & 0.000378 \\ 
		\% reduction & -15\% & -30\% & -44\% & -21\% \\
		\\
		Average social cost	& \$4,004 & \$144,172 & \$7,369,845 & \$7,369,845 \\
		\\
		Expected cost of driving $T=0$ & \$158.44 & \$4,027.06 & \$3,537.53 & \$3,537.53 \\
		Expected cost of driving $T=1$ & \$135.12 & \$2,815.30 & \$1,967.01 & \$2,786.04 \\
		\midrule
		Social saving per person & \$23.32 & \$1,211.76 & \$1,570.51 & \$751.49 \\
		\bottomrule
	\end{tabular}
	\begin{tablenotes}[para,flushleft]
		\footnotesize
		\item Notes: Policy effect estimates correspond to the `best' estimator in Table 6. $T$ means `treated' (i.e. subject to the 50 MSDH policy). Expected cost of driving equals the predicted MVA probability $\times$ the average social cost. All values are expressed in 2019 \$AUD (for references, in 2019 the \$AUD:\$USD exchange rate averaged around 0.7:1).  
	\end{tablenotes}
\end{table} 

Our preferred estimate for the average social benefit is the sum of the average social benefit due to reduced risk of non-injury MVAs (\$23), injury MVAs (\$1,212) and fatalities, assuming that this risk falls by the same percentage as injury MVAs (\$751). This implies an average social benefit of \$2,300. If we ignore the reduction in fatalities, we estimate an average social benefit of \$1,235. 

Under the assumption that on average learners complete 20 hours before the reform and 50 hours after the reform, our estimates imply a social benefit of between \$25-\$46 per hour. By way of comparison, the national minimum wage in 2019-2020 is \$19.49. Given that some supervised driving hours will be for trips that would have been taken anyway, and that there may be positive externalities to supervision (e.g. bonding), it seems likely, based on our estimates, that the 2000 reform was welfare improving.  

\bibliographystyle{apacite}
\bibliography{references}